\documentclass[12pt]{article}
\usepackage[utf8]{inputenc}
\usepackage{authblk}
\usepackage[T1]{fontenc}
\usepackage[a4paper,left=2cm,right=2cm,top=3cm,bottom=3cm]{geometry}
\usepackage[sorting=none]{biblatex}
\usepackage{bbold}
\usepackage[english]{babel}
\usepackage{xcolor}
\usepackage{libertine}
\usepackage{graphicx}
\usepackage{amsmath,amsfonts,amssymb,bm} 
\usepackage{bbm}
\usepackage{array}
\usepackage{enumerate}
\usepackage{verbatim,mathdots,thm-restate}
\usepackage{afterpage}
\usepackage{amsthm}
\usepackage{mathrsfs}
\usepackage{stmaryrd}
\usepackage[export]{adjustbox}
\usepackage{caption}
\usepackage{subcaption}
\usepackage{indentfirst}
\usepackage{slashed}
\usepackage{enumitem}
\usepackage[colorlinks,linkcolor=blue,citecolor=red]{hyperref}
\usepackage[toc]{appendix}
\usepackage{mathtools}

\graphicspath{{figures/}}

\def\R{{\mathbb R}}\def\C{{\mathbb C}}\def\Z{{\mathbb Z}}\def\N{{\mathbb N}}

\usepackage{fancyhdr}
\numberwithin{equation}{section}

\newtheorem{lemma}{Lemma}[section]
\newtheorem*{theorem*}{Nevanlinna-Sokal theorem}
\newtheorem{proposition}{Proposition}[section]
\newtheorem{theorem}{Theorem}[section]
\newtheorem*{remark}{Remark}

\title{Triviality vs perturbation theory: an analysis for mean-field $\varphi^4$-theory in four dimensions}

\author{
  Christoph KOPPER\thanks{christoph.kopper@polytechnique.edu}
  , Pierre WANG\thanks{pierre.wang@polytechnique.edu}
}

\affil{Centre de Physique Théorique (CPHT), CNRS, UMR 7644
Institut Polytechnique de Paris, 91128 Palaiseau, France}

\date{}

\begin{document}
\maketitle

\begin{abstract}
\noindent
 We have constructed the mean-field trivial solution of the $\varphi^4$ theory $O(N)$ model in four dimensions in \cite{mypaper} using the flow equations of the renormalization group. Here we establish a relation between the trivial solutions introduced in \cite{mypaper,Kopper2022} and perturbation theory. We show that if an UV-cutoff is maintained, we can define a renormalized coupling constant $g$ and obtain the perturbative solutions of the mean-field flow equations at each order in perturbation theory. We prove the local Borel-summability of the renormalized mean-field perturbation theory in the presence of an UV cutoff and show that it is asymptotic to the non-perturbative solution.  
\end{abstract}

\section{Introduction}

Perturbative expansions in quantum field theory are supposed to be divergent.
One
manifestation of this divergence is the presence of instanton singularities
which are related to the nontrivial minima of the classical action
as a function of the complex coupling
constant \cite{Lipatov}. In the  expansion in terms of Feynman diagrams,
the divergence is reflected by the fact that the number
of graphs at high orders in perturbation increases rapidly.
In theories like
$\varphi^4$, this number behaves as $K!$ where $\,K\,$
is the order of perturbation
theory. In this paper, we use the notation $\,\varphi^4_d$-theory
for $\,\varphi^4$
theory in $\,d\,$ dimensions. In four dimensions, another possible source of
divergence implied by the need of renormalization is the so-called renormalon
after 't Hooft\cite{tHooft1979}. This singularity is related to the presence
of Feynman graphs with a number of renormalization subtractions proportional
to the order of perturbation theory. For the $\varphi^4_4$-theory, graphs
with $N$ insertions of bubble subgraphs contributing to the
six-point function typically behave as $\,N!\,$, so that the perturbative expansion
is apparently divergent.

Nevertheless, the  $\varphi^4$ Schwinger functions can in some cases be
recovered from the perturbative expansion by Borel resummation.
In $\varphi^4_2$ models \cite{EckmannMagnen}, the $n$-point Schwinger functions
\begin{equation}\label{pert-series}
    S_n(g)\sim\sum_{m\geq 0}a_m\ g^m
\end{equation}
have a divergent perturbative expansion w.r.t. the coupling constant $\,g\,$,
i.e. $\,\vert a_m\vert \,=\, \mathcal{O}(m!)$
\cite{Glimmchap23}. The Borel transform of (\ref{pert-series}) is defined by
\begin{equation}\label{Boreltransform}
    h(t):=\sum_{n\geq 0} \dfrac{a_n}{n!}\ t^n\;.
\end{equation}
It has a finite radius of
convergence around $\,t=0\,$ and an analytic continuation
to a neighbourhood of the positive real $t$-axis.
The Schwinger functions are recovered via
\begin{equation}\label{Laplacetransform}
    S_n(g)=\int_0^{+\infty} e^{-t}\, h(gt)\, dt\;.
\end{equation}

In the seminal work of de Calan and Rivasseau \cite{CalanRiv},
it was proven that even in presence of the two mentioned sources
of divergence in $\varphi^4_4$-theory, the Borel transform of the
perturbative expansion has a finite radius of convergence,
i.e. the perturbative amplitudes at order $K$ do not grow more
rapidly than $ C^K\;K!\,$, where $\,C\,$ is a constant. One of their main
results is the fact that the number of graphs requiring $\,k\leq K\,$
renormalization subtractions is bounded by
\begin{equation}
    C^K\, \frac{K!}{k!}
\end{equation}
 so that the bound on the amplitudes is of the form
\begin{equation}
    C'^K\,  K!\;,
\end{equation}
where $\,C'\,$ is another constant. This implies the local convergence of
the Borel transform of the series. These bounds have been improved and
generalized in \cite{FeldmanMagnenRivSene}. Other results
include the local existence of the Borel transform for quantum electrodynamics (QED) \cite{QED}
and construction and local Borel summability of planar Euclidean
$\varphi^4_4$-theory \cite{RivasseauPlanar}.

The differential flow equations \cite{Wegner1972}  permit to prove perturbative
renormalizability of quantum field theories. Polchinski proved
the perturbative renormalizability of $\varphi^4_4$-theory with
these equations\cite{Polchinski1984}. Instead of analyzing Feynman diagrams,
he derived inductive bounds
on the  regularized perturbative Schwinger functions
with the aid of the flow equations. The bounds are sufficient to prove
renormalizability. With these techniques one can also
prove the renormalizability of
$SU(2)$ Yang-Mills 
theory with \cite{KOPPER_2009_SU} or without the Higgs mechanism
\cite{Efremov_2017} and perturbative renormalizability in 
Minkowski space \cite{keller1996minkowski}. Keller
\cite{Keller_local_Borel} proved in this way the local existence of the
Borel transform of the perturbation series. Kopper \cite{Kopper_Borel}
obtained bounds on the whole set of Schwinger functions
and their behavior at large momenta again implying the local existence of the
Borel transform. Recent results obtained with
the flow equations include the construction of asymptotically free
scalar field theories in the mean-field approximation \cite{Kopper2022},
a new construction of the massive Euclidean Gross-Neveu model
in two dimensions \cite{duch2024constructiongrossneveumodelusing},
a construction of a non-trivial fixed point of the Polchinski equation
for weakly interacting fermionic quantum field theories in $d$
dimensions ($d\in\lbrace 1,2,3\rbrace$) \cite{Greenblatt}, and
the triviality of mean-field $\,\varphi^4_4$-theories
\cite{mypaper,Kopper2022}. In \cite{mypaper}
mean-field $O(N)$ $\,\varphi^4_4$-theories were
constructed non-perturbatively with the flow equations and shown
to be trivial. The analysis of the triviality of
$\,\varphi^4_d\,$-theories goes back in particular
to Aizenman \cite{Aizenman1981}
who proved the triviality of the continuum limit of the lattice
$\,\varphi^4_d\,$-theory for $\,N=1\,$ in $\,d> 4\,$ dimensions. He derived a
crucial bound, called the tree-diagram bound based on random
current representation to obtain triviality. However,
the bound obtained in \cite{Aizenman1981} is not
sufficient  to prove triviality in $d=4$ dimensions.
Fröhlich \cite{Fröhlich1982} proved triviality for
$N\,\leq\,2\,$ and $\,d=4\,$ under the assumption of an infinite wavefunction
renormalization.  In 2021, Duminil-Copin and Aizenman \cite{Copin2021}
completed the triviality proof of $\,\varphi^4_4\,$-theory for $\,N=1\,$,
using a multi-scale analysis to improve the tree-diagram bound
\cite{Aizenman1981}.

The relation between perturbation theory and triviality is not obvious.
An indication of triviality of $\varphi^4_4$-theory is
the presence of the so-called Landau pole. The effective coupling
constant $g(\lambda)$ is a function of the energy scale $\lambda$.
Its behavior is described by the beta function defined by
\begin{equation}\label{beta}
  \beta(g(\lambda)):=\lambda\dfrac{dg}{d\lambda}(\lambda)\ ,
\end{equation}
where the derivative has to be taken at fixed bare coupling.
In practice $\beta(g(\lambda))$ can only be calculated to a
finite order in the perturbative expansion. For non-asymptotically
free theories such as QED or $\varphi^4$-theory, $\beta(g)$ is
positive at lowest order. To this approximation
the solution of (\ref{beta}) grows logarithmically with $\,\lambda\,$.
By extrapolation it diverges at a finite $\,\lambda_L\,$,
called the Landau pole. The location of this
singularity tends to infinity if the renormalized coupling  tends to zero.
The triviality proofs
\cite{mypaper,Aizenman1981,Fröhlich1982,Copin2021,Aizenman2021}
are non-perturbative, there is no assumption on the size of the bare coupling.
If the only renormalized theory that makes sense is the Gaussian one, then
perturbation theory seems irrelevant. But actually triviality does not
rule out the existence of a nontrivial renormalized perturbation theory.
A known model where the exact renormalized field theory is the free field
theory but with a nontrivial renormalized perturbation theory is the
Lee model \cite{LeePT}.
The interacting theory cannot be obtained by any limiting process if the bare
coupling is restricted to the real axis, but is rather obtained by taking limits of
non-hermitian hamiltonians, where the bare coupling is pure imaginary
and vanishes in the  ultraviolet (UV) limit.

In this paper, we are concerned with the relation of 
renormalized perturbation theory and triviality of Euclidean
$\,\varphi^4_4$-theory in the mean-field approximation.
Triviality was proven in
\cite{mypaper,Kopper2022}. 
Our main result  (see Theorems \ref{ThmBoreltransform} and \ref{NS})
  can be roughly stated as follows\,:\\[.1cm] 
  {\bf Theorem} 
    {\it We consider the Schwinger n-point
    functions of massive $\,\varphi_4^4$-theory
    in the mean-field approximation in the presence of an ultraviolet cutoff.
    These functions vanish in the UV-limit for $\,n \ge 4\,$. 
    For (relatively) small bare couplings 
    they can be expanded perturbatively
    w.r.t. a renormalized
    coupling $\,g\,$ related to the truncated four point function,
    which vanishes itself logarithmically in the UV
    limit. The perturbative series is asymptotic to the full
    solution. The Taylor remainder is bounded sufficiently
    strongly and can be continued analytically to a complex
    coupling constant domain, such that the assumptions of the
    Nevanlinna-Sokal Theorem are verified. Therefore the exact trivial
    solutions
    can be uniquely reconstructed from the perturbative series
    for any (sufficiently large)  finite value of the UV cutoff.}

Our paper is organized as follows.
In Sect.\ref{FEs_MFA} we introduce the flow equations and analyze them
in the mean-field approximation. We implement the perturbative expansion
and derive bounds on the perturbative mean-field  Schwinger functions
and their derivatives w.r.t. the logarithmic energy scale.
In Sect.\ref{RelaTriPert} we relate the ansatz for the trivial
solutions studied in \cite{mypaper,Kopper2022} to perturbation theory.
In Sect. \ref{PropTri} we establish a number of bounds and analyticity properties
for the trivial solutions  at the bare scale. 
In Sect.\ref{BPHZrenormalization} we show how to relate the bare two point
function to its value at the physical scale.  
This is a prerequisite  to explicit the renormalization conditions
verified by the trivial solutions.
In Sect.\ref{PerturbExpaning} we analyze the trivial solutions
close to the renormalization scale and verify how the perturbative
expansion can be implemented.
Finally in Sect.\ref{LocBoreSummRenor} we prove local Borel summability
of the regularized renormalized mean-field perturbation theory.
In Sect.\ref{PertMFARem} we introduce the remainders of the perturbative
expansion of the mean-field Schwinger functions and the mean-field flow
equations for the remainders. In addition, we present bounds on the
Taylor remainders for the mean-field two point function. The latter
are crucial to start the inductive scheme that arises in the mean-field
flow equations for the remainders. In Sect.\ref{LocBoreSummRenordef} we
introduce the notion of local Borel summability.
In Sect.\ref{Positiverenormcoupl} we establish the bounds required
for local Borel summability restricting to real couplings and show
that perturbation theory is
asymptotic to the trivial
solution by bounding the Taylor remainders, i.e.  the difference
between the full (trivial) solutions and the truncated perturbative
series, via the flow equations.
Then we show that we can
analytically continue the Schwinger functions
to complex values of the renormalized coupling.
This  establishes local Borel
summability of the perturbative regularized renormalized mean-field
Schwinger functions in the sense of the Nevanlinna-Sokal theorem.


\section{The flow equations and the mean-field approximation}\label{FEs_MFA}

\subsection{The flow equations}\label{FEs}

We consider a theory with a real
one-component self-interacting scalar field $\varphi$ in the
four-dimensional Euclidean space with $\Z_2$ symmetry $\varphi\mapsto-\varphi$.
We adopt the following convention and shorthand notation for the
Fourier transform
\[
f(x)=\int_p e^{ipx}\hat{f}(p),\quad \int_p:=\int\frac{d^4p}{(2\pi)^4}\ .
\]
Therefore the functional derivative $\frac{\delta}{\delta\varphi(x)}$ reads
\[
\dfrac{\delta}{\delta\varphi(x)}
= (2\pi)^4\int_p e^{-ipx}\dfrac{\delta}{\delta\hat{\varphi}(p)}\ .
\]

First, we introduce a regularized propagator in momentum space.
In \cite{Mueller2003}, Müller listed possible choices
for the regularized propagator. Here we choose
as in \cite{Kopper2022,mypaper}
\begin{equation}\label{Propagator_reg_1}
  C^{\alpha_0,\alpha}(p,m):=\dfrac{1}{p^2+m^2}
  \Big(\exp (-\alpha_0(p^2+m^2))-\exp (-\alpha(p^2+m^2))\Big)\;,
\end{equation}
where $m$ is the mass parameter of the field, $\alpha_0>0$ acts
as an ultraviolet cutoff, and $\alpha\in[\alpha_0,+\infty)$ is the flow
  parameter. The regularized propagator (\ref{Propagator_reg_1}) is
  positive and analytic w.r.t. $\alpha$. By taking the limits
  $\alpha_0\to  0$ and $\alpha\to +\infty\,$
  we recover the usual Euclidean propagator in momentum space 
\begin{equation}
  \lim\limits_{\alpha\to +\infty}\,\lim\limits_{\alpha_0\to 0}
  C^{\alpha_0,\alpha}(p,m)=\frac{1}{p^2+m^2}\ .
\end{equation}
We consider bare interaction lagrangians of the form
\begin{equation}\label{interaction_lagrangian}
  L_0^\mathcal{V}(\varphi)=\int_\mathcal{V}
  d^4 x\Big(b_0(\alpha_0)
  (\partial\varphi(x))^2+\sum_{n\in 2\N}c_{0,n}(\alpha_0)\varphi^n(x)\Big)\;,
\end{equation}
where $(\partial \varphi(x))^2=\sum_{\mu=0}^3
(\partial_\mu\varphi(x))^2$ and $\mathcal{V}$ is a finite volume in $\R^4$.
The constants $b_0(\alpha_0)$, $c_{0,n}(\alpha_0)$ are called the bare
couplings.
The quantities in the sum for $n\geq 6$ are the irrelevant terms
while $b_0(\alpha_0),\, c_{0,2}(\alpha_0)$ and $c_{0,4}(\alpha_0)$ are
respectively relevant and marginal terms. They diverge when
$\alpha_0\rightarrow 0$ but they are required to make the renormalized
physical quantities, i.e., the renormalized mass or the renormalized
coupling constant finite upon removing the UV cutoff. They should be
such that for some constant $C^\mathcal{V}\in\R$, depending on $\mathcal{V}$
\begin{equation}
  -\infty<C^\mathcal{V}<L_0^\mathcal{V}(\varphi)<+\infty\;,
  \quad \varphi\in\mbox{supp}(\mu^{\alpha_0,\alpha})\;,
\end{equation}
where $\mu^{\alpha_0,\alpha}$ designates the unique Gaussian measure
associated to the propagator $C^{\alpha_0,\alpha}$. We suppose that
the field $\varphi$ is in the support of the Gaussian measure
$\mu^{\alpha_0,\alpha}$. Since the regularized propagator
$C^{\alpha_0,\alpha}(p,m)$ falls off exponentially in $p^2$ in
momentum space, the support of the Gaussian measure
$\mu^{\alpha_0,\alpha}$ is within the set of functions smooth
in position space, see e.g. \cite{Reed1973}, so that
the products of the fields and the derivatives of the fields in
$L_0^\mathcal{V}$ i.e. $\varphi^2(x),\, \varphi^4(x),\cdots$ are well-defined.

We define the regularized  correlation (or Schwinger) functions in
finite volume by
\begin{equation}
  \langle \varphi(x_1)\cdots\varphi(x_n)\rangle^{\alpha,\alpha_0}_\mathcal{V}
  :=\dfrac{1}{Z^{\alpha,\alpha_0}_\mathcal{V}}
  \int d\mu^{\alpha_0,\alpha}(\varphi)e^{-L_0^\mathcal{V}(\varphi)}
  \varphi(x_1)\cdots\varphi(x_n)\ .
\end{equation}
The normalization factor $Z^{\alpha,\alpha_0}_\mathcal{V}$ is chosen
so that $\langle 1\rangle=1$. The generating functional
of the regularized connected amputated Schwinger functions (CAS) is given
by
\begin{equation}\label{generatingfunctionalKopper}
  e^{-L^{\alpha_0,\alpha}_{\mathcal{V}}(\varphi)}
  :=\dfrac{1}{Z^{\alpha,\alpha_0}_\mathcal{V}}
  \int d\mu^{\alpha_0,\alpha}(\phi)e^{-L_0^{\mathcal{V}}(\varphi+\phi)}\ .
\end{equation}

The flow equations are obtained by taking the $\alpha$-derivative
of the generating functional of the CAS functions. Using the
infinitesimal change of covariance formula in Appendix \ref{appendix_A},
we obtain
\begin{equation}\label{firststep}
\begin{split}
  \partial_\alpha e^{-L_\mathcal{V}^{\alpha_0,\alpha}(\varphi)}
  &=\dfrac{1}{2}\dfrac{1}{Z_\mathcal{V}^{\alpha_0,\alpha}}
  \int d\mu^{\alpha_0,\alpha}(\phi)\Big\langle\dfrac{\delta}{\delta\phi},
  \Dot{C}^\alpha\dfrac{\delta}{\delta\phi}\Big\rangle
  \ e^{-L_0^\mathcal{V}(\phi+\varphi)}
  -\partial_\alpha\log ( Z_\mathcal{V}^{\alpha_0,\alpha})
  \ e^{-L_\mathcal{V}^{\alpha_0,\alpha}(\varphi)} \\
  &= \dfrac{1}{2}\Big\langle\dfrac{\delta}{\delta\varphi},
  \Dot{C}^\alpha\dfrac{\delta}{\delta\varphi}
  \Big\rangle \ e^{-L_\mathcal{V}^{\alpha_0,\alpha}(\varphi)}
  -\partial_\alpha\log ( Z_\mathcal{V}^{\alpha_0,\alpha})\
  e^{-L_\mathcal{V}^{\alpha_0,\alpha}(\varphi)}
\end{split}
\end{equation}
with $\Dot{C}^\alpha:=\partial_\alpha C^{\alpha_0,\alpha}$. In the second step,
we used the fact that $L_0^\mathcal{V}$ depends only on the sum $\phi+\varphi$.
Performing the derivatives on both sides of (\ref{firststep}) gives
the Wilson-Wegner flow equation \cite{Wegner1972} 
\begin{equation}\label{FE-1} 
\begin{split}
  \partial_\alpha L_\mathcal{V}^{\alpha_0,\alpha}
  &=\dfrac{1}{2}\Bigg\langle\dfrac{\delta}{\delta\varphi},
  \Dot{C}^\alpha\dfrac{\delta}{\delta\varphi}
  \Bigg\rangle L_\mathcal{V}^{\alpha_0,\alpha}
  -\dfrac{1}{2}\Bigg\langle\dfrac{\delta}{\delta\varphi}
  L_\mathcal{V}^{\alpha_0,\alpha},
  \Dot{C}^\alpha\dfrac{\delta}{\delta\varphi}
  L_\mathcal{V}^{\alpha_0,\alpha}\Bigg\rangle
  +\partial_\alpha\log (Z_\mathcal{V}^{\alpha_0,\alpha})\;.
\end{split}
\end{equation}

We expand the CAS functions in a formal power series in $\hat{\varphi}$
\begin{equation}\label{momentum_expansion}
  L^{\alpha_0,\alpha}_\mathcal{V}(\varphi)
  =\sum_{n\in 2\N}\int_{p_1,\cdots, p_n}
  \bar{\mathcal{L}}^{\alpha_0,\alpha}_{n,\mathcal{V}}(p_1,\cdots,p_n)\hat{\varphi}(p_1)
  \cdots\hat{\varphi}(p_n)\;.
\end{equation}
In the infinite volume limit the moments
$\bar{\mathcal{L}}^{\alpha_0,\alpha}_{n,\mathcal{V}}$
become distributions. Due to translation invariance
they take then the form (\ref{distrib}) below,
where the ${\mathcal{L}}^{\alpha_0,\alpha}_{n}\,$
are smooth for finite regulators.
Müller\cite{Mueller2003} discussed the infinite volume limit of
(\ref{momentum_expansion}) in more detail.
Subsequently we will drop the subscript $\mathcal{V}\,$,
meaning that we have taken the infinite-volume limit.
So we factorize the infinite volume CAS functions,
 as
\begin{equation}
  \bar{\mathcal{L}}^{\alpha_0,\alpha}_{n}(p_1,\cdots,p_n)
  =\delta^4\Big(\sum_{i=1}^n p_i\Big)
  \mathcal{L}^{\alpha_0,\alpha}_{n}(p_1,\cdots,p_n),\quad p_n=-p_1-\cdots-p_{n-1}\;.
\label{distrib}
\end{equation}
The CAS functions $\mathcal{L}^{\alpha_0,\alpha}_n(p_1,\cdots,p_n)$
are obtained via successive functional derivatives
\begin{equation}
  \dfrac{(2\pi)^{4n}}{n!}\dfrac{\delta}{\delta\hat{\varphi}(p_1)}
  \cdots\dfrac{\delta}{\delta\hat{\varphi}(p_n)}\,
  L^{\alpha_0,\alpha}(\varphi)\vert_{\varphi=0}
  =\delta^4\Big(\sum_{i=1}^n p_i\Big)
  \mathcal{L}^{\alpha_0,\alpha}_n(p_1,\cdots,p_n)\ .
\end{equation}
They are symmetric
under any permutation of the set of the external momenta,
Using (\ref{momentum_expansion}) in (\ref{FE-1}),
we obtain the flow equations in an expanded form as
\begin{equation}\label{FE-2}
        \begin{split}
          &\partial_\alpha \mathcal{L}^{\alpha_0,\alpha}_n(p_1,\cdots,p_n)
          = \binom{n+2}{2}\int_k
          \Dot{C}^\alpha (k,m)\mathcal{L}^{\alpha_0,\alpha}_{n+2}
          (k,-k,p_1,\cdots,p_n) \\
          &-\dfrac{1}{2}\sum_{n_1+n_2=n+2}n_1 n_2\ \mathbb{S}
          \Big(\mathcal{L}^{\alpha_0,\alpha}_{n_1}(p_1,\cdots,p_{n_1-1},q)
          \Dot{C}^\alpha (q,m)\mathcal{L}^{\alpha_0,\alpha}_{n_2}(-q,p_{n_1},
          \cdots,p_{n})\Big)\;,
\end{split}
\end{equation}
with  $q=p_{n_1}+\cdots+p_n=-p_1-\cdots-p_{n_1-1}$. $\mathbb{S}$ is the
symmetrisation operator averaging over the permutations $\sigma$ such
that $\sigma(1)<\sigma(2)<\cdots<\sigma(n_1-1)$ and
$\sigma(n_1)<\sigma(n_1+1)<\cdots<\sigma(n)\,$. Since we considered
a theory with a $\mathbb{Z}_2$-symmetry , only even moments
($n,n_1$ and $n_2\in 2\N$) are nonvanishing as the regularization
does not break this symmetry. The flow equations are an infinite
system of non-linear differential equations, the solutions of
which are the CAS functions. By imposing boundary conditions
for the relevant parameters at the renormalization scale, one
can prove the perturbative renormalizability of the regularized
theory through an inductive scheme which arises from the flow equations,
see \cite{Mueller2003}.

\subsection{The mean-field flow equations}\label{MFFEs}

The mean-field approximation is a tool to simplify the system (\ref{FE-2})
by neglecting fluctuations of the field variable.
Even if this approximation appears to be very drastic at first sight, 
we recall that in statistical physics the mean-field approximation describes
exactly the critical behavior in $\,d>4\,$ dimensions (Ginzburg criterion)
\cite{Fröhlich1982,Aizenman2021}. So essential aspects of the
theory are preserved in this approximation. When fluctuations
are neglected, the $\,n$-point functions $\,\mathcal{L}_n\,$
become  momentum independent
densities. In fact the mean-field flow equations are obtained
by setting all momenta to zero \cite{Kopper2022}. We write
\begin{equation}\label{Ais}
A_n^{\alpha_0,\alpha}=\mathcal{L}_n^{\alpha_0,\alpha}(0,\cdots,0)\;.
\end{equation}
The mean field effective action $L_{mf}^{\alpha_0,\alpha}(\phi)$ takes
the form of a (a priori formal) power series
in the constant (mean) field variable $\,\phi\in\R\,$
\begin{equation}\label{Mean_effective_action}
    L_{mf}^{\alpha_0,\alpha}(\phi)=\sum_{n\in 2\N}A_n^{\alpha_0,\alpha}\phi^n\;.
\end{equation}
An additional technical simplification introduced
in \cite{Kopper2022}
is to set the mass $m\,$ in the propagator $C^{\alpha_0,\alpha}(k,m)$
equal to zero, and to
analyze the theory in the interval
\[
\alpha\in [\alpha_0,\alpha_{\max}]\ ,\quad \alpha_{\max}:=\frac{1}{m^2}\ .
\]
The infrared cutoff $\alpha_{\max}$ then takes the same role
as the infrared cutoff $\,\frac{1}{m^{2}}\,$ in the original theory.
This technical simplification does not change the triviality result,
see \cite{mypaper}.
In this paper we do not study the infrared problem. So we consider
$\,\alpha_{\max}\,$ to be fixed, and we choose units such that 
\[
\alpha_{\max}\,=\, 1\ .
\]
In the mean-field limit the flow equations (\ref{FE-2})
become\cite{Kopper2022}
\begin{equation}\label{MFE-1}
  \partial_\alpha A_n^{\alpha_0,\alpha}=\binom{n+2}{2}c_\alpha\
  A^{\alpha_0,\alpha}_{n+2}-\dfrac{1}{2}\sum_{n_1+n_2=n+2}n_1\,n_2\
  A^{\alpha_0,\alpha}_{n_1}\,A^{\alpha_0,\alpha}_{n_2}\ ,
  \quad\alpha\in[\alpha_0,\alpha_{\max}]\ ,
\end{equation}
where $\,c_\alpha:=\frac{c}{\alpha^2}\,$ with
$\,c:=\frac{1}{16\,\pi^2}\ $. Setting
\begin{equation}\label{definition_cal_A}
\mathcal{A}_n^{\alpha_0,\alpha}:=c^{\frac{n}{2}-1}\;n\;A_n^{\alpha_0,\alpha}\;
\end{equation}
we can rewrite (\ref{MFE-1}) as
\begin{equation}\label{MFE-1bis}
  \partial_\alpha\ \mathcal{A}_n^{\alpha_0,\alpha}=\binom{n+2}{2}\frac{1}{\alpha^2}\
  \mathcal{A}^{\alpha_0,\alpha}_{n+2}-\dfrac{1}{2}\sum_{n_1+n_2=n+2}n_1n_2\
  \mathcal{A}^{\alpha_0,\alpha}_{n_1}\,\mathcal{A}^{\alpha_0,\alpha}_{n_2}\;,
  \quad\alpha\in[\alpha_0,\alpha_{\max}]\ .
\end{equation}

The mean-field flow equations (\ref{MFE-1bis}) can  be analyzed 
\cite{mypaper} as follows:
\begin{itemize}
 \item Fix a bare interaction lagrangian with the mean-field boundary
   conditions corresponding to (\ref{interaction_lagrangian}).
This means we study bare interaction lagrangians without irrelevant terms,
 i.e. setting $\,c_{0,n}=0,\,\  n\geq 6\,$  of the form
\begin{equation}\label{bare_lagrangian_trivial}
  L_{mf}^{\alpha_0,\alpha}(\phi)\,=\,
   c_{0,2}\,\phi^2 \,+\, c_{0,4}\,\phi^4 \ ,
\end{equation}
and the following mean-field boundary conditions following
from (\ref{Ais}), (\ref{Mean_effective_action}),
(\ref{definition_cal_A}),
 (\ref{bare_lagrangian_trivial}):
\begin{equation}\label{BDY_trivial_field}
  \mathcal{A}^{\alpha_0,\alpha_0}_{2} =2(2\pi)^4\, c_{0,2}\;,
  \quad \mathcal{A}^{\alpha_0,\alpha_0}_{4} =4\pi^2\, c_{0,4}\;,
  \quad \mathcal{A}^{\alpha_0,\alpha_0}_{n}=0\ ,\quad n\geq 6\ .
\end{equation}

 \item Define an ansatz for the two pointfunction
   $\mathcal{A}^{\alpha_0,\alpha}_{2}$  and use
   the mean-field flow equations to construct inductively
   smooth solutions $\mathcal{A}^{\alpha_0,\alpha}_{n}$, $\ n\geq 4\ $.
 \end{itemize}


\subsection{The
  perturbative mean-field flow equations}
\label{PertMFA}

In perturbative quantum field theory the Schwinger functions 
are expanded in formal power series w.r.t.
one (or several) renormalized coupling(s) $\,g\,$.
The objects analyzed in Polchinski's flow equation
framework are the connected amputated Schwinger functions
(CAS). The implementation of the perturbative expansion in the flow equation
framework requires boundary conditions which are compatible
with the expansion. For a detailed analysis see
\cite{Mueller2003} and references given there.
In fact the boundary value problem is of mixed type.
According to  (\ref{BDY_trivial_field}) we impose 
\begin{equation}\label{bc}
\qquad \qquad \quad \mbox{ at}\quad  \alpha \,=\, \alpha_0\ : \qquad\qquad
\mathcal{A}^{\alpha_0,\alpha_0}_{n}=0\ ,\quad n\geq 6
\qquad \qquad \qquad \qquad \qquad \quad\, \ .\
\end{equation}
At the renormalization scale we impose
\begin{equation}\label{BPHZgen}
\mbox{at}\quad   \alpha \,=\, \alpha_{max}\,=\,1\ : \quad
 \mathcal{A}_2^{\alpha_{0},1} \,=\,
  \sum_{j=1}^\infty g^j\, \mathcal{A}_{j}\ , \quad
  \mathcal{A}_4^{\alpha_{0},1} \,=\,
  \sum_{j=1}^\infty g^j\, \mathcal{B}_{j}  \ .
\end{equation}
Note that
the perturbative expansion of $\,c_{0,2}\,, \  c_{0,4}\,$  from
(\ref{bare_lagrangian_trivial})
then follows from (\ref{bc}), (\ref{BPHZgen}) and the perturbative
flow equations (\ref{MFE-1bis}).

In renormalized perturbation theory one generally
proves that the perturbative series exists as a
well-defined formal power in the limit when the UV cutoff
is sent to infinity. In this case it is only required
that the coefficients $\,\mathcal{A}_{j}\,,\ \mathcal{B}_{j}\,$
are finite. Since we also want to prove bounds within
and beyond perturbation theory, we will always suppose that
\begin{equation}\label{BPHZgenbd}
  |\mathcal{A}_{j}|\ \le\  K_1\  c^j \ ,
  \quad
  |\mathcal{B}_{j}|\ \le\  K_2\  c^j 
  \end{equation}
for suitable positive constants $\,K_1,\,K_2,\,c\,$.
A particularly simple choice are BPHZ (Bogoliubov-Parasiuk-Hepp-Zimmermann) type conditions
\begin{equation}\label{BPHZ}
  \mathcal{A}_{j}\,=\,0  \ ,\qquad \mathcal{B}_{j}\,=\, \delta_{j,1} 
  \end{equation}
which define the renormalized coupling directly in terms of the
truncated four point function.
Compatibility of perturbation theory with the flow equation
only requires the series in (\ref{BPHZgen}) to start at $\,j \ge 1\,$.
The renormalized coupling $\,g\,$ will be defined
later in (\ref{g}) when we relate
perturbation theory to the trivial mean-field solutions. 

We will now present bounds on the perturbative CAS
based on \cite{Kopper_Borel}. They are not really new,
but presented in a form adapted to the mean-field approximation,
and extended to the CAS derived arbitrarily often w.r.t.
the flow parameter.
Using our boundary conditions
one can consistently expand all CAS 
$\mathcal{A}_n^{\alpha_{0},\alpha}$
 in formal power series w.r.t. $g$
\begin{equation}\label{perturbative_exp}
  \mathcal{A}_n^{\alpha_0,\alpha}
  =\sum_{j=1}^\infty  g^j\  \mathcal{A}_{n,j}^{\alpha_0,\alpha}\ .
\end{equation}
The system of mean-field
flow equations for the $\,\mathcal{A}_{n,j}^{\alpha_0,\alpha}\,$
is obtained by inserting  (\ref{perturbative_exp})
in (\ref{MFE-1bis})
\begin{equation}\label{MFE2_perturbative_part}
  \partial_\alpha \mathcal{A}_{n,j}^{\alpha_0,\alpha}
 \, =\, \frac{n(n+1)}{2\alpha^2}\ \mathcal{A}_{n+2,j}^{\alpha_0,\alpha}
  \ -\ \frac{n}{2}\ \sum_{\substack{
      n_1+n_2=n+2 \\ j_1+j_2=j \\ 2j_i+2\geq n_i   
  }} \mathcal{A}_{n_1,j_1}^{\alpha_0,\alpha}\
  \mathcal{A}_{n_2,j_2}^{\alpha_0,\alpha}\ .
\end{equation} 
 
Using the perturbative flow
equations (\ref{MFE2_perturbative_part}),
one can derive bounds on the
functions $\,\mathcal{A}_{n,j}^{\alpha_0,\alpha}\,$.
For $\,n\geq 6,\  j\geq 1\,$, we 
integrate the flow equations upwards from $\alpha_0$ to $\alpha\,$,
using that 
\begin{equation}\label{BDY_perturb_irrelevant}
  \mathcal{A}_{n,j}^{\alpha_0,\alpha_0}=0\;,\quad n\geq 6,\ \, j\geq 1 \ .
\end{equation}
For $\,n \le 4\,$ the flow equations are integrated downwards
from renormalization scale $\alpha_{\max}\,=1\,$ to $\alpha\,$,
with boundary conditions
(\ref{BPHZgen})  
\begin{equation}\label{BDY_perturb_relevant}
  \mathcal{A}_{2,j}^{\alpha_0,1}=\mathcal{A}_{j} \ ,
  \quad \mathcal{A}_{4,j}^{\alpha_0,1}= \mathcal{B}_{j}\ ,
  \quad  j\,\geq\, 1\ .
\end{equation}
The functions $\mathcal{A}_{n,j}^{\alpha_0,\alpha}$ can also be shown
inductively to satisfy 
\begin{itemize}
\item $\mathcal{A}_{n,j}^{\alpha_0,\alpha}\equiv 0\,,\ $  if
  $\,n\,$ is odd ($\,\Z_2$-symmetry).

\item $\mathcal{A}_{n,j}^{\alpha_0,\alpha}\equiv 0\,,\ $ if $\,n>2j+2\ $
      expressing the fact that only connected amplitudes contribute.
      
\item $\mathcal{A}_{n,j}^{\alpha_0,\alpha} \in C^\infty[\alpha_0,\,1]\ $.
\end{itemize}


\subsection{Bounds on the perturbative mean-field solutions close to the
renormalization scale}
\label{PertMFAbd}

\begin{proposition}\label{Prop_bound_perturbative_solutions_simplified}
  Let $\mathcal{A}_{n,j}^{\alpha_0,\alpha}$ be  solutions of
  the mean-field flow equations (\ref{MFE2_perturbative_part})
  for the boundary conditions (\ref{BDY_perturb_irrelevant}) and the
  renormalization conditions (\ref{BDY_perturb_relevant}).
  For $\,\alpha\in [e^{-1},\,1]\,$.
  They  satisfy the bounds
\begin{equation}\label{bound_perturbative_solutionsn>=2useful}
\Bigl\vert\,\partial_\alpha^k
        \mathcal{A}_{n,j}^{\alpha_0,\alpha} \Bigr\vert\,\leq\,
        \alpha^{\frac{n}{2}-2-k}\ 
 C^{j-\frac{n}{4}+k}\ \dfrac{(j+k+1)!\;  }{(k+1)^2\;(\frac{n}{2})^2\;
          (\frac{n}{2})! }
        \end{equation}
for a suitable constant $\,C>1\,$.
\end{proposition}
\noindent
Proposition \ref{Prop_bound_perturbative_solutions_simplified} follows from 
\begin{restatable}{lemma}{perturbativebounds}
  \label{Prop_bound_perturbative_solutions}
  Let $\,\mathcal{A}_{n,j}^{\alpha_0,\alpha}$ be solutions of
  the mean-field flow equations (\ref{MFE2_perturbative_part})
  for the boundary conditions (\ref{BDY_perturb_irrelevant}) and the
  BPHZ renormalization conditions (\ref{BDY_perturb_relevant}).
  For $\, \alpha\in [\alpha_0,\,1]\,$ 
  they satisfy for  a suitable constant $\,C>1\,$ and $\, k \geq 1\,$
  the bounds 
\begin{equation}\label{bound_perturbative_solutionsn=2}
\begin{split}
  &\Big\vert \mathcal{A}_{2,j}^{\alpha_0,\alpha}\Big\vert\,\leq\,
  \frac{C^{j-\frac{1}{2}}}{\alpha}\
  \dfrac{j!\;  }{(j+1)^2 }\ 
  \sum_{\lambda=0}^{j-1}\dfrac{1}{2^{\lambda}\lambda !}\
  (1-\ln (m^2\alpha))^{\lambda}\;, \\
  & \Big\vert\partial_\alpha^k \ \mathcal{A}_{2,j}^{\alpha_0,\alpha}\Big\vert
  \,\leq\,
  \dfrac{C^{j-\frac{1}{2}+k}}{\alpha^{k+1}}\
  \dfrac{(j+k+1)!\;  }{(j+1)^2\;(k+1)^2  }\ 
  \sum_{\lambda=0}^{j-1}\dfrac{1}{2^{\lambda}\lambda !}\ 
  (1-\ln (m^2\alpha))^{\lambda}\ , \\
\end{split}
\end{equation}
and for $n\geq 4$
\begin{equation}\label{bound_perturbative_solutionsn>=4}
\begin{split}
  &\Big\vert \mathcal{A}_{n,j}^{\alpha_0,\alpha}\Big\vert
  \leq \alpha^{\frac{n}{2}-2} \ C^{j-\frac{n}{4}}\dfrac{j!\;  }{(j-\frac{n}{2}+2)^2\;
   \ (\frac{n}{2})^2(\frac{n}{2})! }
  \sum_{\lambda=0}^{j-\frac{n}{2}+1}\dfrac{1}{2^{\lambda}\lambda !}\
  (1-\ln (m^2\alpha))^{\lambda}, \\
  & \Big\vert\partial_\alpha^k \mathcal{A}_{n,j}^{\alpha_0,\alpha}\Big\vert
  \leq \alpha^{\frac{n}{2}-2-k}\  C^{j-\frac{n}{4}+k}\
  \dfrac{(j+k+1)!\;}{(j-\frac{n}{2}+2)^2\;(k+1)^2
    (\frac{n}{2})^2(\frac{n}{2})! }\
       \sum_{\lambda=0}^{j-\frac{n}{2}+1}\dfrac{1}{2^{\lambda}\lambda !}
       (1-\ln( m^2\alpha))^{\lambda}\ \,.
\end{split}
\end{equation}
\end{restatable}

\begin{proof}
  See \cite{Kopper_Borel} for the case $\,k=0\,$, and for
  the general case $\,k\geq 0\,$, see Appendix \ref{proofperturbativebounds}.
  We remark that the proof in \cite{Kopper_Borel} is written
  for constants $\,\mathcal{A}_j\,=0\,$ and
  $\,\mathcal{B}_j\,=\,  \delta_{j,1}\,$.
  In the proof these constants appear  as integration
  constants 
\begin{equation}
  \mathcal{A}_{2,j}^{\alpha_0,\alpha}\,=\,\mathcal{A}_j
  \,+\,\int_\alpha^{1} d\alpha'\,
  \partial_\alpha\mathcal{A}_{2,j}^{\alpha_0,\alpha'}\ ,
  \quad\mathcal{A}_{4,j}^{\alpha_0,\alpha}\,=\,\mathcal{B}_j
  \,+\,\int_\alpha^{1}
  d\alpha'\, \partial_\alpha\mathcal{A}_{4,j}^{\alpha_0,\alpha'}\ .
\end{equation}
Since  $\,\mathcal{A}_j\,,\ \mathcal{B}_j\,$ obeying
(\ref{BPHZgenbd}) are majorized by our bounds
(\ref{bound_perturbative_solutionsn=2}),
(\ref{bound_perturbative_solutionsn>=4}) for $\,C\,$ large enough,
those bounds can be straightforwardly verified to hold
also for renormalization conditions (\ref{BPHZgenbd}).
\end{proof}
Using Lemma
\ref{Prop_bound_perturbative_solutions}, we can also bound the
derivatives of $\mathcal{A}_{n,j}^{\alpha_0,\alpha}$ w.r.t. $\,\mu\,$, using
standard techniques.
\begin{restatable}{lemma}{perturbativeboundsmu}
  \label{Prop_bound_pertu_mu_gn}
  Under the same assumptions as in Lemma
  \ref{Prop_bound_perturbative_solutions}
  and for $\mu\in[0,\mu_{\max}]$, there exists a constant $C'>1$
  such that the
smooth perturbative solutions $\mathcal{A}_{n,j}^{\alpha_0,\alpha}$
satisfy the bounds
\begin{equation}\label{bound_perturbative_solutions_mu}
  \Big\vert\partial_\mu^m \mathcal{A}_{n,j}^{\alpha_0,\alpha}\Big\vert
  \,\leq\, (\alpha_0\ e^{\mu})^{\frac{n}{2}-2}\ \ C'^{j+\frac{n}{2}+m}\
  \dfrac{(j+m+1)!
    }{(j-\frac{n}{2}+2)^2\ 
    (\frac{n}{2})^2(\frac{n}{2})!}\ \mathcal{F}(j,n,\mu)\ ,\quad m\geq 1\ ,
\end{equation}
where we define 
\begin{equation}
  \mathcal{F}(j,n,\mu):=\sum_{\lambda=0}^{j-\frac{n}{2}+\hat\theta(n)}
  \dfrac{1}{2^{\lambda}\lambda !}\ (1+\mu_{max}-\mu)^{\lambda}\;,\quad
    \hat\theta(n):=\left\{
       \begin{array}{ll}
        1  & \ \mbox{ if }\  n\geq 4 \\
       0 & \ \mbox{ if }\  n=2\ \ .
 \end{array}\right.
\end{equation}
\end{restatable}
\begin{proof}
    See Appendix \ref{proofperturbativebounds}.
\end{proof}


\subsection{Rescaled perturbative mean-field flow equations
  \label{PertMFA2}}

We may scale out the mass dimension of the functions  
 $\,\mathcal{A}_{n}^{\alpha_0,\alpha}\,$ by setting 
\begin{equation}\label{f_n}
 \! \!\! f_n(\mu):=\, \alpha^{2-\frac{n}{2}}\
 \mathcal{A}_n^{\alpha_0,\alpha}\,=\,\alpha^{2-\frac{n}{2}}\, c^{\frac{n}{2}-1}\, n\
{A}_n^{\alpha_0,\alpha}\ ,  \ \,
 f_{n,j}(\mu):=\, \alpha^{2-\frac{n}{2}}\
 \mathcal{A}_{n,j}^{\alpha_0,\alpha}\ ,
 \ \   \mu:=\ln\Bigl(\frac{\alpha}{\alpha_0}\Bigr)\ .
\end{equation}
The mean-field flow equations can be written
equivalently in terms of the functions $\,f_n(\mu)\,$
\begin{equation}\label{MFE-2}
  f_{n+2}(\mu)=\dfrac{1}{n+1}\sum_{n_1+n_2=n+2}f_{n_1}(\mu)f_{n_2}(\mu)
  +\dfrac{n-4}{n(n+1)}f_n(\mu)+\dfrac{2}{n(n+1)}\partial_\mu f_n(\mu)\;,
  \quad \mu\in[0,\mu_{\max}]\;,
\end{equation}
where
\begin{equation}
\mu_{\max}:=\ln\bigl(\frac{1}{\alpha_0}\bigr)\ .
\end{equation}

The (rescaled) perturbative amplitudes $f_{n,j}(\mu)$ satisfy the perturbative mean-field flow equations

\begin{equation}\label{pertMFE-2}
\begin{split}
 f_{n+2,j}(\mu) =\dfrac{1}{n+1}\sum_{\substack{n_1+n_2=n+2\\ j_1+j_2=j}}f_{n_1,j_1}(\mu)f_{n_2,j_2}(\mu)
  +\dfrac{n-4}{n(n+1)}f_{n,j}(\mu)
  +\dfrac{2}{n(n+1)}\partial_\mu f_{n,j}(\mu)\;.
\end{split}
\end{equation}
As a consequence of Lemma \ref{Prop_bound_pertu_mu_gn}
we directly find for the perturbative rescaled functions
$\,f_{n,j}(\mu)\,$
 \begin{proposition}\label{bound_pertur_mu_simpl}
  For $\mu\in [\mu_{\max}-1,\mu_{\max}]$,
  the smooth solutions $f_{n,j}(\mu)$
  satisfy the bounds
\begin{equation}
  \vert\partial_\mu^m f_{n,j}(\mu)\vert\,\leq\,   C_1^{j+\frac{n}{2}+m}\;
  \dfrac{(j+m+1)!\;
   }{(\frac{n}{2})^2\;(\frac{n}{2})!} 
  \quad \mbox{ for a suitable constant }\  C_1>1\ .
\end{equation}
\end{proposition}
\noindent
Proposition \ref{bound_pertur_mu_simpl} follows from
\begin{lemma}\label{bound_pertur_mu_fn}
  For $\,\mu\in [0,\mu_{\max}]\,$, the smooth solutions $\,f_{n,j}(\mu)\,$
  satisfy the bounds
\begin{equation}
  \vert\partial_\mu^m f_{n,j}(\mu)\vert\,
  \leq \  C_1^{j+\frac{n}{2}+m}\ \, \dfrac{(j+m+1)!}
       {(j-\frac{n}{2}+2)^2 (\frac{n}{2})^2(\frac{n}{2})!}\ 
  \mathcal{F}(j,n,\mu)
  \quad \mbox{ for a suitable constant }\  C_1>1\ .
\end{equation}
\end{lemma}

\begin{proof}
Using Leibniz' rule and Lemma \ref{Prop_bound_pertu_mu_gn}, we get
\begin{equation}
\begin{split}
  \vert\partial_\mu^m f_{n,j}(\mu)\vert
  &\,\leq\ \sum_{k=0}^m \binom{m}{k} (\alpha_0 \ e^\mu)^{2-\frac{n}{2}}\ \Big\vert
  \frac{n}{2}-2\Big\vert^k\ \Big\vert\partial_\mu^{m-k}\ 
  \mathcal{A}_{n,j}^{\alpha_0,\alpha}\Big\vert \\
  &\leq \ {C'}^{j+\frac{n}{2}+m}\;\dfrac{ 1}
          {(j-\frac{n}{2}+2)^2\ (\frac{n}{2})^2\ (\frac{n}{2})!}
 \  \mathcal{F}(j,n,\mu)\sum_{k=0}^m \binom{m}{k}\
  \Big\vert \frac{n}{2}-2\Big\vert^k(j+m-k+1)! \\
  &\leq\  2^m\ C'^{j+\frac{n}{2}+m}\; \dfrac{(j+m+1)!\;}
             {(j-\frac{n}{2}+2)^2\ (\frac{n}{2})^2\ (\frac{n}{2})!}\
             \mathcal{F}(j,n,\mu) \\
             &\leq\ \ C_1^{j+\frac{n}{2}+m}\
             \dfrac{(j+m+1)!}{(j-\frac{n}{2}+2)^2
               (\frac{n}{2})^2(\frac{n}{2})!} \ \mathcal{F}(j,n,\mu)
\end{split}
\end{equation}
choosing $\,C_1=2\,C'>\,1\,$.
\end{proof}

The bounds of Sect.\ref{PertMFAbd} or  and Sect.\ref{PertMFA2}
  imply the local existence 
of the Borel transform of the perturbative series for the functions
$\,\mathcal{A}_{n}^{\alpha_0,\alpha}\,$ or $\,f_{n}(\mu)\,$ as stated in
 \cite{Kopper_Borel}.


\section{The trivial solution and  the perturbative expansion}
\label{RelaTriPert}

In this section we relate the trivial solution constructed
in \cite{mypaper,Kopper2022} to perturbation theory. Our main
result in this section is the following: For fixed  UV-cutoff $\,\alpha_0\,$ we
recover the perturbative expansions of  the smooth functions
$\,f_n(\mu)\,$ constituting a trivial solution,
in powers of a renormalized coupling
$\, g\,$. We will show in Sect.
\ref{LocBoreSummRenor} that these perturbation series are locally
Borel summable w.r.t. $\,g\,$
for $\,\mu\,$ close to $\,\mu_{\max}\,$.
To shorten a bit the the notations {\bf we
always assume the UV-cutoff to be sufficiently large such
that $\,\mathbf{\mu_{\max}>6}\,$} from now on.

\subsection{Properties of the trivial solution}
\label{PropTri}

In \cite{Kopper2022} the  trivial solutions of mean-field
$\varphi^4_4$ theories were obtained with the aid of an ansatz
for the mean-field two-point function of the form
\begin{equation}\label{ansatz}
  f_2(\mu)\,=\,  \sum_{n\geq 1} b_n \
  \dfrac{(n \,\mu)^{n-1}}{1+(n \,\mu)^n} \ .
\end{equation}
On expanding $f_n(\mu)$ 
as a power series
around $\,\mu=0$
\begin{equation}
 f_n(\mu)=\sum_{k\geq 0}
  f_{n,k}\ \mu^k\ ,
  \end{equation}
the  Taylor coefficients of $\,f_{2,k}\,$ at $\,\mu=0\,$ can be rewritten as
\begin{equation}\label{b_n_relation}
  f_{2,k}=(k+1)^k\sum_{\rho=1}^{k+1}b_{\lbrace \frac{k+1}{\rho}\rbrace}
 \  (-1)^{\rho-1}\ \dfrac{1}{\rho^k}\;,
\end{equation}
where by convention $\,b_0=0\,$ and
\begin{equation}
    \left\lbrace \frac{m}{n}\right\rbrace:=\left\{
    \begin{array}{ll}
        \frac{m}{n} & \ \mbox{ if } \ \frac{m}{n}\in\N \\
        0 & \ \mbox{ otherwise }\ .
    \end{array}
\right.
\end{equation} 
\begin{proposition}\label{vanishing_solutions_in_the_UV}
$f_2(\mu)\,$ is well defined on $[0,\mu_{\max}]\ $ and
\begin{equation}\label{vanishing_f_2}
  \quad\lim\limits_{\mu_{\max}\rightarrow +\infty}
  \partial_\mu^l f_2(\mu_{\max})\,=\,0\ ,\quad l\geq 0\ .
    \end{equation}
The functions $\partial_\mu^l f_n(\mu)\,$, $\,l\,\geq\, 0,\  n\geq 4\ $,
  are well defined on $\,[0,\mu_{\max}]\,$ and satisfy
\begin{equation}\label{vanishing_f_n}
  \lim\limits_{\mu_{\max}\rightarrow +\infty}
  \partial_\mu^l f_n(\mu_{\max})\,=\,0\ ,\quad n\geq 4\;,\; l\geq 0\;.
\end{equation} 
\end{proposition}

\begin{proof}
    See \cite{mypaper}.
\end{proof}
Proposition \ref{vanishing_solutions_in_the_UV} implies triviality
of the solutions constructed from the ansatz (\ref{ansatz}). The
uniqueness of the trivial solution for fixed mean-field boundary conditions
has been proven in \cite{mypaper}.

The coefficients $\,b_n\,$ in (\ref{ansatz}) are determined as follows:
From (\ref{ansatz})-(\ref{b_n_relation}) we have\\
$\ f_{2,0}\,=\,b_1\ , \quad f_{2,1}\,=\,2b_2\,-\,b_1\ .$
 From (\ref{MFE-2}) it follows that
$\ f_{2,1}\,=\,3\, f_{4,0}\,-\,f_{2,0}(f_{2,0}\,-\,1)\ .$
Therefore
 \begin{equation} \label{b1b2}
   b_1\, =\,f_{2,0} \ ,\quad
 b_2\,=\,  \frac32\,f_{4,0}\,-\,\frac12\, b_1^2\,+b_1 
   \ .
  \end{equation}
   So {\bf the values of $\,\mathbf{b_1}\,$ and $\,\mathbf{b_2}\,$
    are fixed by 
    $\,\mathbf{f_{2,0}}\,$ and $\,\mathbf{f_{4,0}}\,$}, the latter in turn
  being fixed through the choice of the
  terms in (\ref{BDY_trivial_field}).
 {\bf The $\,\mathbf{b_n}\,$'s, $\,\ \mathbf{n\geq 3}\,$,
  are then uniquely determined
  by (\ref{MFE-2}).}
     From (\ref{b_n_relation}) we have for $\,n\,\geq\, 1\,$
\begin{equation}\label{b_n_induction}
  b_{n+1}\,=\,\frac{f_{2,n}}{(n+1)^n}-\sum_{\rho=2}^{n+1}
  b_{\lbrace \frac{n+1}{\rho}\rbrace}\,(-1)^{\rho-1}\dfrac{1}{\rho^n}\ .
\end{equation}
 For further details, see \cite{mypaper,Kopper2022}.
We have established bounds on the coefficients $b_n$
in \cite{mypaper,Kopper2022}. 
\begin{proposition}\label{bounds_b_n_obtained}
  There exists $\,\tilde{C}\,\equiv\, \tilde{C}(c_{0,2},c_{0,4})>1\,$
  and $\,a<1\,$ such that

\begin{equation}
    \vert b_n\vert\, \leq\, \tilde{C}\  n^2\  a^n\ .
\end{equation}    
\end{proposition}

\begin{proof}
   See \cite{mypaper}.
\end{proof}

We  now recall a few results following from the smoothness
of the trivial solution for $\mu \to 0\,$ which were
established in \cite{mypaper,Kopper2022}. 
\begin{lemma}\label{factorization_of_f_n}
  For smooth solutions $f_n(\mu)$ of (\ref{MFE-2}) with boundary
  conditions (\ref{BDY_trivial_field}), we have
    \begin{equation}
      \partial_\mu^l f_n(0)\,=\,0\ ,\quad n\,\geq\, 6,\  0\,\leq\, l \,\leq\,
      \frac{n}{2}\,-\,3\;.
    \end{equation}
\end{lemma}

\begin{proof}
See \cite{Kopper2022}.
\end{proof}
By Lemma \ref{factorization_of_f_n}, we can set 
\begin{equation}\label{fr}
  f_n(\mu)\,=\, \mu^{\frac{n}{2}-2}\ r_n(\mu)\;,\quad n\,\geq\,4
  \ ,
\end{equation}
where the functions $\,r_n(\mu)\,$ are smooth. Note that
 $ r_4(\mu)\,=\,f_4(\mu)\,$.
For $\,n \,\geq\, 4\,$
the mean-field dynamical system
can then be rewritten  
\begin{align}\label{MFE-g_n}
\begin{split}
    \mu^2\ r_{n+2} &=\dfrac{1}{n+1}\sum_{\substack{n_1+n_2=n+2\\
        n_i\geq 4}} r_{n_1}\ r_{n_2}\ +\
    \mu\ \dfrac{1}{n+1}\ r_n\left(2f_2+1-\dfrac{4}{n}\right) \\
    &+\ \dfrac{n-4}{n(n+1)}\ r_n+\ \dfrac{2}{n(n+1)}\
    \mu\partial_\mu r_n\ ,\quad n\,\geq\, 4\ .
\end{split}
\end{align}
Expanding also the $\,r_n(\mu)\,$ in formal
Taylor series around $\,\mu=0$
\begin{equation}\label{power_series}
   r_n(\mu)=\sum_{k\geq 0}r_{n,k}\ \mu^k\;,
\end{equation}
we get from (\ref{MFE-g_n}) and (\ref{fr})
\begin{align}\label{FE_trivial_field1O}
  f_{2,k+1} &=\dfrac{1}{k+1}\Bigl(3\,r_{4,k}\,+\,f_{2,k}
  -\sum_{\nu=0}^k f_{2,\nu}\ f_{2,k-\nu}\Bigr)\ , \\
\label{FE_trivial_g_n,k}
\begin{split}
  r_{n,k+2} &=\ -\dfrac{n-4}{n+2k}\ r_{n,k+1}\,-\, \dfrac{2n}{n+2k}
  \ \sum_{\nu=0}^{k+1}r_{n,\nu}\ f_{2,k+1-\nu}\,-\,
  \dfrac{n}{n+2k}\sum_{\substack{n_1+n_2=n+2\\
      n_i\geq 4}}\sum_{\nu=0}^{k+2}r_{n_1,\nu}\ r_{n_2,k+2-\nu} \\
  &+\dfrac{n(n+1)}{n+2k}\ r_{n+2,k}\ .
\end{split}
\end{align}
 Regularity at $\,\mu=0\,$ implies for $\,n\,\geq\, 4$
\begin{align}\label{FE_trivial_field2_1}
\dfrac{n-4}{n}\ r_{n,0} &+\ \sum_{\substack{n_1+n_2=n+2\\
n_i\geq 4}}r_{n_1,0}\ r_{n_2,0} \,=\,0\;,\\
\label{FE_trivial_field2_2}
\dfrac{n-2}{n}\ r_{n,1} &+\ 2\sum_{\substack{n_1+n_2=n+2\\
    n_i\geq 4}}r_{n_1,0}\ r_{n_2,1}\,+\, r_{n,0}\,\Bigl(2f_{2,0}\,+\,1\,-\,
\dfrac{4}{n}\Bigr)\,=\,0\ .
\end{align}
In \cite{Kopper2022,mypaper} we derived bounds on the
coefficients $\,r_{n,k},\ f_{2,k}\ $. Here we will analyze their dependence
on $\,b_1\,$. First we give closed expressions for $\, r_{n,0}\,,\ r_{n,1}\,$.
\begin{restatable}{lemma}{exact}\label{exactexpressionsgn01}
    We have for $n\geq 4$
    \begin{equation}\label{exactgn0}
      r_{n,0} \,=\,(-1)^{\frac{n}{2}}\ r_{4,0}^{\frac{n}{2}-1}\
      \dfrac{1}{n-1}\ \binom{3(\frac{n}{2}-1)}{\frac{n}{2}-1} \,=\,
      (-1)^{\frac{n}{2}}\ r_{4,0}^{\frac{n}{2}-1}\ C_2\big(\frac{n}{2}-1\big)\ ,
    \end{equation}
  where we introduced the Fuss-Catalan number of parameter $s>0$
    \begin{equation}\label{FussCatalannb}
        C_s(n):=\dfrac{1}{sn+1}\ \binom{(s+1)n}{n}\;.
    \end{equation}
    Moreover we have
    \begin{equation}\label{exactgn1}
      r_{n,1}=(-1)^{\frac{n}{2}-1}\ r_{4,0}^{\frac{n}{2}-1}\
      \Big(\frac{3n-4}{2}b_1+\frac{n-4}{4}\Big)\  C_2\big(\frac{n}{2}-1\big)\ .
    \end{equation}
\end{restatable}

\begin{proof}
    See Appendix \ref{appendixgn01}.
\end{proof}
The expressions in Lemma \ref{exactexpressionsgn01} are exact.They satisfy the
bounds on $\,r_{n,0}\,$ and $\,r_{n,1}\,$ established in
\cite{mypaper,Kopper2022}. Moreover $\,r_{n,0}\,$
and $\,r_{n,1}\,$ are polynomials in $\,b_1\,$. Now we establish in a
fashion similar to \cite{Kopper2022} bounds on
$\,r_{n,k}\,$ and $\,f_{2,k}\,$.

\begin{restatable}{lemma}{BPHZboundsgnkfk}\label{BPHZlemmagnkf2kwhole}
  Let $f_{n}(\mu)$ be the solutions of the flow equations (\ref{MFE-2})
  with the mean-field boundary conditions (\ref{BDY_trivial_field}). For 
    \begin{equation}
      \vert f_{2,0}\vert\,\leq\, K\ , \ \  0\,<\,
      f_{4,0} \,\leq\,\frac{K}{10}\ \Leftrightarrow \ \
    \vert   c_{0,2}\vert \,\leq\, \frac{K}{2(2\pi)^4}\ \frac{1}{\alpha_0} \ , 
     \ \  0\,<\, c_{0,4} \,\leq\, \frac{1}{10}\ \frac{K}{(4\pi)^2}\ ,
            \ \ \, K\,\leq\,\frac{1}{30}\ ,
\end{equation}
we have
\begin{equation}
  \vert r_{n,k}\vert\,\leq\, \Big(\frac{3}{2}\Big)^{k-2}\
  K^{\frac{n}{2}-1}\ \Big(\frac{n-4}{2}+k\Big)!\ \ ,
  \quad \vert f_{2,k}\vert\,\leq\,
  \Big(\frac{3}{2}\ \Big)^{k}K\ \vert k-1\vert !\ \ .
\end{equation}
\end{restatable}

\begin{proof}
    See Appendix \ref{AppendixD}.
\end{proof}
Now we can derive bounds for the coefficients $b_n$
\begin{lemma}\label{BPHZbnwhole}
    Under the assumptions of Lemma \ref{BPHZlemmagnkf2kwhole}, we have
    \begin{equation}   
      \vert b_n\vert\leq \frac{5}{2}\,\Big(\frac{7}{10}\Big)^{n-1}K\;,
      \quad n\geq 1\ .
\end{equation}
\end{lemma}

\begin{proof}
The claim holds obviously for $n=1$. Then we find
\begin{equation}
\left\{
     \begin{array}{ll}
       \vert b_2\vert\leq \frac{3}{2}\,r_{4,0}+\vert b_1\vert
       +\frac{1}{2}\vert b_1\vert^2\,\leq\, K\, \Big(\frac{1}{20}
       \,+\,1\,+\,\frac{K}{2}\Big)\,\leq\, \frac{7}{4}\,K\;,\\
       \vert b_3\vert\,\leq\,\dfrac{\vert f_{2,2}\vert+
         \vert b_1\vert}{9}\,\leq\, K\, \Big(\frac{1}{4}
       +\frac{1}{9}\Big)\,\leq\, \frac{5}{2}\Big(\frac{7}{10}\Big)^{2}K\;, \\
       \vert b_4\vert\,\leq\, \dfrac{\vert f_{2,3}\vert}{64}+
       \frac{\vert b_2\vert}{8}\,\leq\,  K\, \Big(\frac{27}{256}+
       \frac{3}{16}\Big)\,\leq\, \frac{5}{2}\Big(\frac{7}{10}\Big)^{3}K\ .
     \end{array}
\right.
\end{equation}
 
For $\,n\geq 4\,$ we insert the induction hypothesis in the
r.h.s of (\ref{b_n_induction}) to get
\begin{equation}
\begin{split}
  \vert b_{n+1}\vert &\,\leq\, \Big(\frac{3}{2}\Big)^n
  \dfrac{(n-1)!}{(n+1)^n}\ K \,+\, \frac{5}{2}\ K
  \sum_{\rho=2}^{n}\Big(\frac{7}{10}\Big)^{\frac{n+1}{\rho}-1}\frac{1}{\rho^n}
  +\frac{K}{(n+1)^n}\\  
  &\leq\, \Big(\frac{7}{10}\Big)^n\ \frac{3K}{10}
  +\frac{5}{2}\, \frac{2}{2^n}\ K\,+\, \frac{K}{(n+1)^n}
  \,\leq\, \frac{5}{2}\,\Big(\frac{7}{10}\Big)^n\  K\ ,
\end{split}
\end{equation}
where we used successively
\begin{equation}\label{intermediateBPHZ1}
  \Big(\frac{3}{2}\Big)^n\dfrac{(n-1)!}{(n+1)^n}
  \,\leq\, \Big(\frac{7}{10}\Big)^n\frac{3}{10}\;,\quad n\geq 4
\end{equation}
and
\begin{equation}\label{intermediatePBHZ2}
  \frac{3}{10}\,+\, 2\, x\ \Big(\frac{5}{7}\Big)^n
  +\Big(\frac{10}{7(n+1)}\Big)^n\,\leq\, x\ ,
  \quad x\,\geq\, 1,\quad n\,\geq\, 4\ .
\end{equation}
\end{proof}
The bounds established in
Lemmata \ref{BPHZlemmagnkf2kwhole}, \ref{BPHZbnwhole}
are uniform in $\,b_1\,$. Now we analyze the dependence of the constants
$\,b_n\,$ on $\,b_1\,$. From Lemma \ref{exactexpressionsgn01}, 
$\,r_{n,0}\,$ and $\,r_{n,1}\,$ are polynomials in $\,b_1\,$ of degree $0$
and $1$ respectively. More generally 
\begin{lemma}\label{BPHZgnkf2kpolynomials1}
    We have
    \begin{equation}
      r_{n,k}\,=\, \mathcal{P}_{n,k}(b_1)\ ,\quad
      f_{2,k}\,=\, \mathcal{P}_k(b_1)\ ,
    \end{equation}
    where $\,\mathcal{P}_{n,k}\,$ and $\,\mathcal{P}_k\,$  are polynomials with
    real coefficients which depend respectively on $\,n,\,k,\,r_{4,0}\,$ and
    on $\,k,\,r_{4,0}\,$. Furthermore
    $\,\mbox{deg}(\mathcal{P}_{n,k})\,\leq\, k\,$
    and $\,\mbox{deg}(\mathcal{P}_k)\,=\,k+1\,$.
\end{lemma}

\begin{proof}
  For $\,r_{n,k}\,$ we proceed by induction in $\,N\,=\,n+2k\,$, at fixed
  $N$ we  go up in $\,k\,$. By Lemma \ref{exactexpressionsgn01}
  the claim holds for $\,k\,\leq\, 1\,$.
  For $\,k\,\geq\, 0\,$ the claim follows on inserting
  the induction hypothesis on the r.h.s
  of (\ref{FE_trivial_g_n,k}).
    
  As for $\,f_{2,k}\,$ the statement holds for $\,k=0\,$. For $\,k\geq 0\,$ it
  follows when inserting the induction hypothesis in the r.h.s
  of (\ref{FE_trivial_field1O}). In particular, one realizes
  from the inductive proof that the coefficient of the leading term of
  $f_{2,k}$ as a polynomial in $b_1$ is $(-1)^k\,$.
\end{proof}
From Lemma \ref{BPHZgnkf2kpolynomials1}, we can write
\begin{equation}\label{gnkf2kpolynomialsb_1}
  r_{n,k}=\sum_{\nu=0}^k r_{n,k,\nu}\;b_1^\nu\;,
  \quad f_{2,k}=\sum_{\nu=0}^{k+1} f_{2,k,\nu}\; b_1^\nu\;.
    \end{equation}
From Lemma \ref{exactexpressionsgn01} we have
\begin{equation}\label{coeffgn0gn1}
  r_{n,0,0}\,=\,r_{n,0}\ ,\quad r_{n,1,0}
  \,=\,-\,\frac{n-4}{4}\ r_{n,0}\ ,\quad r_{n,1,1}\,=\,-\, \frac{3n-4}{2}
  \ r_{n,0}\ .
\end{equation}
From (\ref{FE_trivial_field1O}), we get 
\begin{equation}\label{coefff201b1}
  f_{2,0,\nu}\,=\,\delta_{1,\nu}\;,\quad f_{2,1,0}\,=\,3\,r_{4,0}\;,\quad
  f_{2,1,1}\,=\,-\,f_{2,1,2}\,=\,1\ .
\end{equation}
We also have from (\ref{FE_trivial_field1O}) and (\ref{gnkf2kpolynomialsb_1})
\begin{equation}\label{coefff22b1}
  f_{2,2,0}=\frac{3}{2}\, r_{4,0}\;,\quad f_{2,2,1}\,=\,-\,9\,r_{4,0}\,+\,
  \frac{1}{2}\ ,
  \quad f_{2,2,2}\,=\,-\,\frac{3}{2}\ ,\quad f_{2,2,3}\,=\,1\ .
\end{equation}
If we insert the polynomial expansion of $\,r_{n,k}\,$ and
$\,f_{2,k}\,$ (\ref{gnkf2kpolynomialsb_1}) in
(\ref{FE_trivial_field1O}), (\ref{FE_trivial_g_n,k}), we
obtain the following inductive systems for the coefficients
$\,r_{n,k,\nu}\,$ and $\,f_{2,k,\nu}\,$

\begin{align}\label{coeffFEsgnkb_1_1}
    \begin{split}
      r_{n,k+2,\nu} &\,=\,-\dfrac{n-4}{n+2k}\ r_{n,k+1,\nu}
      -\dfrac{2n}{n+2k}\sum_{\rho=0}^{k+1}\ \ \
      \sum_{\nu'=\max\lbrace\nu-(k+2-\rho),0\rbrace}^{\min\lbrace \rho,\nu\rbrace}
      r_{n,\rho,\nu'}\ 
      f_{2,k+1-\rho,\nu-\nu'} \\
        &-\dfrac{n}{n+2k}\sum_{\substack{n_1+n_2=n+2\\
          n_i\geq 4}}\sum_{\rho=0}^{k+2}\ \ \
      \sum_{\nu'=\max\lbrace \nu-(k+2-\rho),0\rbrace}^{\min\lbrace\rho,\nu\rbrace}
      r_{n_1,\rho,\nu'}\ r_{n_2,k+2-\rho,\nu-\nu'} \\ &
      \,+\,\dfrac{n(n+1)}{n+2k}\ r_{n+2,k,\nu}
    \end{split}
\end{align}

and
\begin{align}\label{coeffFesf2k1}
    \begin{split}
      f_{2,k+1,\nu} &\,=\,\dfrac{1}{k+1}\Bigl(3r_{4,k,\nu}\,+\,f_{2,k,\nu}
      -\sum_{\rho=0}^k\ \ \
      \sum_{\nu'=\max\lbrace \nu-(k+1-\rho),0\rbrace}^{\min\lbrace\rho+1,\nu\rbrace}
      f_{2,\rho,\nu'}\ f_{2,k-\rho,\nu-\nu'}\Bigr)\ ,
    \end{split}
\end{align}
where  we set for convenience $\,r_{n,k,\nu}=0\,$ for $\ \nu>k\,$
and $\,f_{2,k,\nu}=0\,$ for $\,\nu>k+1\,$.\\
The proof of the following Lemma is similar to the proofs of
Lemmata \ref{BPHZlemmagnkf2kwhole}-\ref{BPHZbnwhole}.

\begin{restatable}{lemma}{BPHZlemma}\label{BPHZgnkf2kpolynomials2}
  Under the assumptions of Lemma \ref{BPHZlemmagnkf2kwhole}, we have
    \begin{equation}
      \vert r_{n,k,\nu}\vert\,\leq\, \frac{1}{4}\ K^{\frac{n}{2}-1}\ 
      \binom{k}{\nu}\ \Big(\frac{n-4}{2}+k\Big)!\ \ ,
      \quad \vert f_{2,k,\nu}\vert\, \leq\,\binom{k+1}{\nu}\
      \vert k-1\vert !\ .
    \end{equation}   
\end{restatable}

\begin{proof}
    See Appendix \ref{AppendixD}.
\end{proof}
Now we determine the dependence of the coefficients $\,b_n\,$
in terms of $\,b_1\,$.
\begin{lemma}\label{BPHZlemmabn}
    We have
    \begin{equation}
        b_n\,=\,p_n(b_1)\;,
    \end{equation}
    where $\,p_n\,$ is a polynomial of
    degree $\,n\,$ with real coefficients
    which  depend on $\,n,\ r_{4,0}\,$. In particular, the leading coefficient
    of $\,p_n\,$ is $\,\frac{(-1)^{n-1}}{n^{n-1}}\,$.
\end{lemma}
\begin{proof}
  The proof proceeds by induction in $\,n\,$. The claim is obvious for
  $\,n=1\,$.
  For $\,n\,\geq\, 1\,$ we insert the induction hypothesis in the r.h.s of
  (\ref{b_n_induction}) to prove our claim.
\end{proof}

From Lemma \ref{BPHZlemmabn}, we write
\begin{equation}\label{bnpolynomial}
  b_n\,=\,\sum_{\nu=0}^n b_{n,\nu}\; b_1^\nu\ .
\end{equation}
Then from (\ref{b_n_induction}) and (\ref{bnpolynomial}) we have
\begin{equation}\label{bnpolynomialinduction}
  b_{n+1,\nu}\,=\, \frac{f_{2,n,\nu}}{(n+1)^n}-
  \sum_{\rho=2}^{n+1}b_{\lbrace \frac{n+1}{\rho}\rbrace,\nu}(-1)^{\rho-1}
  \dfrac{1}{\rho^n}\ ,
\end{equation}
where we set $b_{n,\nu}\,=\,0\,$ if $\ \nu>n\,$. The
coefficients of the polynomials $\,p_n\,$ are bounded through

\begin{restatable}{lemma}{BPHZbn}\label{BPHZbnpolynomialscoeff}
    \begin{equation}\label{boundscoeffbnb1}
      \vert b_{n,\nu}\vert\,\leq\, \frac{1}{n}\ \Big(\frac{3}{4}\ \Big)^{n-2}
      \binom{n}{\nu}\ ,\quad n\geq 1\ ,\quad 0\leq \nu\leq n\ .
\end{equation}
\end{restatable}
\begin{proof}
See Appendix \ref{AppendixD}.
\end{proof}


\subsection{The renormalization conditions corresponding to  the
trivial solution}\label{BPHZrenormalization}

Based on the previous results we now show that the trivial solutions
are compatible
with renormalization conditions (\ref{BPHZgen}), (\ref{BPHZgenbd}).
Recall from (\ref{b_n_induction}) that the first coefficients
of the trivial solution (\ref{ansatz}) satisfy
(\ref{b1b2}).
We restrict to small bare couplings. The following
proposition relates the bare constants of the trivial solution
$\,b_1\,,\ b_2\,$ or equivalently  $\,f_{2,0}\,,\ f_{4,0}\,=\,r_{4,0}\,$ 
to the two-point function at the renormalization scale $\,\mu_{max}\,$.
\begin{proposition}\label{BPHZpossible}
  For any fixed $\,c\,,\ |c| \,\le\, \frac{1}{30}\ $ and
  $\ 0 \,\le\,f_{4,0}\,\leq\, \frac{1}{12}\cdot 10^{-3} \,$,
  there exists a unique (real) $\,b_1\,,\ \, |b_1| \,\leq\, \frac{1}{30}\,$
  such that 
  \begin{equation}
    \label{cv}
        f_2(\mu_{\max})\,=\, \frac{c}{\mu_{\max}}\ .
    \end{equation}
\end{proposition}

\begin{proof}
From  (\ref{ansatz}) 
 we have
\begin{equation}\label{fmaxb1}
 f_2(\mu_{\max})\,=\,\frac{b_1}{1+\mu_{\max}}
  \,+\,b_1\,\dfrac{2\mu_{\max}}{1+4\mu_{\max}^2}
\,+\,(3f_{4,0}-b_1^2)\,\frac{\mu_{\max}}{1+4\mu^2_{\max}}\,+\,\sum_{n\ge 3}
  b_n\, \dfrac{(n\,\mu_{\max})^{n-1}}{1+(n\,\mu_{\max})^n}\ .
\end{equation}
Solving for the linear term in $\,b_1\,$
and imposing  the  renormalization condition for
the two pointfunction (\ref{cv}),
one realizes that this condition is equivalent to  
\begin{equation}
    \mathcal{G}(b_1)=b_1\ ,
\end{equation}
where 
\begin{equation}\label{gmaxb1}
  \mathcal{G}(b_1):= 
  \Big[\frac{c}{\ \mu_{\max}}\,-\,(3f_{4,0}-b_1^2)
    \frac{\mu_{\max}}{1+4\mu^2_{\max}}
    \,-\,\sum_{n\geq 3} b_n\,\dfrac{(n\,\mu_{\max})^{n-1}}{1+(n\,\mu_{\max})^n}\Big]
   \
  F(\mu_{\max})
\end{equation}
with
\begin{equation}
  F(x):=\dfrac{(1+x)(1+4x^2)}{1+2x+6x^2}\ ,\quad x\geq 0\ .   
\end{equation}
This means that  $\,\mathcal{G}(b_1)\,$ has a fixed point
in $\,\R\,$. We first prove 

\begin{lemma}\label{Gb1continderiv}
  The function $\mathcal{G}(b_1)$ from (\ref{gmaxb1})
  is differentiable on an interval  $[-a,a]\,$, 
  for  $a\leq\frac{1}{30}\,$. Moreover
    \begin{equation}
      \big\vert\mathcal{G}(b_1)\big\vert \,<\, a\ ,
      \quad
      \Big\vert
      \dfrac{\partial \mathcal{G}}{\partial b_1}(b_1)\Big\vert \,<\, 1\ ,
      \quad b_1\in [-a,a\,]\ .
    \end{equation}
\end{lemma}

\begin{proof}
  It follows from Lemma \ref{BPHZlemmabn},
  that the coefficients $\,b_n\,$ are smooth
  functions of $\,b_1\,$.
  The bounds from
  Lemma \ref{BPHZbnwhole} imply that
    \begin{equation}
    \begin{split}
      \big\vert \mathcal{G}(b_1)\big\vert&
      \leq \ a\ \Big[\frac{c}{\ \mu_{\max}}
        \,+\, \Big(\frac{1}{10}+a\Big)\frac{1}{2\mu_{\max}}
        \,+\,\frac{1}{\mu_{\max}}\sum_{n\geq 3} \frac{\vert b_n\vert}{n}\Big]\
      \big\vert F(\mu_{\max})\big\vert   \\
      &\leq\, \frac{3a}{4}\Big[\frac{1}{2}+\frac{1}{15}
        \,+\,\frac{5}{2}\sum_{n\geq 3}\frac{1}{n}\Big(\frac{7}{10}\Big)^{n-1}
        \Big] \\
      &\leq\, \frac{3a}{4}\Big[\frac{1}{3}+\frac{1}{15}
        +\frac{5}{2}\frac{10}{7}\Big(\ln\Big(\frac{10}{3}\Big)\,-\,
        \frac{189}{200}\Big)\Big]\,<\,a\ .
    \end{split}
    \end{equation}
    From Lemma \ref{BPHZbnpolynomialscoeff}
    \begin{equation}\label{boundderivativebn}
        \begin{split}
          \Big\vert\frac{\partial b_n}{\partial b_1}\Big\vert
          &\,\leq\,  \Big(\frac{3}{4}\Big)^{n-2}
          \sum_{\nu=1}^n \frac{\nu}{n}\binom{n}{\nu}\ \vert b_1\vert^{\nu-1}
          \,=\, \Big(\frac{3}{4}\Big)^{n-2}\sum_{\nu=0}^{n-1}\binom{n}{\nu}
          \ \vert b_1\vert^\nu\,=\,\frac{4}{3}
          \,\Big(\frac{3(1+\vert b_1\vert)}{4}
          \Big)^{n-1}\ .
        \end{split}
    \end{equation}
    For $\,\vert b_1\vert\,\leq\, a\,$ the bounds (\ref{boundderivativebn})
    imply that the series of functions
\begin{equation}
  \Big(\sum_{n=1}^N  \frac{\partial b_n}{\partial b_1}
  \   \dfrac{(n\,\mu_{\max})^{n-1}}{1+(n\,\mu_{\max})^n}\Big)_{N\in\N}
\end{equation}
converges uniformly on $[-a,a]$ so that $\,\mathcal{G}(b_1)\,$
is differentiable w.r.t. $\,b_1\in [-a,a\,]\,$.
Then we can bound the derivative
of $\,\mathcal{G}_{\mu_{\max}}(b_1)\,$
\begin{equation}
          \Big\vert\frac{\partial \mathcal{G}}{\partial b_1}(b_1)
      \Big\vert \, \leq\,  \Big[\, \frac{a}{2\mu_{\max}}
        \,+\,\frac{4}{3\mu_{\max}}\, \sum_{n\geq 3}\frac{1}{n}\ 
        \Big(\frac{3(1+K)}{4}\Big)^{n-1}\,\Big]\
      \big\vert F(\mu_{\max})\big\vert
      \,<\,1\ .
    \end{equation}
\end{proof}

\noindent
 {\it Proof of Proposition \ref{BPHZpossible} continued}. 
  From Lemma \ref{Gb1continderiv}, the function
  $\,\mathcal{G}(b_1)\,$ satisfies the assumptions of the Banach-Picard
  fixed point theorem \cite{Picard,Banach}. Therefore the unique fixed point
  of $\,\mathcal{G}(b_1)\,$ is found by iteration: define
  $\,u_0:=b\,$ for an arbitrary $\,b\in [-a,a\,]\,$. Then for
  $\,n\in\N_0\,$, $\,u_{n+1}:=\mathcal{G}(u_n)\,$. The sequence
  $\,(u_n)_{n\in\N_0}\,$   converges to the unique fixed point of
  $\,\mathcal{G}(b_1)\,$   in $\,[-a,a\,]\,$.
\end{proof}
So we have shown that there exists $\,b_1\,$ such
the two pointfunction $\,f_2(\mu)\,$
equals $\,\frac{c}{\ \mu_{\max}}\,$ for given sufficiently small
$\,c\,$. The mean-field flow equations (\ref{MFE-2}) then imply that
\begin{equation}
\label{cv4}
f_4(\mu_{\max})\,=\,-\,\frac{1}{3}\,\frac{c}{\ \mu_{\max}}
\,+\,\mathcal{O}((\frac{c}{\ \mu_{\max}})^2)\ .
\end{equation}
So we have determined the renormalization conditions
for the two and  four point functions in dependence of the bare parameters,
to leading order in $\,\frac{1}{\ \mu_{\max}}\,$ .
In the next section we will be more precise on the four point
function and on the relation with perturbation theory.


\subsection{Analyticity properties of the trivial
  solution close to the renormalization scale
  and the renormalized conditions} \label{PerturbExpaning}

The function $\,f_2(\mu)\,$ from (\ref{ansatz})
 depends on the parameters $b_1$ and $b_2$. This
 dependence can be reexpressed as a dependence
 on the perturbative renormalization conditions for
 $\,f_{2,j}(\mu_{max})\,$ and $\,f_{4,j}(\mu_{max})\,$.
 It turns out that we can make this dependence
 explicit in terms of a convergent expansion
 for $\,\mu\,$ sufficiently close to  $\,\mu_{max}\,$.
 For $\,\mu>1\,$ we can write 
\begin{equation}
  f_2(\mu)\,=\,\frac{1}{\mu}\sum_{n\geq 1}\dfrac{b_n}{n}\
  \dfrac{1}{1+\frac{1}{\mu^n n^n}}\ .
\end{equation}
We define the function 
\begin{equation}
  \tilde{f}_2(z):=\, z\sum_{n\geq 1}\dfrac{b_n}{n}\
  \dfrac{1}{1+\frac{z^n}{ n^n}}
  \ ,  \quad z\in (-1,1]
\end{equation}
so that $\,\tilde{f}_2(\frac{1}{\mu})\,=\,f_2(\mu)\,$.
For $\,z\in [0,1]$, $\,\tilde{f}_2(z)\,$ is well-defined from Proposition
\ref{bounds_b_n_obtained}. 
In \cite{mypaper}, we have proven that $\,f_2(\mu)\,$
is locally
analytic w.r.t. $\,\mu\,$ for $\,1<\mu\leq \mu_{\max}\,$. Actually
$\,\tilde{f}_2(z)\,$
has an analytic continuation
\begin{proposition}\label{tildef2analytic}
  $\tilde{f}_2$ is analytic w.r.t. $\,z\,$ in the disk
  $\,D(0,\frac{1}{2}):=\, \lbrace z\in \C\;\vert
  \ \, \vert z\vert <\frac{1}{2}\rbrace\ $. 
\end{proposition}

\begin{proof}
 First note that
 \begin{equation}
     \tilde{f}_n(z):= \dfrac{b_n}{n}\dfrac{1}{1+\frac{z^n}{n^n}}
 \end{equation}
  is analytic w.r.t.
 $z\,$ in $\,D(0,\frac{1}{2})\,$. Then
\begin{equation}
 \Big\vert \sum_{n\geq m+1} \dfrac{b_n}{n}\
  \dfrac{1}{1+\frac{z^n}{n^n}}\Big\vert
  \,\leq\, \sum_{n\geq m+1}\frac{\vert b_n\vert}{n}
  \frac{1}{\vert 1+\frac{z^n}{n^n}\vert}
  \,\leq\, \sum_{n\geq m+1}\frac{\vert b_n\vert}{n}\
  \frac{1}{1-\frac{1}{2^n q^n}}
    \,\leq\, 2\, \sum_{n\geq m+1}\frac{\vert b_n\vert}{n}\ .
\end{equation}
Since $\,\sum_{n\geq 1}\frac{b_n}{n}\,$ is absolutely convergent,
 $\,\sum_{1\leq n\leq N}\; z\;\tilde{f}_n(z)\,$  converges uniformly to
$\,\tilde{f}_2(z)\,$ on $\,D(0,\frac{1}{2})\,$,
and $\,\tilde{f}_2(z)\,$
is analytic in the disk $\,D(0,\frac{1}{2})\ $. 
\end{proof}

For $\vert z\vert<\frac{1}{2}\,$ we expand
\begin{equation}\label{tildef2expansionz}
       \tilde{f}_2(z)= \sum_{m\geq 1} c_m z^m\;,
    \end{equation}
where 
\begin{equation}\label{definec_m}
        c_m:=\sum_{\substack{k\geq 0, n\geq 1 \\ nk+1=m}}\dfrac{(-1)^k b_n}{n^m}\ .
    \end{equation}
    In particular we have
    \begin{equation}
        c_1=\sum_{n\geq 1}\dfrac{b_n}{n}\ ,
    \end{equation}
    while for $\,c_m\,$, $\,m\geq 2\,$, the sum in (\ref{definec_m}) is finite.
    From Proposition \ref{bounds_b_n_obtained} we have uniformly in $\,m\,$
    \begin{equation}\label{culz}
      \vert c_m\vert\, \leq\, C_3\  \ \mbox{ for a suitable constant }
      \ \ C_3>0\ .
    \end{equation}
       Now we set 
    \[
    \lambda(\mu) :=\, \mu_{\max}-\mu\quad \mbox{for} \quad
    \mu \in (\mu_{max}-1,\mu_{max}]
\quad \mbox{which implies} \quad \lambda \in [0, 1)\ . 
    \]
    And we define the {\bf renormalized coupling}
    \begin{equation}\label{g}
    g\,=\, \frac{1}{\mu_{max}}\ .
    \end{equation}
    We fix  $\,c\,\in (0, \frac{1}{3})\,$ 
    and choose $\,b_1\,$ such that
    $\,f_2(\mu_{\max})\,=\,\frac{c}{\mu_{\max}}\,$.
   From the formal expansion
\begin{equation}
  \frac{1}{\mu}\,=\,\frac{1}{\mu_{\max}- \lambda}
  =\sum_{k=1}^{+\infty}\, \dfrac{\lambda^{k-1}}{\mu_{\max}^k}
  = \sum_{k=1}^{+\infty}\, \lambda^{k-1}  g^k
\end{equation}
we get formally
\begin{equation}\label{perturbf2}
  f_2(\mu) \,=\, \sum_{j=1}^{+\infty}c_j
\  \Big(\sum_{k=1}^{+\infty}\lambda^{k-1}\ g^k\Big)^j
  \,=\, \sum_{j=1}^{+\infty}\  \sum_{\alpha=1}^j
  \ \ \sum_{\substack{k_1+\cdots+k_{\alpha}=j \\
      k_i\geq 1}} c_\alpha \ \lambda^{j-\alpha} g^j \ .   
\end{equation}
We define for $\,z \in \C,\ |z|\,<\,\frac{1}{6}$
\begin{equation}\label{pmi}
F_2(\mu,z):=\sum_{j=1}^{+\infty}a_j(\mu)\ z^j \ ,\quad
  a_j(\mu):
  =\,\sum_{\alpha=1}^j c_\alpha\ \lambda(\mu)^{j-\alpha}\ \binom{j-1}{\alpha-1}\ ,
  \quad j\geq 1
\end{equation}
so that
\begin{equation}\label{F2}
F_2(\mu, g)\,=\, f_2(\mu)\  .
\end{equation}
The perturbative expansion (\ref{perturbf2})
and the mean-field flow equations (\ref{MFE-2})  both imply that
all $\,f_n(\mu)\,$
have a (formal) perturbative expansion w.r.t. $\, g\,$.

\begin{lemma}\label{amanalytic}
  The functions $\,a_j(\mu)\,$, $\,j\geq 1\,$, are analytic on
  $\,(\mu_{max}-1,\mu_{max}]\,$ and 
\begin{equation}\label{bj}
  \vert a_j(\mu)\vert\leq C_3(1+\lambda(\mu))^{j-1}\;,
  \quad \vert\partial_\mu a_j(\mu)\vert
  \,\leq\, C_3(j-1)(1+\lambda(\mu))^{j-2}\ ,
\end{equation}
where the constant $\,C_3\,$ is the one introduced in (\ref{culz}).
\end{lemma}
\begin{proof}
  It is clear that the functions $\,a_j(\mu)\,$ are analytic w.r.t
  $\mu \in (\mu_{max}-1,\mu_{max}]\,$. From (\ref{culz}), we get
    \begin{equation}
      \vert a_j(\mu)\vert\,\leq\, C_3
      \sum_{j=1}^\alpha\binom{j-1}{\alpha-1}\,\lambda(\mu)^{j-\alpha}
      =C_3(1+\lambda(\mu))^{j-1}
    \end{equation}
    and
    \begin{equation}
      \vert\partial_\mu a_j(\mu)\vert
      \,\leq\, \sum_{\alpha=1}^{j-1}\binom{j-1}{\alpha-1}\,
      \vert c_\alpha\vert (j-\alpha)\,
      (\lambda(\mu))^{j-\alpha-1}
      \,\leq\, C_3\sum_{\alpha=1}^{j-1}\binom{j-1}{\alpha}\alpha\;
      (\lambda(\mu))^{\alpha-1} 
    \end{equation}
    \[
      \,=\, C_3 (j-1)(1+\lambda(\mu))^{j-2}\ .  
      \] 
\end{proof}
\vspace{-.3cm}
\noindent
Since $\,\lambda(\mu)<1\,$ we get uniformly in $\,\mu\,$
\begin{equation}
  \label{ambd}
  \vert a_j(\mu)\vert\,\leq\, C_3\ 2^{j-1}\ ,\quad
  \vert \partial_\mu a_j(\mu)\vert \,\leq\, C_3\ (j-1)\ 2^{j-2}\ .
\end{equation}
From Lemma \ref{amanalytic}, the series (\ref{perturbf2}) converges
for $\, g<\frac{1}{6}\,$, and the function $F_2(\lambda,z)\,$ is analytic
w.r.t. $(\lambda,z)\,\in\,
\Omega:=(\mu_{max}-1,\mu_{max}]\times \,D(0,\frac{1}{6})\,$. Remark that
the perturbative expansion (\ref{perturbf2}) starts at $\,j=1\,$.
From (\ref{perturbf2}) (noting that $\lambda(\mu_{max}) =0\,$)
we obtain the renormalization conditions for the
mean-field two-point function  
\begin{equation}\label{two}
    f_{2,j}(\mu_{\max})\,=\,a_j(\mu_{\max})\,=\,c_j\,=\,c\;\delta_{j,1}\ .
\end{equation}
From the mean-field flow equations (\ref{MFE-2}) and the perturbative
expansion (\ref{perturbf2}), the perturbative
renormalization conditions for the
mean-field  four-point function are
\begin{equation}\label{four}
  f_{4,1}(\mu_{\max})\,=\, -\frac{c}{3}\ ,\quad f_{4,j}(\mu_{\max})
  \,=\,
  \frac{1}{3}\Big(\partial_\mu a_j(\mu_{\max})\,+\,c^2\,\delta_{j,2}\Big)\ ,
  \quad j\geq 2\ .
\end{equation}
Due to (\ref{ambd})
these conditions (\ref{two}) and (\ref{four}) satisfy
(\ref{BPHZgen}) and (\ref{BPHZgenbd}). Therefore
the inductive scheme from Sect.\ref{PertMFA} applies.
So we have shown   
  \begin{proposition}\label{pertr}
    Under the assumptions of Proposition \ref{BPHZpossible}
    the trivial solution has a perturbative expansion w.r.t.
    the renormalized coupling $\,g\,$ (\ref{g}).
    The boundary conditions at $\,\mu_{max}\,$ obey
    (\ref{two}), (\ref{four}) and (\ref{ambd}) so that the four-point function
    $\,f_4(\mu_{max})\,$ is given in terms of a power series in terms
    of $\,g\,$ with a radius of convergence $\,\ge\, \frac{1}{2}\,$.
    \end{proposition}

\textbf{Remarks:}
  It is possible to redefine the renormalized coupling $\,g\,$ from (\ref{g})
  without changing our main results. If we set
\begin{equation}\label{newseries}
    \tilde g(g)\,=\, \kappa g+\sum_{j=2}^\infty g^j\; \tilde{a}_j
\end{equation}
with $\,\kappa \,>\, 0\,$, assuming that the formal power series
(\ref{newseries}) is locally Borel summable, then the relation
(\ref{newseries}) can be
inverted
\begin{equation}
    g(\tilde g)=\kappa' \;\tilde g+\sum_{j=2}^\infty\tilde g^j\; a_j\;,
\end{equation}
with $\,\kappa'\,=\,\kappa^{-1}\,$. Since local Borel summability is preserved
by the composition of locally Borel summable functions~\cite{Auberson1},
the local Borel summability of the perturbative series w.r.t. $\,g\,$
implies the local Borel summability of the perturbative series w.r.t.
$\,\tilde g\,$.
\newline
The case $\,c=0\,$ in (\ref{two}) and (\ref{four})
corresponds   to the renormalization condition (\ref{BPHZ}).
  In this case we have  $\,f_{4,1}(\mu_{\max})\,=\,0\,$, and
  the expansion starts at $\,1/\mu_{max}^2\,$ which then is to
  be identified (possibly up to a multiplicative constant)
  with the renormalized coupling in standard language.
We will not analyze this particular (nongeneric)  case here in detail. 
We also remark that our results are not sharp enough to determine the
sign of $\,f_{4}(\mu_{\max})\,$, even if we suspect that $\,f_4(0)\,>\,0\,$
implies $\,f_{4}(\mu_{\max})\,>0\,$ as suggested by the lowest
order perturbative relation
\begin{equation}\label{pico}
     f_4(0)\,=\, f_{4}(\mu_{\max}) \,+\, \mathcal{O}(f_{4}(\mu_{\max})^2) 
 \end{equation}
 following from (\ref{perturbative_exp}),
 (\ref{MFE2_perturbative_part}), and (\ref{BDY_perturb_relevant}).
  In constructive field theory positivity of the
 renormalized coupling for positive bare coupling follows
 from the analysis of the
 functional integral through discrete renormalization group steps
 \cite{Bauerschmidt2019}
 between $\,\alpha_0$ and $\,\alpha_{\max}\,$. In this case
 one finds for (very) small bare couplings $\,f_4(0)\,$
 that the renormalized coupling decreases but stays positive, and tends
 logarithmically to zero for $\,\mu_{\max}\rightarrow\, +\infty\,$.


\section{Borel summability of the mean-field regularized
  renormalized perturbation theory}\label{LocBoreSummRenor}

Local Borel summability, see section \ref{LocBoreSummRenor} below,
implies that the perturbative expansion
is asymptotic to a function  which can be uniquely constructed from it 
without requiring convergence of the expansion.
Here we  need not construct the function
because it is the trivial solution  already known. 
But we want to elucidate the status of the perturbative expansion
with respect to this solution.


\subsection{Mean-field flow equations for the Taylor remainders
  $\,\Delta f_{n}^{J+1}(\mu,g)$}\label{PertMFARem}

Since  the global existence of the trivial solution
is established and since this solution can be expanded
in a perturbation series  (\ref{expansionfn(mu)})
w.r.t. $g\,$ (\ref{g}) as shown in the previous
section, we can write for any $\,\mu\,$
\begin{equation}\label{expansionfn(mu)}
  f_n(\mu)\,=\,
  \sum_{j=1}^J g^j\,f_{n,j}(\mu)\,+\, g^{J+1}\, \Delta f_{n}^{J+1}(\mu,g)\ .
\end{equation}
 We will show that for $\mu\,$ close
to $\,\mu_{\max}\,$, $\,\Delta f_n^{J+1}(\mu,g)$ is not singular when
$\,g\rightarrow 0\,$. From the mean-field flow equations (\ref{MFE-2}) and the
perturbative expansion (\ref{perturbative_exp}),
we find the mean-field flow equations
satisfied by the remainder $\,\Delta f_n^{J+1}(\mu,g)$
\begin{equation}\label{MFE1_remainder_ugly_form}
\begin{split}
  \Delta f_{n+2}^{J+1}(\mu,g) &\, =\,\dfrac{2}{n(n+1)}\
  \partial_\mu \Delta f_n^{J+1}(\mu,g)+\dfrac{n-4}{n(n+1)}\
  \Delta f_n^{J+1}(\mu,g) \\
  &+\, \dfrac{1}{n+1}\sum_{n_1+n_2=n+2}\Big[g^{J+1}\, \Delta f_{n_1}^{ J+1}(\mu,g)\,
    \Delta f_{n_2}^{J+1}(\mu,g)\,+\, \Delta f_{n_1}^{J+1}(\mu,g)
    \sum_{j=1}^J g^jf_{n_2,j}(\mu) \\
    &\,+\, \Delta f_{n_2}^{J+1}(\mu,g)\sum_{j=1}^J g^j\,f_{n_1,j}(\mu)\ +
    \sum_{\substack{J<j_1+j_2\leq 2J \\
        1\leq j_i\leq J} }g^{j_1+j_2-(J+1)}
    \ f_{n_1,j_1}(\mu)\ f_{n_2,j_2}(\mu)\Big]\ .   
\end{split}
\end{equation}
In this form (\ref{MFE1_remainder_ugly_form})
the flow equations  are inconvenient
for our analysis
because the
dynamical system (\ref{MFE1_remainder_ugly_form})
is not homogeneous w.r.t.
$g\,$.
But  (\ref{MFE1_remainder_ugly_form}) can be recast 
into a more suitable form.
The sum of the first and the third term in square brackets
give $\,f_{n_1}(\mu)\ \Delta f_{n_2}^{J+1}(\mu,g)\,$.
The second plus fourth term give
\begin{equation}
\begin{split}
  &\quad \, \sum_{j=1}^J f_{n_1,j}(\mu)\sum_{s=1}^j g^{j-s}\;f_{n_2,J+1-s}(\mu)
  +\Delta f_{n_1}^{J+1}(\mu,g)\sum_{j=1}^J g^j\;f_{n_2,j}(\mu) \\
  &= \sum_{s=1}^J f_{n_2,J+1-s}(\mu)\Big(\sum_{j=s}^J g^{j-s}
  \; f_{n_1,j}(\mu)+g^{J+1-s}\ \Delta f_{n_1}^{J+1}(\mu,g)\Big) \\
&= \sum_{s=1}^J f_{n_2,J+1-s}(\mu)\ \Delta f_{n_1}^s(\mu,g)\ ,
\end{split}
\end{equation}
where we used the relation
\begin{equation}
  g^{J+1}\ \Delta f_n^{J+1}(\mu,g)\,=\,
  \sum_{i=J+1}^{J'}g^i \,f_{n,i}(\mu)\,+\, g^{J'+1}\,
  \Delta f_n^{J'+1}(\mu,g)\ ,\quad J'>J\,\geq\, 0\ .
\end{equation}
Therefore (\ref{MFE1_remainder_ugly_form}) can be rewritten as
\begin{equation}\label{MFE2_remainder_good_form}
\begin{split}
  \Delta f_{n+2}^{J+1}(\mu,g) &\,=\, \dfrac{2}{n(n+1)}\
  \partial_\mu \Delta f_n^{J+1} (\mu,g)\,+\,\dfrac{n-4}{n(n+1)}\
  \Delta f_n^{J+1} (\mu,g) \\
  &\,+\,\dfrac{1}{n+1}\sum_{n_1+n_2=n+2}
  \Big[\sum_{j=1}^J f_{n_2,J+1-j}(\mu)\ \Delta f_{n_1}^j (\mu,g)\,+\,f_{n_1}(\mu)\;
    \Delta f_{n_2}^{J+1} (\mu,g)\Big]\ .
\end{split}
\end{equation}
We will use the flow equations (\ref{MFE2_remainder_good_form}) 
to prove Borel summability of the perturbation series
of the regularized renormalized mean-field CAS. The boundary conditions
for the remainders are determined by the boundary conditions for
the $\,f_{n,j}(\mu)\,$ and for $\,f_n(\mu)\,$.
The bounds are established using the following induction scheme:
\begin{itemize}
\item We start from the remainders $\,\Delta f_2^{J+1}(\mu,g)\,$  for an
  arbitrary value of $\,J\geq 1\,$.
\item From (\ref{MFE2_remainder_good_form}) we can compute
  $\Delta f_{n+2}^{J+1}(\mu,g)$ from the remainders $\,\Delta f_{n'}^{J'}(\mu,g)\,$
  for $n'\leq n\,$ and $\,J'\leq J+1\,$, from the perturbative solutions
  $\,f_{m,j}(\mu)\,$ for $\,m\,\leq\, n\,$ and
  $\,j\leq J+1\,$, and from the global
  solutions
  $\,f_{n''}(\mu)\,$ for $\,n''\leq n\,$.
\end{itemize} 
From Lemma \ref{amanalytic} we have for $\,j\geq 2\,$ 
\begin{equation}\label{qfdf}
 \vert f_{4,j}(\mu_{\max})\vert\,\leq\, C_4 \;j\; 2^j    
\end{equation}
for a constant $\,C_4\,$ that does not depend on $\,j\,$. 
Since $\,F_2(\lambda,z)\,$ from (\ref{pmi})
is analytic for $\,|z| \,<\, \frac{1}{6}\,$,
we find for  $\,g < \frac16\,$ 
\begin{equation}\label{taylor}
  f_2(\mu)=F_2(\mu,g)
  =\sum_{j=1}^J  g^j\; a_j(\mu)+ g^{J+1}\ 
  \Delta f_2^{J+1}(\mu, g)\ ,
\end{equation}
where the remainder of the perturbative expansion of the 
two point function is given by 
\begin{equation}\label{reste}
  \Delta f_2^{J+1}(\mu, z)\,=\,\frac{1}{J!}
  \int_0^1 dt\;(1-t)^J\;\partial_z^{J+1} F_2(\mu,t\,z)\ .  
\end{equation}

\begin{proposition}\label{boundsremainderf2}
We have for $\,l\geq 0\,$ and $\vert z\vert<\frac{1}{6}$
\begin{equation}
  \vert \partial_\mu^l \Delta f_2^{J+1}(\mu, z)\vert
  \,\leq\,C_5^{J+1+l}\  \dfrac{(J+1+l)!}{(J+1)!}
\end{equation}
for a suitable constant $\,C_5 >0\,$.
\end{proposition}

\begin{proof}
  Since $F_2(\mu,z')$ is analytic w.r.t. $(\mu,z')\in\Omega\,$, and
 since  for $\,t\,\in\, [0,1]\,$ and $\vert z\vert<\frac{1}{6}$ we have
  $\,(\mu,t  z)\in\Omega\,$, we get the following bounds 
\begin{equation}\label{fk}
 \big\vert\, \partial_\mu^l\partial_{ z}^{J+1} F_2(\mu,t z)\,\big\vert
  \,\leq\, C_5^{J+1+l}\,(J+1+l)!
    \end{equation}
    for a suitable constant $\,C_5\,$.
    From the uniform bounds (\ref{fk}) 
    \begin{equation}
      \big\vert \,\partial_\mu^l \Delta f_2^{J+1}(\mu, z)\,\big\vert
      \,\leq\,
     C_5^{J+1+l}\ \dfrac{(J+1+l)!}{J!}\int_0^1dt\;(1-t)^J
            \,=\, C_5^{J+1+l}\ \dfrac{(J+1+l)!}{(J+1)!}\ .
\end{equation}
\end{proof}

Proposition \ref{boundsremainderf2} implies that the Borel transform
of the perturbative series (\ref{perturbf2}) w.r.t. $\, g<\frac{1}{6}\,$ exists
on the whole complex plane.
Subsequently we analyze the remainders $\,\Delta f_n^{J+1}(\mu,g)\,$
for $n\geq 4\,$. They are constructed from the remainder
$\,\Delta f_2^{J+1}(\mu, g)\,$ using the flow equations
(\ref{MFE2_remainder_good_form}). 


\subsection{The definition of local Borel summability}
\label{LocBoreSummRenordef}

We recall the definition of local Borel summability.
Let $F(t)$ be a formal power series
\begin{equation}
   F(t):= \sum_{n\geq 0} a_n\ t^n\ .
\end{equation}
We say that the formal power series $F(t)$ is locally Borel-summable if
\begin{itemize}
\item $B(t):=\,\sum_{n\geq 0}\frac{a_n}{n!}\,\,t^n\,$ converges in a
  circle of radius $\,r>0\,$.
    \item $B(t)$ can be analytically continued to a neighborhood of the
      positive real axis.
    \item The function 
    \begin{equation}
        g(z):=\frac{1}{z}\int_0^{+\infty}dt\; e^{-\frac{t}{z}}\ B(t)
    \end{equation}
    converges for some $\,z\neq 0\,$.
\end{itemize}
$B(t)$ is called the Borel transform of the power series $\,F(t)\,$ and
$g(z)$ is called its Borel sum. One sees that $\,g(z)\,$ is a Laplace
transform of the Borel transform of $\,F(t)\,$. It is known that the
Laplace transform converges in right half-planes \cite{Laplacetransform}.
Theorems on local Borel summability of quantum field theories usually
rely on Watson's theorem \cite{Watson} which gives a sufficient condition
for local Borel summability. Sokal pointed out that an improved version
has been established by Nevanlinna \cite{Nevanlinna}. Here we will state
the theorem proven by Sokal \cite{SokalNevanlinna}, giving a necessary
and sufficient condition for local Borel summability.

\begin{theorem*}\label{ThmNevanlinnaSokal}
  Let $f$ be analytic in the circle
  $C_R:=\lbrace z\in \C,\quad \mbox{Re}(z^{-1})>R^{-1}\rbrace$ such that
    \begin{equation}\label{SokalNevanlinnaremainder}
      f(z)=\sum_{k=0}^{N-1}a_k\, z^k\,+\,R_N(z)\;,
      \quad \vert R_N(z)\vert\,\leq \,A\,
      \sigma^N N!\;\vert z\vert ^N\;,\quad z\in C_R\;,
    \end{equation}
    uniformly in $N$ and for suitable constants $\,A,\sigma\,$.
    Then the Borel transform $\,B(t)\,$ converges for
    $\,\vert t\vert\leq\frac{1}{\sigma}\,$ and can be continued analytically
    to the striplike region
    $S_\sigma:=\lbrace t\in\C\;\vert\ \ d(t,\R_+)\,<\,
    \frac{1}{\sigma}\rbrace$
    and satisfies the bound
    \begin{equation}\label{boundexponenNevanlinnaSokal}
        \vert B(t)\vert\,\leq\, K\ e^{\frac{\vert t\vert}{R}}
    \end{equation}
    uniformly in every strip $S_{\sigma'}\,$ with $\sigma'>\sigma\,$.
    Moreover, $f(z)$ can be recovered and represented by the absolutely
    convergent integral
    \begin{equation}\label{Borelsum}
      f(z)=\frac{1}{z}\int_0^{+\infty}dt\; e^{-\frac{t}{z}}\;B(t)\;,
      \quad z\in C_R\;.
    \end{equation}
    Conversely, if $\,B(t)\,$ is analytic in a strip $\,S_{\sigma''}\,$ for
    $\,\sigma''<\sigma\,$ and satisfies the bound
    (\ref{boundexponenNevanlinnaSokal}),
    then the function $\,f(z)\,$ defined in (\ref{Borelsum}) is analytic in the
    circle $\,C_R\,$ and (\ref{SokalNevanlinnaremainder}) holds with
    $\,a_n\,=\,\frac{d^n}{dt^n}B(t)\vert_{t=0}\,$ uniformly in the set of
    circles $\,C_{R'}\,$ with $\,R'<R\,$.
\end{theorem*}

\begin{figure}[h!]
    \centering
    \includegraphics[width=1\linewidth]{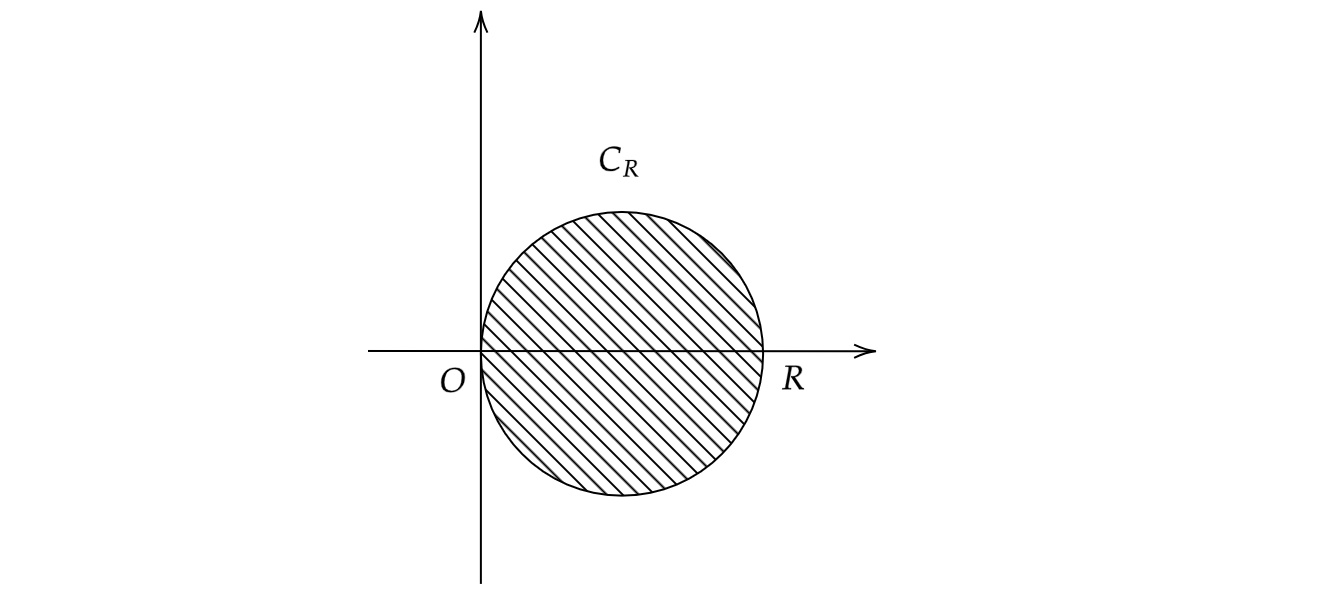}
    \caption{The region of analyticity of the Borel-summable function.}
    \label{circle}
\end{figure}

\begin{figure}[h!]
    \centering
    \includegraphics[width=0.9\linewidth]{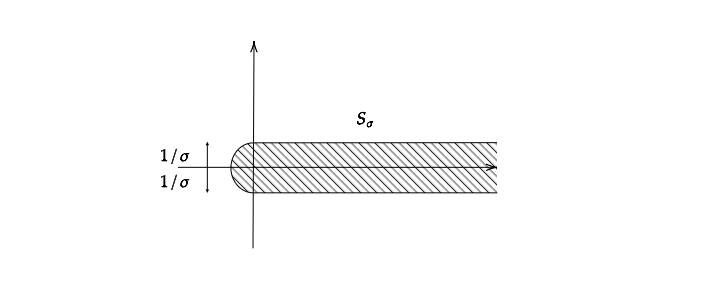}
    \caption{The region of analyticity of the Borel transform of a
      function satisfying the assumptions of Nevanlinna-Sokal theorem.}
    \label{strip}
\end{figure}
\newpage


\subsection{Asymptoticity of the perturbative expansion
  and local Borel summability}\label{Positiverenormcoupl}

We suppose as before 
\begin{equation}
 \vert\, c_{0,2}\, \vert
 \,\leq\,\frac{K}{2^5\pi^4}\ \Lambda_0^2\;,\quad K\,\leq\,\frac{1}{30}
 \ ,\quad 
 0\,<\,c_{0,4}\,\leq\, \frac{K}{40\pi^2}
    \end{equation}
and consider the  renormalization conditions
(\ref{two}), (\ref{four}).
The corresponding renormalization constants $\,\mathcal{A}_j\,$ are
\begin{equation}\label{behaviorcomplexrenormcondiAj}
 \mathcal{A}_j= m^2\, c\;\delta_{j,1} \;.
\end{equation}
From (\ref{qfdf})  we have
\begin{equation}\label{behaviorcomplexrenormcondiBj}
    \big\vert\, \mathcal{B}_j\, \big\vert\,\leq\, C_4 \;j\; 2^j\;.
\end{equation}
We now prove bounds on the remainders $\,\Delta f_n^{J+1}(\mu,g)\,$.
We assume $\,\mu_{\max}>6\,$ and we fix
$\,\mu\,>\,\mu_{\max}\,-\,1\,$.
In \cite{mypaper} we derived bounds for the smooth
solutions $f_n(\mu)\,$: 
\begin{lemma}\label{more_precise_bounds_derivatives_f_2}
For a constant $K_1$
\begin{equation}\label{bounds_generic_derivatives_f_2}
  \big\vert\,\partial_\mu^l f_2(\mu)\,\big\vert
 \, \leq \,  \frac{C_6^{l+1}}{\,M_l(\mu)}\ l!\;\;,
  \quad l\geq 0\, ,\quad \mu\in (0,\mu_{\max}]
\end{equation}
for a suitable constant $\,C_6\,$
with the definition
\begin{equation}
    M_l(\mu):=\,\min\lbrace \mu^{2l+1},\mu^l\rbrace\ .
\end{equation}
\end{lemma}

\begin{proof}
    See \cite{mypaper}.
\end{proof}

\begin{lemma}\label{bounds_f_n_more_precise}
  Let $\,f_n(\mu)\,$ be smooth mean-field solutions of the flow equations
  (\ref{MFE-2}). If the derivatives of the two point
  function $\,\partial_\mu^l f_2(\mu)\,$ satisfy the bounds
  (\ref{bounds_generic_derivatives_f_2}),
  we have for a constant $\,C_7\,>\,C_6\,$
    \begin{equation}
       \big\vert\, \partial_\mu^l f_n(\mu)\,\big\vert
      \,\leq\, \dfrac{C_7^{n+l-1}}{(l+1)^2}\ \dfrac{(n+l)!\;}{n!}\ 
      \dfrac{1}{\mu^{2l+n-1}}\ ,\quad n\geq 2\ ,\ \; l\geq 0\ ,\ \; \mu<1\ ,
    \end{equation}
    and
    \begin{equation}
      \big\vert\,\partial_\mu^l f_n(\mu) \,\big\vert\,\leq\,
      \dfrac{C_7^{n+l-1}}{(l+1)^2}\dfrac{(n+l)!\;}{n!}\ ,
      \quad n\geq 2\ ,\;\  l\,\geq\, 0\ ,\;\  \mu\geq 1\ \,.
    \end{equation}
\end{lemma}

\begin{proof}
    See \cite{mypaper}.
\end{proof}

Now we turn to the main result regarding the local Borel summability
of the regularized renormalized mean-field perturbation theory,
in the case of a real coupling.

\begin{lemma}\label{Borel_summability_f_n}
    The remainders $\,\Delta f^{J+1}_n(\mu,g)\,$ satisfy the following bounds
    \begin{equation}\label{boundsBorelsumm}
      \big\vert\, \partial_\mu ^l \Delta f^{J+1}_n(\mu,g)\,\big\vert\,\leq\,
      C_8^{J+n+l-1}\
      \dfrac{(n+J+l)!}{(n-1)!}\, ,\quad n\geq 2\  ,
      \;\  l\geq 0\ ,\;\  J\geq 0
    \end{equation}
for a suitable  constant $\,C_8 >0\,$.
\end{lemma}

\begin{proof}
  The proof is done by induction in $\,n+J+l\,$,
  going up in $n,\ J\,$ at a fixed
  value of $\,n+J+l\,$.
  For $\,n=2\,$ the bounds follow from Proposition \ref{boundsremainderf2}.
  The bounds (\ref{boundsBorelsumm}) can be checked explicitly for
  $\,n=4\,$.
  To prove the statement for  $\,n\geq 4\,$ 
  we differentiate (\ref{MFE2_remainder_good_form})
  $l$ times w.r.t. $\mu$ to obtain
\begin{equation}\label{wholeeqn_Borel_f_n}
\begin{split}
&\partial_{\mu}^l\Delta f_{n+2}^{J+1}(\mu,g) =\, \dfrac{2}{n(n+1)}
  \partial_\mu^{l+1}\Delta f_n^{J+1} (\mu,g)+\, \dfrac{n-4}{n(n+1)}\partial_\mu^l \Delta f_n^{J+1}(\mu,g) \\
  &+\, \dfrac{1}{n+1}\sum_{\substack{n_1+n_2=n+2 \\
      l_1+l_2=l}}\binom{l}{l_1} \bigg[\sum_{j=1}^J
    \partial_\mu^{l_2} f_{n_2,J+1-j}(\mu)\;\partial_\mu^{l_1}
    \Delta f_{n_1}^{j}(\mu,g)\,+\, \partial_\mu^{l_1} f_{n_1}(\mu)
    \;\partial_\mu^{l_2} \Delta f_{n_2}^{J+1}(\mu,g)\bigg]\ .
\end{split}
\end{equation}

We analyze each term in the r.h.s of (\ref{wholeeqn_Borel_f_n}):

\begin{itemize}
    \item First term: we insert the induction hypothesis, it is bounded
    \begin{equation}\label{Borel_first}
      \dfrac{2}{n(n+1)}\, C_8^{J+n+l}\,\dfrac{(n+J+l+1)!}{(n-1)!}\,\leq\,
      C_8^{J+n+l+1}\ \dfrac{(n+J+l+2)!}{(n+1)!}\ \dfrac{2}{C_8\,(n+J+l+2)}\ .
    \end{equation}
    \item Second term: it is bounded by
    \begin{equation}\label{Borel_second}
    \begin{split}
      \dfrac{n-4}{n(n+1)}\ C_8^{J+n+l}\ \dfrac{(n+J+l)!}{(n-1)!} &\,\leq\,
      C_8^{J+n+l+1}\  \dfrac{(n+J+l+2)!}{(n+1)!}\ \dfrac{(n-4)}{C_8\,n^2}\;.   
    \end{split}
    \end{equation}
  \item Third term: we use the induction hypothesis and Proposition
    \ref{bound_pertur_mu_simpl} to bound the third term by
    \begin{equation}
      \dfrac{1}{n+1}\sum_{\substack{n_1+n_2=n+2 \\
          l_1+l_2=l}}\binom{l}{l_1}
      \sum_{j=1}^J C_8^{j+n_1+l_1-1}\ C'^{J+1-j+\frac{n_2}{2}+l_2}\
      \dfrac{(n_1+j+l_1-1)!}{(n_1-1)!}\ 
      \dfrac{(J+1-j+l_2+1)! }{(\frac{n_2}{2})^2\;(\frac{n_2}{2})!}\ .
    \end{equation}
    We use the crude bound
\begin{equation}
  \frac{1}{(\frac{n_2}{2})!}\,\leq\,
  \dfrac{3}{2}\,\dfrac{(n_2-2)!}{(n_2-1)!},\quad n_2\in 2\N\ ,
\end{equation}
and  the Vandermonde inequality (\ref{bound_Vandermonde})
together with  $\,m!\;n!\leq (m+n)!\,$ to obtain
\begin{equation}
  \dfrac{1}{n!}\binom{J}{j}\,\binom{l}{l_1}\,
  \binom{n}{n_1-1}(n_1+j+l_1-1)!\; (n_2+J-j+l_2)! \,\leq\,
  \dfrac{(n+J+l+1)!}{n!}\ .
\end{equation}

Choosing $\,C_8\,>\,2\pi\, C'\,$ and using
\begin{equation}
    \sum_{j=1}^J \binom{J}{j}^{-1}\,\leq\, 6
\end{equation}
 we can bound  the third term by
\begin{equation}\label{Borel_third}
  \dfrac{1}{4}\,  C_8^{J+n+l+1}\
 \dfrac{(n+J+l+2)!\; l}{(n+1)!\;(n+J+l+2)}\,\leq\, 
 \dfrac{1}{4}\,  C_8^{J+n+l+1}\ \dfrac{(n+K+l+2)!}{(n+1)!}\ .
\end{equation}
\item Fourth term: we use Lemma \ref{bounds_f_n_more_precise}
  and we insert the induction hypothesis to obtain
    \begin{equation}
      \frac{1}{n+1}\sum_{\substack{n_1+n_2 \\
          l_1+l_2=l}}\binom{l}{l_1}
      \dfrac{C_7^{n_1+l_1-1}(n_1+l_1)!}{n_1!\;
        (l_1+1)^2}\ C_8^{J+n_2+l_2-1}\ \dfrac{ (n_2+J+l_2)!}{(n_2-1)!}\ .
    \end{equation}

We use again (\ref{bound_Vandermonde}) to obtain
\begin{equation}
\begin{split}
 \binom{l}{l_1}\dfrac{(n_2+J+l_2)!\; (n_1+l_1)!}{n_1!\;(n_2-1)!} 
 &=\,\binom{J}{0}\binom{l}{l_1}\binom{n+1}{n_1}\dfrac{1}{(n+1)!}
 (n_2+J+l_2)!\; (n_1+l_1)! \\
 &\leq\,\dfrac{1}{(n+1)!}\,\binom{n+J+l+1}{n_1+l_1}\,(n_2+J+l_2)!\;
 (n_1+l_1)! \\
 &\leq\, (n-2+J+l)\,\dfrac{(n+l+J+1)!}{(n+1)!}\,\leq\,
 \dfrac{(n+J+l+2)!}{(n+1)!}\;.
\end{split} 
\end{equation}
The fourth term is then bounded by
\begin{equation}\label{Borel_fourth}
C_8^{J+n+l+1}\ \dfrac{(n+J+l+2)!}{(n+1)!}\ \frac{\pi^2}{12\,C_8}    
\end{equation}
choosing $\,C_8\,\ge\,C_7\,$.
\end{itemize}
Summing together (\ref{Borel_first}), (\ref{Borel_second}),
(\ref{Borel_third}) and (\ref{Borel_fourth}) we finally obtain
\begin{equation}
\begin{split}
  \big\vert\,\partial_\mu^l\Delta f_{n+2}^{J+1}(\mu,g)\,\big\vert &\,\leq\,
  \bigg[\dfrac{1}{C_8}+\dfrac{1}{C_8}\,+\,\dfrac{1}{4}\,+\,
    \dfrac{\pi^2}{12\,C_8}\bigg]\,
  C_8^{J+n+l+1}\ \dfrac{(n+J+l+2)!}{(n+1)!} \\
   &\,\leq\, C_8^{J+n+l+1}\ \dfrac{(n+J+l+2)!}{(n+1)!}\ ,
\end{split}
\end{equation}
if we choose $\,C_8\,>\, \max\lbrace C_7,4\rbrace\,$.
\end{proof}
\noindent
We collect our findings from Propositions 
  \ref{BPHZpossible} and \ref{pertr},
and from Lemma \ref{Borel_summability_f_n} in
 \begin{theorem}[The renormalized perturbative expansion is asymptotic]
  \label{ThmBoreltransform}
  Consider the bare interaction lagrangian (\ref{bare_lagrangian_trivial})
  of mean-field $\,\varphi_4^4$-theory corresponding to the boundary
  conditions  (\ref{BDY_trivial_field}) for the solutions (\ref{Ais})
  of the  flow equations (\ref{MFE-1}).
  We assume 
\begin{equation}
 \big\vert\, c_{0,2}\,\big\vert
 \,\leq\, \frac{K}{2^5\pi^4}\Lambda_0^2\ ,\quad
 0\, <\, c_{0,4}\,\leq\, \frac{K}{40\pi^2}\ , \quad K\,\leq\,\frac{1}{30}\ \,.
\end{equation}
The mean-field solutions $\,A_n^{\alpha_0,\alpha}\,, \ n\ge 4\ ,$
vanish logarithmically in the UV-limit
\begin{equation}
  \lim\limits_{\alpha_0\rightarrow 0}  A_n^{\alpha_0,\alpha}\,=\,0
  \ ,\quad n\,\geq\, 4\ .
\end{equation} 
The renormalized coupling  $\,g\,$ (\ref{g})
also vanishes logarithmically in this limit.
The (rescaled) mean-field (connected amputated) Schwinger
functions $\,f_n(\mu)\,$ (\ref{f_n}) have a perturbative expansion
in powers of $\,g\,$
\begin{equation}
  f_n(\mu)\,=\,\sum_{j=1}^J g^j\, f_{n,j}(\mu)\,+\,g^{J+1}\,
  \Delta f_{n}^{J+1}(\mu,g)\ ,
  \quad \mu\in \big(\mu_{\max}\,-\, 1,\,\mu_{\max}\big]\ .
\end{equation}
The perturbative series is asymptotic to the trivial solution $\,f_n(\mu)\,$
obeying the same boundary conditions:
\begin{equation}\label{estimatesremainder}
 \Big\vert  f_n(\mu)-\sum_{j=1}^J g^j\; f_{n,j}(\mu)\Big\vert\,\leq\, g^{J+1}\;
  {C}^{J+n}\ \dfrac{(n+J)!}{(n-1)!}\ ,\quad n\geq 2\,,\ \, J\geq 0\ ,
  \quad \mu\in \big(\mu_{\max}-1,\,\mu_{\max}\big]   
\end{equation}
for a suitable constant $\,{C}>0\,$.
\end{theorem}
   
In order to be able to apply the Nevanlinna-Sokal Theorem
we now analyze the extension to complex couplings.
We still assume
$\,\mu\,>\, \mu_{\max}-1\,$ and
$\,\mu_{\max}>6\,$.
From the perturbative expansion (\ref{perturbf2}) and Lemma \ref{amanalytic}
in Sect.\ref{PerturbExpaning}, $\,F_2(\mu,z)\,$ from
(\ref{pmi}) can be analytically
continued to $\,(\mu_{max}-1,\mu_{max}]\,\times\, D(0,\frac{1}{6})\,$. We
choose $\,0 < R < 1/6\,$ implying $\,C_R\subset D(0,\frac{1}{6})\,$,
and we assume $\, z\,\in\, C_R\,$.
\begin{remark}
  Due to triviality the (real) renormalized coupling $\,g\,$ is small
  for large values of the cutoff. Thus the condition $\,z\in C_R\,$ is in
  fact not very stringent from the point of view of application.
  \end{remark}

\begin{itemize}
\item
  We can then analytically extend (\ref{expansionfn(mu)}) to complex values
  of the coupling (remember (\ref{F2}))
  \begin{equation}\label{expansionF2(mu)}
  F_2(\mu,z)\,=\,
  \sum_{j=1}^J z^j\,f_{2,j}(\mu)\,+\, z^{J+1}\, \Delta f_{2}^{J+1}(\mu,z)\ .
  \end{equation}
  and then also via the mean-field flow equations (\ref{MFE-2}) the n-point functions $F_n(\mu,z)$ which are constructed from $F_2(\mu,z)$.

  \begin{lemma}\label{analyticallyF_n(mu,z)}
      We have
    \begin{equation}\label{expansionFn(mu)}
  F_n(\mu,z)\,=\,
  \sum_{j=1}^J z^j\,f_{n,j}(\mu)\,+\, z^{J+1}\, \Delta f_{n}^{J+1}(\mu,z)\,
  \end{equation}
  where the remainders $\Delta f_{n}^{J+1}(\mu,z)$ satisfy the mean-field flow equations for the remainders (\ref{MFE2_remainder_good_form}).
  \end{lemma}

  \begin{proof}
      The proof is done by induction in $n+J$ going up in $n$. The claim holds for $n=2$. For $n\geq 2$, we insert (\ref{expansionFn(mu)}) in the mean-field flow equations (\ref{MFE-2}). Using the perturbative mean-field flow equations (\ref{pertMFE-2}), we have
      \begin{equation}
          F_{n+2}(\mu,z)=\sum_{j=1}^J z^j\; f_{n+2,j}(\mu)+z^{J+1}\Delta f_{n+2}^{J+1}(\mu,z)\;.
      \end{equation}
      Proceeding as in Sect. \ref{PertMFARem}, the remainders $\Delta f_{n+2}^{J+1}(\mu,z)$ satisfy (\ref{MFE2_remainder_good_form}).
  \end{proof}
  
    The bounds from  Lemma \ref{more_precise_bounds_derivatives_f_2}
  extended to the $\,\mu$-derivatives of $\,F_2(\mu,z)\,$ for $\mu\in (\mu_{\max}-1,\mu_{\max}]$
 remain valid since the function $F_2(\mu,z)$ is analytic w.r.t. $\mu\in (\mu_{\max}-1,\mu_{\max}]$. Then the bounds from Lemma \ref{bounds_f_n_more_precise} can be extended to the $\mu$-derivatives of $F_n(\mu,z)$ without any change in the proof (see \cite{mypaper}).

 \item From Proposition \ref{boundsremainderf2} and Lemma \ref{analyticallyF_n(mu,z)} , the bounds in Lemma \ref{Borel_summability_f_n} can be extended to the $\mu$-derivatives of the remainders $\Delta f_n^{J+1}(\mu,z)$ without any change in the proof.
\item The first part of the Taylor expansion in the r.h.s.
  of (\ref{expansionF2(mu)}) is clearly analytic w.r.t. $\,{z}\,$.
\item To conclude with the Nevanlinna-Sokal theorem, we verify that the remainders $\Delta f_2^{J+1}(\mu,z)$ are analytic w.r.t. $\,{z}\,$.

\begin{lemma}\label{remainderanalytic}
 The remainder $\,\Delta f_{2}^{J+1}(\mu,z)\,$ is analytic
  w.r.t. $\,z\in C_{R}\,$.
\end{lemma}

\begin{proof}
  For $\,t\in [0,1]\,$, the integrand in (\ref{reste}) is analytic
  w.r.t. $z\,$ due to Lemma \ref{amanalytic} and the definition
  of $F_2$ (\ref{pmi}). We fix a closed curve $\,\gamma\in C_{R}\,$.
  From the uniform bounds (\ref{fk}), Fubini's theorem yields
\begin{equation}
  \oint_{\gamma} dz\  \Delta f_{2}^{J+1}(\mu,z)
  \,=\,\frac{1}{J!} \,\int_0^1 dt\, (1-t)^J\oint_\gamma d z\;
  \partial_z^{J+1}F_2\big(\mu,t\,z\big)\,=\,0\ .
\end{equation} 
We conclude with Morera's theorem.
\end{proof}

\end{itemize}
 
From the mean-field  flow equations (\ref{MFE-2}) and the
mean-field flow equations for the remainders
(\ref{MFE2_remainder_good_form}),
the analytically continued
mean-field trivial solutions $\,F_n(\mu,z)\,$ satisfy the assumptions of the
first statement of the Nevanlinna-Sokal theorem\,:
\begin{theorem}[Local Borel summability - Nevanlinna-Sokal]
  \label{NS}
        Under the same assumptions and with the same notations
        as in Theorem \ref{ThmBoreltransform},
        the analytically extended
        trivial solutions $\,F_n(\mu,z)\,$ of the mean field flow equations
        are the Borel sums of their perturbative series in the sense of the
        Nevanlinna-Sokal theorem. They thus can be uniquely recovered from
        their perturbative expansion w.r.t. $\,z\ $. The
        solutions of the mean-field flow equations are given by
        \begin{equation}
          f_n(\mu) \,=\, F_n(\mu,g)\ .
          \end{equation}
\end{theorem}

\newpage

\appendix

\section{Generalities}\label{Appendix A}

\subsection{Properties of Gaussian measures}\label{appendix_A}
We consider a Gaussian probability measure $d\mu$ on the space of 
continuous  real-valued functions $C(\Omega)$, 
where $\Omega$ is a finite (simply connected compact)
volume in $\R^d$, $d\geq 1\,$.

\subsubsection{Covariance of a Gaussian measure} 

We recall here the definition of the covariance of a Gaussian measure,
for details, see \cite{Glimm1987}.

A Gaussian measure of mean zero is uniquely characterized by its 
covariance $C(x,y)$
    \begin{equation}
        \int d\mu_C(\phi)\,\phi(x)\phi(y)=\tilde C(x,y)= \tilde C(y,x)\;.
    \end{equation}
$\tilde C\,$ is a positive non-degenerate bilinear form defined 
on $\mathcal{C}^\infty(\Omega)\times\mathcal{C}^\infty(\Omega)\,$. 
We assume that $\tilde C(x,y)$ is translation invariant, then 
$C(z):= \tilde C(x,y)\,, \ z=x-y\,$, is well defined. 
Using the notations
\begin{equation}
    \langle\phi,J\rangle=\int_\Omega d^dx\,\phi(x)J(x)\;,
\quad \langle J,CJ\rangle=\int_\Omega d^dx d^dy\, J(x)C(x-y)J(y)
\end{equation}
with $J\in\mathcal{C}^\infty (\Omega)$, 
the generating functional of the correlation functions is
\begin{equation}
    \int d\mu_C(\phi)e^{\langle\phi,J\rangle}=e^{\frac{1}{2}\langle J,CJ\rangle}\;.
\end{equation}
The generating functional is also called the characteristic functional 
of the Gaussian measure $\mu_C$.
For $C=(-\Delta+I)^{-1}$, where $\Delta$ denotes the Laplacian operator 
in $\R^d$, the corresponding Gaussian measure $\mu_C$ is supported on 
distributions with $1-\frac{d}{2}-\varepsilon$ continuous derivatives, 
$\varepsilon>0$. For a regularized propagator, the Fourier transform
of which falls off rapidly in momentum space, 
the Gaussian measure is supported on smooth functions. 

\subsubsection{Properties of Gaussian measures}
 We list here some properties of Gaussian measures. Proofs can be found in 
\cite{Glimm1987}.
 \begin{itemize}
     \item Integration by parts: Let $A(\phi)$ be a polynomial in 
$\phi(x)$ and its derivatives $\partial_\mu\phi(x)$. 
     \begin{equation}\label{IPP}
        \int d\mu_C(\phi) \phi(x)A(\phi)
=\int d\mu_C(\phi)\int_\Omega dy\;C(x-y)\dfrac{\delta}{\delta\phi(y)}A(\phi)\;.
    \end{equation}
    \item Translation of a Gaussian measure: Let $C$ be a covariance. 
Under a change of variable $\phi=\varphiÂ¨+\psi$ for 
$\varphi\in\mbox{supp}(\mu_C)$ and $\psi$ such that its Fourier transform
$\hat \psi(p)$ is compactly supported.
    \begin{equation}\label{Translation_GM}
    d\mu_C(\phi)
=e^{-\frac{1}{2}\langle\psi,C^{-1}\psi\rangle}
e^{-\langle C^{-1}\psi,\varphi\rangle}d\mu_C(\varphi)\;.
\end{equation}
\item Decomposition of the covariance: Assume that
\[C=C_1+C_2\;,\quad C_i>0\;.\]
Then for $A(\phi)$ as in (\ref{IPP})
\begin{equation}
    \int d\mu_C(\phi)A(\phi)
=\int d\mu_{C_1}(\phi_1)\int d\mu_{C_2}(\phi_2)A(\phi_1+\phi_2)\;.
\end{equation}
\item Infinitesimal change of covariance: We assume the covariance 
depends on a parameter $t$, and is differentiable w.r.t. $t$
\[
C(x-y)\equiv C_t(x-y)\;,\quad \Dot{C}_t(x-y):=\dfrac{d}{dt}C_t(x-y)
\;.\]
Let $F(\phi)$ be a smooth functional, integrable w.r.t. 
$\mu_{C_t}$ $\forall t\,$. We have
\begin{equation}\label{differentiation}
    \dfrac{d}{dt}\int d\mu_{C_t}(\phi)F(\phi)
=\dfrac{1}{2}\int d\mu_{C_t}(\phi)
\left\langle\dfrac{\delta}{\delta\phi},
\Dot{C}_t\dfrac{\delta}{\delta\phi}\right\rangle F(\phi)\;.
\end{equation}
 \end{itemize}

\subsection{Faà  di Bruno's formula}

Here we recall the Faà  di Bruno formula, discovered first by Faà  di Bruno \cite{FaadiBruno}.

\begin{proposition}\label{FaadiBruno1}
  Let $I,J,K$ intervals in $\R$, $g:I\rightarrow J$ and $f:J\rightarrow K$
  such that $g$ has derivatives up to order $n\in \N_0$ at $x\in I$, $y=g(x)\in J$
  and $f$ has derivatives up to order $n$ at $y=g(x)$. Then $f\circ g$ has
  derivatives up to order $n$ at $x$ and 

\begin{equation}\label{FaadiBrunoformula}
  \dfrac{d^n}{dx^n}(f\circ g)(x)=\sum_{k=1}^n\dfrac{d^k}{dy^k}f(y)
  \sum_{p(n,k)} n!\prod_{j=1}^{n-k+1}
  \dfrac{(g^{(j)}(x))^{\lambda_j}}{\lambda_j!\; (j!)^{\lambda_j}}\;,
\end{equation}
where $g^{(j)}(x)$ denotes $\frac{d^j}{dx^j}g(x)$ and the set $p(n,k)$
is defined as follows

\begin{equation}
  p(n,k):=\left\lbrace(\lambda_1,\cdots,\lambda_{n-k+1})\in\N_0^{n-k+1},
  \quad \sum_{j=1}^{n-k+1} \lambda_j=k,\quad\sum_{j=1}^{n-k+1} j\lambda_j=n \right\rbrace\;.
\end{equation}

The formula (\ref{FaadiBrunoformula}) can be rewritten as

\begin{equation}\label{Faadibrunobell}
  \dfrac{d^n}{dx^n}(f\circ g)(x)=\sum_{k=1}^n\dfrac{d^k}{dy^k}f(y)\;
  B_{n,k}(g'(x),g''(x),\cdots,g^{(n-k+1)}(x))\;,
\end{equation}

where we introduced the Bell polynomials

\begin{equation}\label{Bellpolynomials}
  B_{n,k}(x_1,x_2,\cdots,x_{n-k+1}):= \sum_{p(n,k)} n!
  \prod_{j=1}^{n-k+1} \dfrac{x_j^{\lambda_j}}{\lambda_j!\; (j!)^{\lambda_j}},\quad n\geq k\;.
\end{equation}
\end{proposition}

\subsection{Derivatives of 
$\frac{f}{g}$}\label{l-th_derivative_f/g}

We prove
\begin{proposition}
    For $f,g$ smooth with $g>0$,
    \begin{equation}\label{formula_appendix_D}
        \left(\dfrac{f}{g}\right)^{(l)}=\frac{1}{g}
\left[f^{(l)}-l!\;\sum_{j=1}^{l}\frac{g^{(l+1-j)}}{(l+1-j)!}
\frac{1}{(j-1)!}\left(\dfrac{f}{g}\right)^{(j-1)}\right]\;.
    \end{equation}   
\end{proposition}

\begin{proof}
The proof is done by induction in $l\in\N$. For $l=1$, the statement 
is easily verified. 
Then differentiating (\ref{formula_appendix_D}) and using the 
induction hypothesis, we obtain
\begin{equation}
    \begin{split}
       \left(\dfrac{f}{g}\right)^{(l+1)} 
&= \dfrac{f^{(l+1)}}{g}-\dfrac{g' f^{(l)}}{g^2}+\frac{g'}{g^2}
\sum_{j=1}^l \binom{l}{j-1}g^{(l+1-j)}\left(\dfrac{f}{g}\right)^{(j-1)} \\
       &-\frac{1}{g}\sum_{j=1}^l \binom{l}{j-1}\Big(g^{(l+2-j)}
\left(\dfrac{f}{g}\right)^{(j-1)}+g^{(l+1-j)}\left(\dfrac{f}{g}\right)^{(j)}
\Big) \\
       &= \dfrac{f^{(l+1)}}{g}-\frac{g'}{g} 
\left(\dfrac{f}{g}\right)^{(l)}-\frac{g^{(l+1)}}{g}\dfrac{f}{g}-l 
\frac{g'}{g} \left(\dfrac{f}{g}\right)^{(l)} \\
       &-\frac{1}{g}\sum_{j=2}^l \Bigg[\binom{l}{j-1}
+\binom{l}{j-2}\Bigg]g^{(l+2-j)}\left(\dfrac{f}{g}\right)^{(j-1)} \\
       &= \frac{1}{g}\left[f^{(l+1)}-(l+1)!\;\sum_{j=1}^{l+1}
\frac{g^{(l+2-j)}}{(l+2-j)!}\frac{1}{(j-1)!}
\left(\dfrac{f}{g}\right)^{(j-1)}\right]\;,
    \end{split}
\end{equation}
where we used 
\begin{equation}
    \binom{n}{k}+\binom{n}{k-1}=\binom{n+1}{k},\quad n\in\N_0\,,\ k\in\N\;.
\end{equation}  
\end{proof}

\section{Proof of the bounds of the mean-field perturbative CAS-functions}\label{appendixB}

\subsection{Useful inequalities}\label{usefulinequalities}

In order to derive bounds on the derivatives $\partial_\alpha^k
\mathcal{A}_{n,j}^{\alpha_0,\alpha}$, we will first prove useful and elementary
bounds which we will use in the proof of Lemma
\ref{Prop_bound_perturbative_solutions}.

\begin{lemma}\label{inversesquare}
    For $n\geq 12$
\begin{equation}
    \frac{n}{n-2}\sum_{\substack{n_1+n_2=n+2\\
n_i\geq 4,n_i\in 2\N}}\dfrac{1}{n_1^2(n+2-n_1)^2}\leq\frac{1}{n^2}\;.
\end{equation}
\end{lemma}

 \begin{proof}
     First we have for $n\geq 12$
\[\sum_{\substack{n_1+n_2=n+2\\
n_i\geq 4,n_i\in 2\N}}\dfrac{1}{n_1^2(n+2-n_1)^2}\leq 
\dfrac{1}{16}\sum_{\substack{n_1+n_2=\frac{n}{2}+1\\
n_i\geq 2,n_i\in\N}}\dfrac{1}{n_1^2(\frac{n}{2}+1-n_1)^2}\;.\]
We use the decomposition
\[\dfrac{1}{X^2(X-A)^2}=\dfrac{1}{A^2}\left(\dfrac{1}{X^2}
+\dfrac{1}{(X-A)^2}+\dfrac{2}{AX}-\dfrac{2}{A(X-A)}\right),\quad A>0\;.\]
We get
\begin{equation*}
\begin{split}
&\sum_{\substack{n_1+n_2=n+2\\
n_i\geq 4,n_i\in 2\N}}\dfrac{1}{n_1^2(n+2-n_1)^2}\\
&\leq\dfrac{1}{4(n+2)^2}\sum_{2\leq n_1\leq\frac{n}{2}-1}
\Bigg(\dfrac{1}{n_1^2}+\frac{1}{(\frac{n}{2}+1-n_1)^2}
+\dfrac{2}{(\frac{n}{2}+1)n_1} 
+\dfrac{2}{(\frac{n}{2}+1)(\frac{n}{2}+1-n_1)}\Bigg)\\
&\leq \dfrac{1}{2(n+2)^2}\left(\zeta(2)-1+\frac{n-4}{n+2}\right)
\leq \frac{5}{6(n+2)^2}\ ,
\end{split}
\end{equation*}
where we used the fact that 
$\sum_{2\leq n_1\leq \frac{n}{2}-1}\frac{1}{n_1}\leq \frac{n-4}{4}\,$. 
Therefore we have for $n\geq 12$
\begin{align*}
\dfrac{n}{n-2}\sum_{\substack{n_1+n_2=n+2\\
n_i\geq 4}}\dfrac{1}{n_1^2(n+2-n_1)^2} 
&\leq\dfrac{5}{6(n+2)^2}\frac{n}{n-2} 
\leq \frac{5}{6n^2}\frac{n^2}{(n+2)^2}\frac{n}{n-2}\leq\frac{1}{n^2}\;.
\end{align*}
\end{proof}

\begin{lemma}\label{lemma_inequalities1}
    For $l\in \N_0$, $n\in\N$,
\begin{equation}\label{mk}
\begin{split}
  &\sum_{\substack{l_1+l_2=l \\ l_i\geq 0}}\dfrac{1}{(l_1+1)^2(l_2+1)^2}\leq
  \frac{5}{(l+1)^2},
  \quad \sum_{\substack{l_1+l_2=l \\ l_i\geq 1}}\dfrac{1}{(l_1+1)^2(l_2+1)^2}
  \leq \frac{3}{(l+1)^2}  \\
&\sum_{\substack{n_1+n_2=n+1 \\ n_i\geq 1}}\dfrac{1}{n_1^3 n_2^3}\leq \frac{4}{n^3}
    \;.     
\end{split}
\end{equation}
\end{lemma}
\begin{proof}
    
 For $l\leq 5$, the inequality can be verified by hand. For $l>5$, we have
        \begin{equation}
            \begin{split}
              \sum_{\substack{l_1+l_2=l \\ l_i\geq 0}}\dfrac{1}{(l_1+1)^2 (l_2+1)^2}
              &=\dfrac{2}{(l+1)^2}+\sum_{k=1}^{l-1}\dfrac{1}{(k+1)^2(l-k+1)^2} \\
              &\leq \dfrac{2}{(l+1)^2}+\int_0^l\dfrac{dx}{(x+1)^2(l-x+1)^2}
              = \dfrac{2}{(l+1)^2} \\
              &+\int_1^{l+1}dx\Big(\dfrac{a+bx}{x^2}
              +\dfrac{c-bx}{(l+2-x)^2}\Big)\;,
            \end{split}
        \end{equation}
    where
    \begin{equation}
      a=\dfrac{1}{(l+2)^2},\quad b=\dfrac{2}{(l+2)^3},
      \quad c=\dfrac{3}{(l+2)^2}\;.
    \end{equation}
    Then the integral equals
    \begin{equation}\label{bound_proof_lemma_inequalities1}
      \dfrac{1}{(l+2)^2}\left(2\left[1-\frac{1}{l+1}\right]
      +\frac{4}{l+2}\ln(l+1)\right)\leq\frac{3}{(l+1)^2},\quad l>5\;.
    \end{equation}
    The second statement in (\ref{mk}) is a consequence of the first one,
    since one has to subtract $\frac{2}{(l+1)^2}$ in the l.h.s.
    
    Again we can verify the inequality for $n\leq 5$. Assuming now that $n>5$,
    we proceed as before and we obtain
    \begin{equation}\label{proof_inequalities1_n}
        \begin{split}
          \sum_{\substack{n_1+n_2=n+1 \\ n_i\geq 1}}\dfrac{1}{n_1^3 n_2^3}
          &= \sum_{\substack{n_i\geq 0 \\ n_1+n_2=n-1}}\dfrac{1}{(n_1+1)^3 (n_2+1)^3} \\
          &\leq \dfrac{2}{n^3}+\sup_{1\leq n_1\leq n-1}
          \dfrac{1}{(n_1+1)(n-n_1)}\sum_{\substack{1\leq n_i \\
              n_1+n_2=n-1}}\dfrac{1}{(n_1+1)^2(n_2+1)^2} \\
          &\leq\dfrac{2}{n^3}+\dfrac{1}{2(n-1)}
          \sum_{1\leq n_1\leq n-2}\dfrac{1}{(n_1+1)^2(n-n_1)^2}
          \leq \dfrac{2}{n^3}+\dfrac{1}{2(n-1)}\frac{3}{n^2}  \\
            \leq\dfrac{4}{n^3}\;,
        \end{split}
    \end{equation}
    where we used (\ref{bound_proof_lemma_inequalities1})
    on (\ref{proof_inequalities1_n}) in the second to last inequality.
\end{proof}

\begin{lemma}\label{lemma_convolutions_inequalities}
 \begin{itemize}
       \item For integers $n\geq 3, l\geq 0,\lambda\geq 0$ 
    \begin{equation}\label{first_inequality_convol}
        \sum_{\substack{
      n_1+n_2=n+1 \\  n_i\geq 1 \\ l_1+l_2=l   
      \\ \lambda_1\leq l_1,\lambda_2\leq l_2 \\
      \lambda_1+\lambda_2=\lambda}}
        \dfrac{1}{(l_1+1)^2 (l_2+1)^2 n_1^2 n_2^2 }
        \dfrac{n!}{n_1!\; n_2!}\dfrac{\lambda !}{\lambda_1!\;
          \lambda_2!}\dfrac{(n_1+l_1-1)!\; (n_2+l_2-1)!}{(n+l-1)!}
        \leq K_0\dfrac{1}{(l+1)^2}\frac{1}{n^2}\;,
    \end{equation}
    where we may choose $K_0=20$.
    \item For $n\geq 1$, $n_1=1$, $n_2=n$
    \begin{equation}\label{second_inequality_conv}
    \sum_{\substack{
      l_1+l_2=l   
      \\ \lambda_1\leq l_1,\lambda_2\leq l_2 \\ \lambda_1+\lambda_2=\lambda}}
    \dfrac{1}{(l_1+1)^2 (l_2+1)^2 n_1^2 n_2^2 }
    \dfrac{n!}{n_1!\; n_2!}\dfrac{\lambda !}{\lambda_1!\;\lambda_2!}
    \dfrac{(n_1+l_1-1)!\; (n_2+l_2-1)!}{(n+l-1)!}\leq K'_0\dfrac{1}{(l+1)^2}\frac{1}{n^2}\;,
    \end{equation}
     where we may choose $K'_0=5$.

   \item For integers $n\geq 3, l\geq 0,\lambda\geq 0, k\geq \alpha$, $\alpha\in\N_0$. 
    \begin{equation}\label{third_inequality_convol}
    \begin{split}
      & \sum_{\substack{
      n_1+n_2=n+1 \\  n_i\geq 1 \\ l_1+l_2=l   
      \\ \lambda_1\leq l_1,\lambda_2\leq l_2 \\ \lambda_1+\lambda_2=\lambda \\
      k_1+k_2=k-\alpha}}  \dfrac{(k-\alpha)!}{k_1!\; k_2!}
      \dfrac{(n_1+l_1+k_1)!\; (n_2+l_2+k_2)!\;}{(l_1+1)^2 (l_2+1)^2  n_1^2 n_2^2\;
        (n+l+k-\alpha+1)!}\dfrac{(n+1)!}{n_1!\; n_2!}
      \dfrac{\lambda !}{\lambda_1!\;\lambda_2!}\dfrac{1}{(k_1+1)^2(k_2+1)^2} \\
     &\leq K_0''\dfrac{1}{(l+1)^2}\frac{1}{n^2}\dfrac{1}{(k-\alpha+1)^2}\;,   
    \end{split}
    \end{equation}
    where we may choose $K_0''=75$.

    \item For integers $n\geq 1, k\geq 0, n_1=1, n_2=n$ 
    \begin{equation}\label{fourth_inequality_convol}
    \begin{split}
      & \sum_{\substack{
      l_1+l_2=l   
      \\ \lambda_1\leq l_1,\lambda_2\leq l_2 \\ \lambda_1+\lambda_2=\lambda \\
      k_1+k_2=k-\alpha}}
      \dfrac{(k-\alpha)!}{k_1!\; k_2!}
      \dfrac{(l_1+k_1+1)!\; (n+l_2+k_2)!\;}{(l_1+1)^2 (l_2+1)^2 n^2\;
        (n+l+k-\alpha+1)!}\dfrac{(n+1)!}{ n!}\dfrac{\lambda !}
            {\lambda_1!\;\lambda_2!}\dfrac{1}{(k_1+1)^2(k_2+1)^2} \\
     &\leq K_0'''\dfrac{1}{(l+1)^2}\frac{1}{n^2}\frac{1}{(k-\alpha+1)^2}\;,   
    \end{split}
    \end{equation}
    where we may choose $K_0'''=25$.
    \end{itemize}
\end{lemma}

\begin{proof}
First for $n_1,n_2\geq 1,\;l_1,l_2,\lambda_1,\lambda_2\geq 0$
\begin{equation}
        \begin{split}
          & \dfrac{n!}{n_1!\;n_2!}\dfrac{\lambda!}{\lambda_1!\;\lambda_2!}
          \dfrac{(n_1+l_1-1)!\; (n_2+l_2-1)!}{(n+l-1)!} \\
          &= \dfrac{n}{n_1 n_2}\binom{n-1}{n_1-1}\binom{\lambda}{\lambda_1}
          \left[\binom{n+l-1}{n_1+l_1-1}\right]^{-1}\;.
        \end{split}
    \end{equation}
From the Vandermonde identity, we have the following inequality
 \begin{equation}\label{bound_Vandermonde}
        \binom{a}{b}\binom{c}{d}\leq \binom{a+c}{b+d},\quad a,b,c,d\in\N_0\;.
    \end{equation}
    
Then we show that for $l=l_1+l_2$,
    \begin{equation}\label{small_lemma_convolution}
      \sum_{\substack{\lambda_1\leq l_1,\lambda_2\leq l_2, \\
          \lambda_1+\lambda_2=\lambda}}\dfrac{\lambda!}{\lambda_1!\;\lambda_2!}\leq \binom{l}{l_1}\;.
    \end{equation}
    We proceed as follows: we assume that $l\geq 1$ and
    without loss $l_2\leq l_1$. By induction on $0\leq a\leq l_2$ we prove that
    \begin{equation}
      A_a:=\left[\binom{l}{l_1}\right]^{-1} \sum_{\substack{\lambda_1\leq l_1,\lambda_2\leq l_2, \\
          \lambda_1+\lambda_2=\lambda-a}}\dfrac{(l-a)!}{\lambda_1!\;\lambda_2!}\leq 1\;.
    \end{equation}
    We start from $A_0=1$ since in the sum, only $\lambda_2=l_2$ and
    $\lambda_1=l_1$ are allowed when $a=0$. Assuming that for
    $a\geq 1$, $A_{a-1}\leq 1$, we find
    \begin{equation}\label{A_k_upperbound}
        \begin{split}
          A_a &=\dfrac{l_1-(a-1)}{l-(a-1)}A_{a-1}+\left[\binom{l}{l_1}\right]^{-1}
          \binom{l-a}{l_1}\leq 1-\dfrac{l_2}{l-(a-1)} \\
            &+\dfrac{l_2}{l}\dfrac{(l_2-1)\cdots (l_2-(a-1))}{(l-1)\cdots (l-(a-1))}\;.
        \end{split}
    \end{equation}
    The latter expression equals $1$ for $a=1$. For $a>1$, we can bound the upper
    bound in (\ref{A_k_upperbound}) by 
    \begin{equation}
      1-\dfrac{l_2}{l-(a-1)}\left(1-\dfrac{(l_2-1)(l_2-2)
        \cdots(l_2-(a-1))}{l(l-1)\cdots (l-(a-2))}\right)\leq 1\;.
    \end{equation}
    For $l_2<a\leq l$, the sum in $A_a$ does not contain more non-vanishing
    terms than the one in $A_{a-1}$ and we can bound them as follows:
    \begin{equation}
      \dfrac{(l-a)!}{\lambda_1!\; \lambda_2!}
      \leq \dfrac{(l-(a-1))!}{(\lambda_1+1)!\;\lambda_2!}\;.
    \end{equation}
    Therefore we have in that case $A_a\leq A_{a-1}$.

 Now from (\ref{bound_Vandermonde}) and (\ref{small_lemma_convolution}) we have
    \begin{equation}\label{useful_inequality}
      \sum_{\substack{\lambda_1\leq l_1,\lambda_2\leq l_2 \\
          \lambda_1+\lambda_2=\lambda}}\dfrac{n}{n_1n_2}
      \dfrac{(n_1+l_1-1)!}{(n_1-1)!\;\lambda_1!}
      \dfrac{(n_2+l_2-1)!}{(n_2-1)!\;\lambda_2!}
      \dfrac{(n-1)!\;\lambda!}{(n+l-1)!}\leq \dfrac{n}{n_1 n_2}\;.
    \end{equation}
    Using Lemma \ref{lemma_inequalities1} we obtain statement
    (\ref{first_inequality_convol}). Proof of statement
    (\ref{second_inequality_conv}) follows the proof of (\ref{first_inequality_convol}).

    To prove statements (\ref{third_inequality_convol})-(\ref{fourth_inequality_convol}),
    we use that for $n_1,n_2\geq 1$,
    $k_1,k_2,l_1,l_2,\lambda_1,\lambda_2\geq 0$ and $0\leq \alpha\leq k$
    \begin{equation}
        \begin{split}
          & \dfrac{(k-\alpha)!}{k_1!\;k_2!}\dfrac{(n+1)!}{n_1!\; n_2!}
          \dfrac{\lambda!}{\lambda_1!\;\lambda_2!}\dfrac{(n_1+l_1+k_1)!\;
            (n_2+l_2+k_2)!}{(n+l+k-\alpha+1)!} \\
          &= \binom{k-\alpha}{k_1}\binom{n-1}{n_1-1}
          \binom{\lambda}{\lambda_1}\left[\binom{n+l+k-\alpha+1}{n_1+l_1+k_1}\right]^{-1}\;.
        \end{split}
    \end{equation}
    Then from (\ref{bound_Vandermonde}) we have
    \begin{equation}
      \binom{k-\alpha}{k_1}\binom{n+1}{n_1}
      \binom{l}{l_1}\leq\binom{n+l+k-\alpha+1}{n_1+l_1+k_1}\;.
    \end{equation}
    Then the rest of the proof is identical to the proof
    of (\ref{first_inequality_convol}). Proof of statement
    (\ref{fourth_inequality_convol}) follows from the proof
    of (\ref{third_inequality_convol}). 

\end{proof}

\begin{lemma}\label{bound_sum_log}
    For $s\in\N$, $l\in\N_0$ and $\alpha\geq \alpha_0$,
    \begin{equation}
      \sum_{\lambda=0}^l \dfrac{1}{2^\lambda \lambda!}
      \int_{\alpha_0}^\alpha d\alpha' \alpha'^{s-1}(1-\ln (m^2\alpha'))^\lambda
      \leq \dfrac{2\alpha^{s}}{s}
      \sum_{\lambda=0}^l \dfrac{1}{2^\lambda \lambda!}(1-\ln(m^2\alpha))^\lambda\;.
    \end{equation}
\end{lemma}

\begin{proof}
    Through successive integration by parts, we obtain for $0\leq \lambda\leq l$
    \begin{equation}
        \begin{split}
          \int_{\alpha_0}^\alpha d\alpha' \alpha'^{s-1}(1-\ln(m^2\alpha'))^\lambda
          &\leq \dfrac{\alpha^s}{s}(1-\ln(m^2\alpha))^\lambda
          +\dfrac{\lambda}{s}\int_{\alpha_0}^\alpha d\alpha'
          \alpha'^{s-1}(1-\ln(m^2\alpha'))^{\lambda-1} \\
          &\leq \dfrac{\alpha^s}{s}\;\lambda!\;\sum_{\nu=0}^\lambda
          \dfrac{(1-\ln(m^2\alpha))^\nu}{\nu!}\dfrac{1}{s^{\lambda-\nu}}\;.
        \end{split}
    \end{equation}
    Summing over $\lambda$, we get
    \begin{equation}
        \begin{split}
          \sum_{\lambda=0}^l \dfrac{1}{2^\lambda \lambda!}
          \int_{\alpha_0}^\alpha d\alpha' \alpha'^{s-1}(1-\ln(m^2\alpha'))^\lambda
          &\leq \dfrac{\alpha^s}{s} \sum_{\lambda=0}^l\dfrac{1}{2^\lambda}
          \sum_{\nu=0}^\lambda  \dfrac{(1-\ln(m^2\alpha)^\nu}{\nu!}\dfrac{1}{s^{\lambda-\nu}} \\
          &= \dfrac{\alpha^s}{s} \sum_{\nu=0}^l
          \dfrac{(1-\ln(m^2\alpha))^\nu}{2^\nu \nu!}
          \sum_{\lambda=0}^{l-\nu}\dfrac{1}{(2s)^\lambda} \\
          &\leq \dfrac{2\alpha^s}{s} \sum_{\nu=0}^l
          \dfrac{(1-\ln(m^2\alpha))^\nu}{2^\nu \nu!}\;.
        \end{split}
    \end{equation}
\end{proof}

\subsection{Proof of the mean-field perturbative bounds}\label{proofperturbativebounds}

\perturbativebounds*
\begin{proof}
We proceed by induction as follows:
\begin{itemize}
    \item we go up in $j\in\N$.
    \item at a fixed value of $j$, we go downwards from $n=2j+2$ to $n=2$.
    \item at a fixed value of $j,n$ we go up in $m$.
\end{itemize}

We start the induction at $j=1$. The non-linear term in the r.h.s of
(\ref{MFE2_perturbative_part}) vanishes. Direct computation shows that
\begin{equation}
  \mathcal{A}_{4,1}^{\alpha_0,\alpha}=1,
  \quad \mathcal{A}_{2,1}^{\alpha_0,\alpha}=3\Big(m^2-\frac{1}{\alpha}\Big) \;,
\end{equation}
therefore the bounds (\ref{bound_perturbative_solutionsn=2})
-(\ref{bound_perturbative_solutionsn>=4}) are satisfied.
For a fixed $j>1$, we start at $n=2j+2$ and we go downwards to $n=2$.
The induction hypothesis holds for the set 
\begin{equation}
\begin{split}
&\Bigg\lbrace (j',n',k')\in\N\times (2\N\cap [1,2j+2])\times\N_0, \\
  &\quad \Bigg(\Big(\lbrace j'=j\rbrace\cap
  \lbrace n'>n\rbrace\Big)\cup
  \Big(\lbrace j'<j\rbrace\cap\lbrace n'\in 2\N\cap [1,2j'+2]\rbrace \Big)\Bigg)
  \cap\lbrace k'\leq k\rbrace\Bigg\rbrace\;.    
\end{split} 
\end{equation}

For $n>2$, we proceed as follows 
\begin{itemize}
\item $k=0$: We integrate the l.h.s of (\ref{MFE2_perturbative_part})
  upwards from $\alpha_0$ to $\alpha$ for $n>4$ and downwards from
  $\alpha=\frac{1}{m^2}$ to $\alpha$ for $n=4$. We bound the r.h.s of
  (\ref{MFE2_perturbative_part}) with the induction hypothesis.
  We first start with the linear term.

    \begin{itemize}
    \item $n>4$: The linear term is non-zero as long as $n+2\leq 2j+2$.
      We use Lemma \ref{bound_sum_log} to obtain
        \begin{equation}
        \begin{split}\label{first_term_m=0}
          &\frac{n(n+1)}{2} \int_{\alpha_0}^\alpha d\alpha'
          \frac{\vert\mathcal{A}_{n+2,j}^{\alpha_0,\alpha'}\vert}{\alpha'^2} \\
          &\leq \frac{n(n+1)}{2}\int_{\alpha_0}^\alpha d\alpha' \alpha'^{\frac{n}{2}-3}
          \dfrac{C^{j-\frac{n}{4}-\frac{1}{2}}\;j!}{(j-\frac{n}{2}+1)^2
            (\frac{n}{2}+1)^2(\frac{n}{2}+1)!}
          \sum_{\lambda=0}^{j-\frac{n}{2}}\dfrac{1}{2^{\lambda}\lambda !}
          (1-\ln(m^2\alpha'))^{\lambda} \\
          &\leq  \alpha^{\frac{n}{2}-2}C^{j-\frac{n}{4}}
          \dfrac{j!}{(j-\frac{n}{2}+2)^2 (\frac{n}{2})^2(\frac{n}{2})!}
          \sum_{\lambda=0}^{j-\frac{n}{2}+1}
          \dfrac{1}{2^{\lambda}\lambda !}(1-\ln(m^2\alpha))^{\lambda}\;
          \frac{4(n+1)}{\sqrt{C}(n-4)}\;.
        \end{split}
    \end{equation}

      The non-linear term is always non-zero, we bound it first by
     \begin{equation}\label{first_bound_nonlinear_term}
       \frac{n}{2} \sum_{\substack{
      n_1+n_2=n+2 \\ j_1+j_2=j \\ 2j_i+2\geq n_i  
       }} \int_{\alpha_0}^\alpha d\alpha'
       \vert \mathcal{A}_{n_1,j_1}^{\alpha_0,\alpha'}\; \mathcal{A}_{n_2,j_2}^{\alpha_0,\alpha'}\vert\;.
     \end{equation}
     It is convenient to distinguish $n_1=2$ or $n_1=n$ from $n_i\geq 4$.
     We find for $n_i\geq 4$, $j_i\geq \frac{n_i}{2}-1$,
     \begin{equation}
         \begin{split}
           &\int_{\alpha_0}^\alpha d\alpha'
           \vert \mathcal{A}_{n_1,j_1}^{\alpha_0,\alpha'}
           \vert \vert \mathcal{A}_{n_2,j_2}^{\alpha_0,\alpha'}\vert \\
           &\leq \dfrac{C^{j-\frac{n}{4}-\frac{1}{2}} j_1!\; j_2!\;}
                 {(j_1-\frac{n_1}{2}+2)^2 (j_2-\frac{n_2}{2}+2)^2
                   (\frac{n_1}{2})^2 (\frac{n_2}{2})^2 (\frac{n_1}{2})!\;
                   (\frac{n_2}{2})!}\times \\
                 &\int_{\alpha_0}^\alpha d\alpha' \alpha'^{\frac{n}{2}-3}
                 \sum_{\lambda_1=0}^{j_1-\frac{n_1}{2}+1}\dfrac{1}{2^{\lambda_1}\lambda_1 !}
                 (1-\ln(m^2\alpha'))^{\lambda_1} \sum_{\lambda_2=0}^{j_2-\frac{n_2}{2}+1}
                 \dfrac{1}{2^{\lambda_2}\lambda_2 !}(1-\ln(m^2\alpha'))^{\lambda_2}\;.
         \end{split}
     \end{equation}
     
     Setting the loop numbers $l_k=j_k-\frac{n_k}{2}+1$ for $k=1,2$
     and $l=j-\frac{n}{2}+1$, and summing over the even integers $n_i\geq 4$,
     we get the following bound
     \begin{equation}\label{bound_nonlinear_term_n_i}
     \begin{split}
       & C^{j-\frac{n}{4}-\frac{1}{2}} \sum_{\lambda=0}^{l}\sum_{\substack{
           n_1+n_2=n+2 \\ n_i\geq 4, n_i\in 2\N \\  l_1+l_2=l \\
           \lambda_1\leq l_1,\lambda_2\leq l_2 \\
           \lambda_1+\lambda_2=\lambda 
       }} \dfrac{(\frac{n_1}{2}+l_1-1)!\;
         (\frac{n_2}{2}+l_2-1)!}{(\frac{n_1}{2})!\;
         (\frac{n_2}{2})!}\dfrac{1}{(l_1+1)^2
         (l_2+1)^2 (\frac{n_1}{2})^2 (\frac{n_2}{2})^2}
       \frac{\lambda!}{\lambda_1!\;\lambda_2!}\times \\
       &\int_{\alpha_0}^\alpha d\alpha' \alpha'^{\frac{n}{2}-3}
       \dfrac{1}{2^\lambda\lambda!} (1-\ln(m^2\alpha'))^\lambda \;.
     \end{split}      
     \end{equation}
     Using Lemma \ref{lemma_convolutions_inequalities} (\ref{first_inequality_convol})
     and Lemma \ref{bound_sum_log}, (\ref{bound_nonlinear_term_n_i}) is bounded by
     \begin{equation}
       \alpha^{\frac{n}{2}-2}C^{j-\frac{n}{4}}
       \dfrac{j!}{(j-\frac{n}{2}+2)^2
         (\frac{n}{2})^2 (\frac{n}{2})!}
       \sum_{\lambda=0}^{j-\frac{n}{2}+1}
       \dfrac{1}{2^\lambda \lambda!}(1-\ln(m^2\alpha))^{\lambda}\dfrac{4K_0}{\sqrt{C}(n-4)}\;.
     \end{equation}
    
     For $n_1=2$ or $n_2=2$, we use again Lemma
     \ref{lemma_convolutions_inequalities} (\ref{second_inequality_conv})
     and Lemma \ref{bound_sum_log} to obtain the bound
    \begin{equation}
      \alpha^{\frac{n}{2}-2}C^{j-\frac{n}{4}}
      \dfrac{j!}{(j-\frac{n}{2}+2)^2
        (\frac{n}{2})^2 (\frac{n}{2}+1)!}
      \sum_{\lambda=0}^{j-\frac{n}{2}}
      \dfrac{1}{2^\lambda \lambda!}(1-\ln(m^2\alpha))^{\lambda}\dfrac{4K'_0}{\sqrt{C}(n-4)}\;.
    \end{equation}
    Since $\ln(m^2\alpha)<0$, the summand is
    positive and (\ref{first_bound_nonlinear_term}) is bounded by
    \begin{equation}\label{second_term_m=0}
      \alpha^{\frac{n}{2}-2}C^{j-\frac{n}{4}}
      \dfrac{j!}{(j-\frac{n}{2}+2)^2 (\frac{n}{2})^2
        (\frac{n}{2})!}\sum_{\lambda=0}^{j-\frac{n}{2}+1}
      \dfrac{1}{2^\lambda \lambda!}(1-\ln(m^2\alpha))^{\lambda}
      \;2(K_0+2K'_0)\dfrac{n}{\sqrt{C}(n-4)}\;.
    \end{equation}

    Summing together (\ref{first_term_m=0}) and (\ref{second_term_m=0}), we have
    \begin{equation}
    \begin{split}
      \vert \mathcal{A}_{n,j}^{\alpha_0,\alpha}\vert &
      \leq \Bigg[\frac{14}{\sqrt{C}}+\dfrac{6(K_0+2K'_0)}{\sqrt{C}}\Bigg]
      \dfrac{ \alpha^{\frac{n}{2}-2}C^{j-\frac{n}{4}}j!}{(j-\frac{n}{2}+2)^2
        (\frac{n}{2})^2 (\frac{n}{2})!}\sum_{\lambda=0}^{j-\frac{n}{2}+1}
      \dfrac{1}{2^\lambda \lambda!}(1-\ln(m^2\alpha))^{\lambda} \\
      &\leq \alpha^{\frac{n}{2}-2}C^{j-\frac{n}{4}}
      \dfrac{j!}{(j-\frac{n}{2}+2)^2 (\frac{n}{2})^2 (\frac{n}{2})!}
      \sum_{\lambda=0}^{j-\frac{n}{2}+1}\dfrac{1}{2^\lambda \lambda!}(1-\ln(m^2\alpha))^{\lambda}\;,
    \end{split}
    \end{equation}
    choosing $C$ sufficiently large. We may choose $\sqrt{C}=194$.

  \item $n\leq 4$: We integrate the flow equations downwards from
    $\alpha_{\max}$ to $\alpha$. We start with $n=4$. The linear term
    is non-zero if $j\geq 2$.  Inserting the induction hypothesis,
    the linear term is bounded by
     \begin{equation}
     \begin{split}
       &10\int_{\alpha}^{\alpha_{\max}} d\alpha' \dfrac{1}{\alpha'}
       \dfrac{C^{j-\frac{3}{2}}\;j!}{(j-1)^2 3^2 3!}
       \sum_{\lambda=0}^{j-2}\dfrac{1}{2^\lambda \lambda!}(1-\ln(m^2\alpha'))^{\lambda} \\
       &\leq C^{j-1}\dfrac{j!}{j^2 2^2 2!}
       \sum_{\lambda=0}^{j-1}\dfrac{1}{2^\lambda \lambda!}
       (1-\ln(m^2\alpha))^{\lambda}\;\dfrac{80}{3\sqrt{C}}\;,   
     \end{split}
     \end{equation}
     where we used
     \begin{equation}\label{bound_sum_log_n=4}
     \begin{split}
       \sum_{\lambda=0}^{j-2}\frac{1}{2^\lambda\lambda!}
       \int_\alpha^{\alpha_{\max}} d\alpha'
       \dfrac{(1-\ln(m^2\alpha'))^\lambda}{\alpha'}
       &\leq\sum_{\lambda=0}^{j-2}\dfrac{1}{2^\lambda (\lambda+1)!}
       (1-\ln(m^2\alpha))^{\lambda+1} \\
       &\leq 2\sum_{\lambda=0}^{j-1}\dfrac{1}{2^\lambda \lambda!}
       (1-\ln(m^2\alpha))^{\lambda} \;.   
     \end{split}
     \end{equation}

     In the non-linear term, we have $n_1=2,\; n_2=4$ or $n_1=4,\; n_2=2$.
     The non-linear term is non-zero if $j\geq 2$.
     Therefore we can bound it by
     \begin{equation}
         \begin{split}
           &4\sum_{\substack{j_1+j_2=j \\ j_i\geq 1}}\int_\alpha^{\alpha_{\max}}
           d\alpha'\vert\mathcal{A}_{2,j_1}^{\alpha_0,\alpha'}
           \mathcal{A}_{4,j_2}^{\alpha_0,\alpha'}\vert \\
           &\leq 4C^{j-\frac{3}{2}}\sum_{\lambda=0}^{j-2}\sum_{\substack{j_1+j_2=j \\
               j_i\geq 1 \\
               \lambda_1+\lambda_2=\lambda}}\dfrac{j_1!\;j_2!}{(j_1+1)^2j_2^2\;
             2!\; 2^2}\frac{\lambda!}{\lambda_1!\;\lambda_2!}\frac{1}{2^\lambda\lambda!}
           \int_{\alpha}^{\alpha_{\max}}d\alpha' (1-\ln(m^2\alpha'))^\lambda\;.
         \end{split}
     \end{equation}
     Using Lemma \ref{lemma_convolutions_inequalities}
     (\ref{second_inequality_conv}) and (\ref{bound_sum_log_n=4}),
     these contributions are bounded by
     \begin{equation}
       C^{j-1}\dfrac{4  K_0' j!}{\sqrt{C} m^2 j^2\; 2^2\; 2!}
       \sum_{\lambda=0}^{j-1}\dfrac{1}{2^\lambda \lambda!} (1-\ln(m^2\alpha))^{\lambda}\;.
     \end{equation}
    We may choose $\sqrt{C}\geq 194$ such that
    \begin{equation}
        \frac{80}{3\sqrt{C}}+\frac{4K_0'}{m^2\sqrt{C}}<1
    \end{equation}
    so that we obtain the claim for $n=4$.

     For $n=2$, we use the bounds established for $n=4$. The linear term is then bounded by
     \begin{equation}
     \begin{split}
       3 C^{j-1}\dfrac{j!}{j^2 2^2 2!}\int_{\alpha}^{\alpha_{\max}}
       \dfrac{d\alpha'}{\alpha'^2}\sum_{\lambda=0}^{j-1}(1-\ln(m^2\alpha'))^{\lambda}
       \leq \dfrac{3}{2\alpha} C^{j-1}\dfrac{j!}{(j+1)^2}
       \sum_{\lambda=0}^{j-1}(1-\ln(m^2\alpha))^{\lambda}\;\;.
     \end{split}
     \end{equation}

     The non-linear term in the r.h.s of (\ref{MFE2_perturbative_part})
     only contains terms corresponding to $n_1=n_2=2$. Since for $n_k=2$, $l_k=j_k$,
     the non-linear term is bounded by
     \begin{equation}\label{exp}
     \begin{split}
         &C^{j-1}\sum_{\lambda=0}^{j-2}\sum_{\substack{
      j_1+j_2=j   
      \\ \lambda_1\leq j_1,\lambda_2\leq j_2 \\ \lambda_1+\lambda_2=\lambda}}
       \dfrac{j_1!\; j_2!\;\lambda!}{(j_1+1)^2 (j_2+1)^2 \lambda_1!\;\lambda_2!}
       \int_{\alpha}^{\alpha_{\max}}  \dfrac{d\alpha'}{\alpha'^2}
       \dfrac{1}{2^\lambda\lambda!}(1-\ln(m^2\alpha'))^{\lambda}\;.
     \end{split}       
     \end{equation}
     We use Lemma \ref{lemma_convolutions_inequalities}
     (\ref{second_inequality_conv}) to bound (\ref{exp}) by
\begin{equation}
\begin{split}
  &C^{j-1} K_0'\dfrac{j!}{(j+1)^2}\int_\alpha^{\alpha_{\max}}
  \frac{d\alpha'}{\alpha'^2}\sum_{\lambda=0}^{j-2}
  \dfrac{1}{2^\lambda \lambda!}(1-\ln(m^2\alpha'))^\lambda \\
  &\leq \dfrac{C^{j-1} K_0'}{\alpha}\dfrac{j!}{(j+1)^2}
  \sum_{\lambda=0}^{j-1}\dfrac{1}{2^\lambda \lambda!}(1-\ln(m^2\alpha))^\lambda\;,   
\end{split}
\end{equation}
\end{itemize}
because $\ln(m^2\alpha)<0$. Choosing $\sqrt{C}>194$ such that 
\begin{equation}
    \frac{3}{2\sqrt{C}}+\frac{K_0'}{\sqrt{C}}\leq 1\;,
\end{equation}
we obtain the claim for $n=2$.

\item $k\geq 1$

  To obtain the bounds, we multiply (\ref{MFE2_perturbative_part})
  by $\alpha^2$ and differentiate $k$ times w.r.t. $\alpha$. Then
  we solve $\partial_\alpha^{k+1} \mathcal{A}_{n,j}^{\alpha_0,\alpha}$ to get
\begin{equation}\label{pert.eqn_m>=1}
       \begin{split}
         \partial_\alpha^{k+1} \mathcal{A}_{n,j}^{\alpha_0,\alpha}
         &=-\frac{2k}{\alpha}\partial_\alpha^k \mathcal{A}_{n,j}^{\alpha_0,\alpha}
         -\frac{k(k+1)}{\alpha^2}\partial_\alpha^{k-1} \mathcal{A}_{n,j}^{\alpha_0,\alpha}
         +\dfrac{n(n+1)}{2\alpha^2}\partial_{\alpha}^{k} \mathcal{A}_{n+2,j}^{\alpha_0,\alpha} \\
           &-\frac{n}{2}\sum_{\substack{
      n_1+n_2=n+2 \\ j_1+j_2=j \\ 2j_i+2\geq n_i \\ k_1+k_2=k   
         }}\dfrac{k!}{k_1!\; k_2!}\partial_\alpha^{k_1}
         \mathcal{A}_{n_1,j_1}^{\alpha_0,\alpha} \partial_\alpha^{k_2}
         \mathcal{A}_{n_2,j_2}^{\alpha_0,\alpha} \\
     &-\frac{n k}{\alpha}\sum_{\substack{
      n_1+n_2=n+2 \\ j_1+j_2=j \\ 2j_i+2\geq n_i \\ k_1+k_2=k-1   
         }}\dfrac{(k-1)!}{k_1!\; k_2!}\partial_\alpha^{k_1}
         \mathcal{A}_{n_1,j_1}^{\alpha_0,\alpha} \partial_\alpha^{k_2}
         \mathcal{A}_{n_2,j_2}^{\alpha_0,\alpha} \\
     &-\frac{n k(k+1)}{2\alpha^2}\sum_{\substack{
      n_1+n_2=n+2 \\ j_1+j_2=j \\ 2j_i+2\geq n_i \\ k_1+k_2=k-2   
         }}\dfrac{(k-2)!}{k_1!\; k_2!}\partial_\alpha^{k_1}
         \mathcal{A}_{n_1,j_1}^{\alpha_0,\alpha} \partial_\alpha^{k_2}
         \mathcal{A}_{n_2,j_2}^{\alpha_0,\alpha}\;.
       \end{split}
   \end{equation}

We follow the convention that an empty sum is zero. We successively bound the terms
in the r.h.s of (\ref{pert.eqn_m>=1}). For $n>2$, we successively obtain 
\begin{itemize}
    \item First term:
    \begin{equation}\label{first_term_m>=1}
    \begin{split}
      &\left\vert\frac{2k}{\alpha}\partial_\alpha^k
      \mathcal{A}_{n,j}^{\alpha_0,\alpha}\right\vert \\
      &\leq \dfrac{2k}{\alpha} \alpha^{\frac{n}{2}-2-k}
      \dfrac{C^{j-\frac{n}{4}+k}\; (j+k+1)! }{(j-\frac{n}{2}+2)^2\;(k+1)^2
        (\frac{n}{2})^2(\frac{n}{2})!}\sum_{\lambda=0}^{j-\frac{n}{2}+1}
      \dfrac{1}{2^{\lambda}\lambda !}(1-\ln(m^2\alpha))^{\lambda} \\
      &\leq  \alpha^{\frac{n}{2}-2-k-1}
      \dfrac{C^{j-\frac{n}{4}+k+1}\;(j+k+2)! }{(j-\frac{n}{2}+2)^2 (k+2)^2
        (\frac{n}{2})^2(\frac{n}{2})! }\sum_{\lambda=0}^{j-\frac{n}{2}+1}
      \dfrac{1}{2^{\lambda}\lambda !}(1-\ln(m^2\alpha))^{\lambda}\;\dfrac{8}{C}\;.
    \end{split}
    \end{equation}
    \item Second term\footnote{This term is non-zero if $k\geq 1$.}:
    \begin{equation}\label{second_term_m>=1}
        \begin{split}
          &\left\vert\frac{k(k+1)}{\alpha^2}\partial_\alpha^{k-1}
          \mathcal{A}_{n,j}^{\alpha_0,\alpha}\right\vert  \\
          &\leq\frac{k(k+1)}{\alpha^2}\alpha^{\frac{n}{2}-2-k+1}
          \dfrac{C^{j-\frac{n}{4}+k-1}\;(j+k)!}{(j-\frac{n}{2}+2)^2 k^2
            (\frac{n}{2})^2(\frac{n}{2})!}\sum_{\lambda=0}^{j-\frac{n}{2}+1}
          \dfrac{1}{2^{\lambda}\lambda !}(1-\ln(m^2\alpha))^{\lambda} \\
          &\leq  \alpha^{\frac{n}{2}-2-k-1} \dfrac{C^{j-\frac{n}{4}+k+1}\;
            (j+k+2)!}{(j-\frac{n}{2}+2)^2 (k+2)^2
            (\frac{n}{2})^2(\frac{n}{2})!}
          \sum_{\lambda=0}^{j-\frac{n}{2}+1}\dfrac{1}{2^{\lambda}\lambda !}
          (1-\ln(m^2\alpha))^{\lambda}\frac{9}{C^2}\;.
        \end{split}
    \end{equation}
    
    \item Third term:
    \begin{equation}\label{third_term_m>=1}
    \begin{split}
      &\left\vert\dfrac{n(n+1)}{2\alpha^2}\partial_{\alpha}^{k}
      \mathcal{A}_{n+2,j}^{\alpha_0,\alpha}\right\vert  \\
      &\leq \dfrac{n(n+1)}{2\alpha^2} \alpha^{\frac{n}{2}-1-k}
      \dfrac{C^{j-\frac{n}{4}-\frac{1}{2}+k}\;(j+k+1)!}
            {(j-\frac{n}{2}+1)^2 (k+1)^2(\frac{n}{2}+1)^2(\frac{n}{2}+1)!}
            \sum_{\lambda=0}^{j-\frac{n}{2}-1+1}\dfrac{1}{2^{\lambda}\lambda !}
            (1-\ln(m^2\alpha))^{\lambda} \\
            &\leq \alpha^{\frac{n}{2}-2-k-1}
            \dfrac{C^{j-\frac{n}{4}+k+1}\;(j+k+2)!}
                  {(j-\frac{n}{2}+2)^2 (k+2)^2 (\frac{n}{2})^2(\frac{n}{2})!}
                  \sum_{\lambda=0}^{j-\frac{n}{2}+1}\dfrac{1}{2^{\lambda}\lambda !}
                  (1-\ln(m^2\alpha))^{\lambda}\dfrac{32}{C^{\frac{3}{2}}}\;,
    \end{split}       
    \end{equation}
    since we recall that $j\geq\frac{n}{2}$.
  \item Fourth term: We proceed as in the case $k=0$. We use together
    Lemma \ref{lemma_convolutions_inequalities}, inequalities
    (\ref{third_inequality_convol})-(\ref{fourth_inequality_convol}) to get 
    \begin{equation}\label{fourth_term_m>=1}
    \begin{split}
      & \alpha^{\frac{n}{2}-3-k}C^{j-\frac{n}{4}+k+1}
      \dfrac{n(j+k+2)!}{2(j-\frac{n}{2}+2)^2 (k+1)^2
        (\frac{n}{2})^2 (\frac{n}{2}+1)!}\times \\
      &\sum_{\lambda=0}^{j-\frac{n}{2}+1}
      \dfrac{1}{2^{\lambda}\lambda !}(1-\ln(m^2\alpha))^{\lambda}
      \dfrac{(K_0''+2K_0''')}{C^{\frac{3}{2}}} \\
      &\leq \alpha^{\frac{n}{2}-3-k}C^{j-\frac{n}{4}+k+1}
      \dfrac{(j+k+2)!}{(j-\frac{n}{2}+2)^2 (k+2)^2 (\frac{n}{2})^2 (\frac{n}{2})!}
      \times \\
      &\sum_{\lambda=0}^{j-\frac{n}{2}+1}\dfrac{1}{2^{\lambda}\lambda !}
      (1-\ln(m^2\alpha))^{\lambda}\dfrac{4(K_0''+2K_0''')}{C^{\frac{3}{2}}}\;.
    \end{split}
    \end{equation}

  \item Fifth term\footnote{This term is non-zero if $k\geq 1$.}:
    Again, we use together Lemma (\ref{lemma_convolutions_inequalities})
    inequalities (\ref{third_inequality_convol})-(\ref{fourth_inequality_convol}) to get
     \begin{equation}\label{fifth_term_m>=1}
    \begin{split}
      &\alpha^{\frac{n}{2}-3-k}C^{j-\frac{n}{4}+k+1}
      \dfrac{nm (j+k+1)!}{(j-\frac{n}{2}+2)^2 k^2 (\frac{n}{2})^2 (\frac{n}{2}+1)! }
      \times \\
      &\sum_{\lambda=0}^{j-\frac{n}{2}+1}\dfrac{1}{2^{\lambda}\lambda !}
      (1-\ln(m^2\alpha))^{\lambda}\dfrac{K_0''+2K_0'''}{C^\frac{5}{2}} \\
      &\leq \alpha^{\frac{n}{2}-3-k}
      C^{j-\frac{n}{4}+k+1}\dfrac{(j+k+2)!}{(j-\frac{n}{2}+2)^2 (k+2)^2
        (\frac{n}{2})^2 (\frac{n}{2})! }\times \\
      &\sum_{\lambda=0}^{j-\frac{n}{2}+1}
      \dfrac{1}{2^{\lambda}\lambda !}(1-\ln(m^2\alpha))^{\lambda}
      \dfrac{18(K_0''+2K_0''')}{C^\frac{5}{2}}\;.
    \end{split}
    \end{equation}
   
   \item Sixth term\footnote{This term is non-zero if $k\geq 2$.}: we repeat
     the previous steps when dealing with the fourth and fifth terms. This
     leads to the following bound
     \begin{equation}\label{sixth_term_m>=1}
    \begin{split}
      &\alpha^{\frac{n}{2}-3-k}C^{j-\frac{n}{4}+k+1} \frac{nk(k+1)}{2}
      \dfrac{(j+k)!}{(j-\frac{n}{2}+2)^2 (k-1)^2 (\frac{n}{2})^2 (\frac{n}{2}+1)! }
      \times \\
      &\sum_{\lambda=0}^{j-\frac{n}{2}+1}\dfrac{1}{2^{\lambda}\lambda !}
      (1-\ln(m^2\alpha))^{\lambda}\dfrac{(K_0''+2K_0''')}{C^\frac{7}{2}} \\
      &\leq \alpha^{\frac{n}{2}-3-k}C^{j-\frac{n}{4}+k+1}
      \dfrac{(j+k+2)!}{(j-\frac{n}{2}+2)^2 (k-1)^2 (\frac{n}{2})^2 (\frac{n}{2})! }
      \times \\
      &\sum_{\lambda=0}^{j-\frac{n}{2}+1}\dfrac{1}{2^{\lambda}\lambda !}
      (1-\ln(m^2\alpha))^{\lambda}\dfrac{16(K_0''+2K_0''')}{C^\frac{7}{2}} \;.
    \end{split}
    \end{equation}
    
\end{itemize}
Adding together (\ref{first_term_m>=1})-(\ref{sixth_term_m>=1}), we find 
\begin{equation}
    \begin{split}
        &\vert \partial_\alpha^{k+1}\mathcal{A}_{n,j}^{\alpha_0,\alpha}\vert 
 \\
 &\leq \Bigg[\frac{8}{C}+\frac{9}{C^2}+\frac{32}{C^{\frac{3}{2}}}
   +\frac{38}{C^{\frac{3}{2}}}(K_0''+2K_0''')\Big)\Bigg]
 \alpha^{\frac{n}{2}-2-k} \dfrac{C^{j-\frac{n}{4}+k+1}
   \;(j+k+2)!}{(j-\frac{n}{2}+2)^2 (k+2)^2 (\frac{n}{2})^2(\frac{n}{2})!}\times \\
 &\sum_{\lambda=0}^{j-\frac{n}{2}+1}\dfrac{1}{2^{\lambda}\lambda !}
 (1-\ln(m^2\alpha))^{\lambda} \\
 &\leq \alpha^{\frac{n}{2}-2-k}
 \dfrac{C^{j-\frac{n}{4}+k+1}\;(j+k+2)!}{(j-\frac{n}{2}+2)^2
   (\frac{n}{2})^2(\frac{n}{2})!\; (\frac{n}{2}-1)!}
 \sum_{\lambda=0}^{j-\frac{n}{2}+1}\dfrac{1}{2^{\lambda}\lambda !}(1-\ln(m^2\alpha))^{\lambda}\;,
    \end{split}
\end{equation}
choosing $C$ such that 
\begin{equation}
    \frac{8}{C}+\frac{1}{C^2}\Big(9+32+38(K_0''+2K_0''')\Big)\leq 1\;.
\end{equation}
\end{itemize}

For $n=2$, we repeat the same steps above. The essential difference w.r.t.
the case $n>2$ is that in the r.h.s of (\ref{bound_perturbative_solutionsn=2}),
the sum runs over $0\leq\lambda\leq j-1$. Not to overload the proof, we will
only present the non-trivial terms.

\begin{itemize}
\item The first and second term in the r.h.s of (\ref{pert.eqn_m>=1}) are
  treated as above so that they are bounded by terms similar to
  (\ref{first_term_m>=1}) and (\ref{second_term_m>=1}) with the aforementioned change.
    
    \item Third term: Inserting the induction hypothesis, we find
    \begin{equation}\label{third_term_m>=1,n=2}
    \begin{split}
      &\left\vert\dfrac{3}{\alpha^2}\partial_{\alpha}^{k}
      \mathcal{A}_{4,j}^{\alpha_0,\alpha}\right\vert  \\
      &\leq \dfrac{3}{\alpha^{k+2}}  \dfrac{C^{j-1+k}\;
        (j+k+1)!}{j^2 \,(k+1)^2\,4\times 2}\sum_{\lambda=0}^{j-1}
      \dfrac{1}{2^{\lambda}\lambda !}(1-\ln(m^2\alpha))^{\lambda} \\
      &\leq \frac{1}{\alpha^{k+2}} \dfrac{C^{j+k+\frac{1}{2}}
        \;(j+k+2)!}{(j+1)^2 (k+2)^2 } \sum_{\lambda=0}^{j-1}\dfrac{1}{2^{\lambda}
        \lambda !}(1-\ln(m^2\alpha))^{\lambda}\dfrac{6}{C^{\frac{3}{2}}}\;,
    \end{split}       
    \end{equation}
    \item Fourth term: The terms are of the form
      \begin{equation}    \partial_\alpha^{k_1}
        \mathcal{A}_{2,j_1}^{\alpha_0,\alpha}\,
        \partial_\alpha^{k_2}\mathcal{A}_{2,j_2}^{\alpha_0,\alpha},\quad k_1+k_2=k,\, j_1+j_2=j\;.
    \end{equation}
    
Therefore, we can bound these terms by
\begin{equation}\label{fthtermn=2n,m>=1}
\begin{split}
  &\frac{1}{\alpha^{k+1}}C^{j-1+k}\sum_{\lambda=0}^{j-2}
  \frac{1}{2^\lambda\,\lambda!}(1-\ln(m^2\alpha))^\lambda \times \\
  &\sum_{\substack{j_1+j_2=j \\  \lambda_1\leq j_1-1,\,\lambda_2\leq j_2-1 \\
      k_1+k_2=k}}\dfrac{k!}{k_1!\, k_2!}\dfrac{\lambda!}{\lambda_1!\,\lambda_2!}
  \dfrac{(j_1+k_1+1)!\,(j_2+k_2+1)!}{(j_1+1)^2\,(j_2+1)^2\,(k_1+1)^2\,(k_2+1)^2}\;.
\end{split}
\end{equation}
Using Lemma \ref{lemma_convolutions_inequalities} (\ref{fourth_inequality_convol}),
(\ref{fthtermn=2n,m>=1}) is bounded by
\begin{equation}
  \frac{1}{\alpha^{k+1}}C^{j+k+\frac{1}{2}}
  \dfrac{(j+k+2)!}{(j+1)^2  (k+1)^2}\sum_{\lambda=0}^{j-1}\frac{1}{2^\lambda\,\lambda!}
  (1-\ln(m^2\alpha))^\lambda\,\frac{K_0'''}{2C^{\frac{3}{2}}}\;.
\end{equation}
\item The remaining terms in the r.h.s of (\ref{pert.eqn_m>=1})
  can be treated analogously. They are bounded by terms similar
  to (\ref{fifth_term_m>=1})-(\ref{sixth_term_m>=1}) with the
  aforementioned changes. 
\end{itemize}
Summing the different bounds, we obtain the claim for $n=2$.
\end{proof}

\perturbativeboundsmu*
\begin{proof}
  We use Faà  di Bruno's formula (see Appendix \ref{FaadiBruno1})
  and Lemma \ref{Prop_bound_perturbative_solutions} to obtain
    \begin{equation}
        \begin{split}
          \vert\partial_\mu^m \mathcal{A}_{n,j}^{\alpha_0,\alpha}\vert
          &\leq\sum_{k=1}^{m} \vert\partial_\alpha^k
          \mathcal{A}_{n,j}^{\alpha_0,\alpha}\vert
          \sum_{p(m,k)}m!\prod_{j=1}^{m-k+1}
          \dfrac{(\alpha_0 e^\mu)^{\lambda_j}}{\lambda_j!\; (j!)^{\lambda_j}} \\
          &\leq \sum_{k=1}^m (\alpha_0 e^\mu)^{\frac{n}{2}-2} \dfrac{(j+k+1)!
            \; C^{j-\frac{n}{4}+k}}{(j-\frac{n}{2}+2)^2
            (\frac{n}{2})^2(\frac{n}{2})!}\mathcal{F}(j,n,\mu)S^k_m\;,
        \end{split}
    \end{equation}
    where $S^k_m$ is the Stirling number of the second kind whose expression
    is (see e.g. \cite{Stirlingsecondkind})
\begin{equation}
    S^k_m:=\sum_{p(m,k)}m!\prod_{j=1}^{m-k+1}\dfrac{1}{\lambda_j!\;
            (j!)^{\lambda_j}}\;.
\end{equation}
Then we have
\begin{equation}
    \begin{split}
      \vert\partial_\mu^m \mathcal{A}_{n,j}^{\alpha_0,\alpha}\vert
      &\leq   (\alpha_0 e^\mu)^{\frac{n}{2}-2}
      \dfrac{C^{j-\frac{n}{4}+m}}{(j-\frac{n}{2}+2)^2
        (\frac{n}{2})^2(\frac{n}{2})!}  \mathcal{F}(j,n,\mu) \sum_{k=1}^m (j+k+1)!\; S^k_m \\
      &\leq (\alpha_0 e^\mu)^{\frac{n}{2}-2}
      \dfrac{ C^{j-\frac{n}{4}+m}}{(j-\frac{n}{2}+2)^2
        (\frac{n}{2})^2(\frac{n}{2})!}
      \mathcal{F}(j,n,\mu) \sum_{k=1}^m (j+1)!\;k! \binom{j+k+1}{k}\; S^k_m \\
      &\leq (\alpha_0 e^\mu)^{\frac{n}{2}-2}
      \dfrac{ C^{j-\frac{n}{4}+m}}{(j-\frac{n}{2}+2)^2 (\frac{n}{2})^2(\frac{n}{2})!}
      \mathcal{F}(j,n,\mu) (j+1)!\; 2^{j+m+1} a(m)\;,
    \end{split}
\end{equation}

where we introduced the ordered Bell number $a(n)$
(see e.g. \cite{Ordered_BN1,Ordered_BN2})
\begin{equation}
    a(n):=\sum_{k=0}^n k!\; S^k_n\;.
\end{equation}

The ordered Bell numbers $a(n)$ obey the following
formula \cite{Ordered_BN3,Ordered_BN4}
\begin{equation}\label{Orde_BN_rec}
    a(n)=\sum_{i=1}^n \binom{n}{i}a(n-i)\;.
\end{equation}

From (\ref{Orde_BN_rec}), one can prove inductively
that $\vert a(n)\vert\leq e^n\; n!$\;. Then 
\begin{equation}
\begin{split}
  \vert\partial_\mu^m \mathcal{A}_{n,j}^{\alpha_0,\alpha}\vert
  &\leq (\alpha_0 e^\mu)^{\frac{n}{2}-2} \dfrac{(j+1)!\; m!\;
    C'^{j+\frac{n}{2}+m}\; }{(j-\frac{n}{2}+2)^2 (\frac{n}{2})^2(\frac{n}{2})!}
  \mathcal{F}(j,n,\mu) \\
  &\leq (\alpha_0 e^\mu)^{\frac{n}{2}-2}
  \dfrac{(j+m+1)!\;  C'^{j+\frac{n}{2}+m}\; }{(j-\frac{n}{2}+2)^2
    (\frac{n}{2})^2(\frac{n}{2})!}  \mathcal{F}(j,n,\mu)\;,   
\end{split}    
\end{equation}
where we can choose for instance $C'=2eC>C>1$.
\end{proof}

\section{Useful Lemmata used to prove the renormalization
  conditions compatibility}\label{appendixBPHZ}

\subsection{Exact expressions of $r_{n,0}$ and $r_{n,1}$}\label{appendixgn01}

\exact*
\begin{proof}
  First we prove (\ref{exactgn0}) by induction in
  $n\geq 4$. For $n=4$, the result is obvious. For
  $n\geq 6$ we use (\ref{FE_trivial_field2_1}) to obtain
    \begin{equation}
    \begin{split}
      r_{n,0} &= -\frac{n}{n-4}\sum_{\substack{n_1+n_2=n+2 \\
          n_i\geq 4 \\ n_i\in 2\N}}(-1)^{\frac{n}{2}-1}r_{4,0}^{\frac{n}{2}-1}
      C_2\Big(\frac{n_1}{2}-1\Big) C_2\Big(\frac{n_2}{2}-1\Big) \\
      &=(-1)^{\frac{n}{2}}r_{4,0}^{\frac{n}{2}-1}\dfrac{n}{n-4}
      \sum_{\substack{n_1+n_2=n+2 \\ n_i\geq 4\\ n_i\in 2\N}}C_2\Big(\frac{n_1}{2}-1\Big)
      C_2\Big(\frac{n_2}{2}-1\Big)\\
      &=(-1)^{\frac{n}{2}}r_{4,0}^{\frac{n}{2}-1}
      \dfrac{n}{n-4}\sum_{\substack{n_1+n_2=\frac{n}{2}-1 \\ n_i\geq 1}}C_2(n_1) C_2(n_2)\;.
    \end{split}
    \end{equation}
We use the convolution identity \cite{catalan_convolution}
\begin{equation}\label{Fusscatalanconvol}
  \sum_{\substack{i_1+i_2=m \\
      i_j\geq 0}}C_s(i_1)C_s(i_2)=\dfrac{2}{(s+1)m+2}\binom{(s+1)m+2}{m}\;,
  \quad s\geq 1,\; m\geq 0\;,
\end{equation}
to obtain

\begin{equation}
    \begin{split}
      r_{n,0} &=(-1)^{\frac{n}{2}}r_{4,0}^{\frac{n}{2}-1}
      \dfrac{n}{n-4}\Big[-2C_2\Big(\frac{n}{2}-1\Big)
        +\dfrac{2}{\frac{3n}{2}-1}\binom{\frac{3n}{2}-1}{\frac{n}{2}-1}\Big] \\
         &= (-1)^{\frac{n}{2}}r_{4,0}^{\frac{n}{2}-1} C_2\Big(\frac{n}{2}-1\Big)\;.
    \end{split}
\end{equation}

To prove (\ref{exactgn1}) we proceed by induction in $n$.
The claim is true for $n=4$. Then we have
\begin{equation}
\begin{split}
    r_{n,1} &= -\frac{2n}{n-2}\sum_{\substack{n_1+n_2=n+2\\
n_i\geq 4}}r_{n_1,0}r_{n_2,1}-\frac{n}{n-2}r_{n,0}\left(2f_{2,0}+1-\dfrac{4}{n}\right) \\
&= (-1)^{\frac{n}{2}}r_{4,0}^{\frac{n}{2}-1}\frac{n}{n-2}\Big[2\sum_{\substack{n_1+n_2=n+2\\
          n_i\geq 4}} C_2\Big(\frac{n_1}{2}-1\Big)C_2\Big(\frac{n_2}{2}-1\Big)
      \Big(\frac{3n_2-4}{2}b_1+\frac{n_2-4}{4}\Big) \\
&+\Big(2b_1+1-\frac{4}{n}\Big)C_2\Big(\frac{n}{2}-1\Big)\Big] \\
    &=(-1)^{\frac{n}{2}-1}r_{4,0}^{\frac{n}{2}-1}C_2\Big(\frac{n}{2}-1\Big)
    \Big(\frac{3n-4}{2}b_1+\frac{n-4}{4}\Big)\;,
\end{split}
\end{equation}
where we used the following identity
\begin{equation}
    \sum_{\substack{n_1+n_2=n+2\\
        n_i\geq 4}} n_2C_2\Big(\frac{n_1}{2}-1\Big)C_2\Big(\frac{n_2}{2}-1\Big)
    =\frac{(n-4)(n+2)}{2n}C_2\Big(\frac{n}{2}-1\Big)\;,
\end{equation}
which can be derived from (\ref{Fusscatalanconvol}).
\end{proof}

\subsection{Behavior of the coefficients $r_{n,k}, f_{2,k},b_n$
  in terms of $b_1$}\label{AppendixD}

\BPHZboundsgnkfk*
\begin{proof}
  The proof is done by induction in $N=n+2k$; we go up in $N$
  and at a fixed value of $N$ we go up in $k$. For $k\leq 1$,
  we use the bounds in Lemma \ref{exactexpressionsgn01} to obtain successively
   \begin{equation}\label{reuse}
     \vert r_{n,0}\vert\leq \dfrac{K^{\frac{n}{2}-1}}{30^{\frac{n}{2}-1}}
     \dfrac{1}{n-1}\binom{3(\frac{n}{2}-1)}{\frac{n}{2}-1}
     \leq \Big(\frac{4K}{15}\Big)^{\frac{n}{2}-1}\frac{1}{n-1}
     \leq \frac{4}{9} K^{\frac{n}{2}-1}\Big(\frac{n-4}{2}\Big)!\;.
\end{equation}

\begin{equation}
\begin{split}
  \vert r_{n,1}\vert &\leq \dfrac{K^{\frac{n}{2}-1}}{30^{\frac{n}{2}-1}}
  \dfrac{1}{n-1}\binom{3(\frac{n}{2}-1)}{\frac{n}{2}-1}
  \Big(\frac{n-4}{4}+\frac{3n-4}{2}\vert b_1\vert\Big) \\
  &\leq \Big(\frac{4K}{15}\Big)^{\frac{n}{2}-1} K^{\frac{n}{2}-1}
  \Big(\frac{1}{4}+\frac{3K}{2}\Big)\leq \frac{2}{3}
  K^{\frac{n}{2}-1}\Big(\frac{n-2}{2}\Big)!\;.
\end{split}
\end{equation}
For $k\geq 0$ we insert the induction hypothesis
in the r.h.s of (\ref{FE_trivial_g_n,k}) to obtain

\begin{equation}
        \begin{split}
          \vert r_{n,k+2}\vert &\leq
          \dfrac{n-4}{n+2k}\vert r_{n,k+1}\vert +\dfrac{2n}{n+2k}
          \sum_{\nu=0}^{k+1}\vert r_{n,\nu}f_{2,k+1-\nu}\vert
          +\dfrac{n}{n+2k}\sum_{\substack{n_1+n_2=n+2\\
              n_i\geq 4}}\sum_{\nu=0}^{k+2}\vert r_{n_1,\nu}r_{n_2,k+2-\nu}\vert \\
          &+\dfrac{n(n+1)}{n+2k}\vert r_{n+2,k}\vert \\
          &\leq \Big(\frac{3}{2}\Big)^{k}K^{\frac{n}{2}-1}
          \Bigg[\dfrac{2(n-4)}{3(n+2k)}
            \left(k+1+\dfrac{n-4}{2}\right)!+\dfrac{4n K}{3(n+2k)}
            \sum_{\nu=0}^{k+1}\left(\nu+\dfrac{n-4}{2}\right)!\;\vert k-\nu\vert! \\
&+\dfrac{4n}{9(n+2k)}\sum_{\substack{n_1+n_2=n+2\\
                n_i\geq 4}}\sum_{\nu=0}^{k+2}\left(\nu+\dfrac{n_1-4}{2}\right)!
            \left(k+2-\nu+\dfrac{n_2-4}{2}\right)! \\
&+\dfrac{4n(n+1)K}{9(n+2k)}\left(k+\dfrac{n-2}{2}\right)!\Bigg] \\
          &\leq \Big(\frac{3}{2}\Big)^{k}K^{\frac{n}{2}-1}\left(k+2
          +\dfrac{n-4}{2}\right)!
          \Bigg[\dfrac{4(n-4)}{3n^2}+\dfrac{16K}{3n}+\dfrac{4}{9}
            +\dfrac{8(n+1)}{9n^2}K\Bigg] \\
&\leq \Big(\frac{3}{2}\Big)^{k}K^{\frac{n}{2}-1}\left(k+2+\dfrac{n-4}{2}\right)! \;,
        \end{split}
    \end{equation}
where we used
\begin{equation*}
    \sum_{\nu=0}^{n-a} (n-\nu)!\; \nu!\leq 2\; n!\;,\quad a\in\N,\quad a\leq n\;.
\end{equation*}

Now we bound $f_{2,k}$. The bound obviously holds for $k=0$. Then we have
\begin{equation}
  \vert f_{2,1}\vert\leq 3r_{4,0}+\vert f_{2,0}\vert
  (1+\vert f_{2,0}\vert)\leq \frac{17}{15}K\leq\frac{3}{2}K\;.
\end{equation}
Then we have for $k\geq 1$ by inserting the induction
hypothesis in the r.h.s of (\ref{FE_trivial_field1O})
\begin{equation}
\begin{split}
  \vert f_{2,k+1}\vert &\leq\frac{1}{k+1}
  \Big(3 \Big(\frac{3}{2}\Big)^{k-2} K\; k!
  + \Big(\frac{3}{2}\Big)^k K ( k-1) !\;
  +  \Big(\frac{3}{2}\Big)^k K^2 \sum_{\nu=0}^k \vert \nu-1\vert!\;
  \vert k-\nu-1\vert!\Big) \\
  &\leq \Big(\frac{3}{2}\Big)^{k+1}\frac{1}{2}
  \Big(\frac{8}{9} K\; k!+\frac{2}{3}K k!+4K^2\frac{2}{3}k!\Big) \\
    &\leq \Big(\frac{3}{2}\Big)^{k+1} K\; k!\;.
\end{split}
\end{equation}
\end{proof}

\BPHZlemma*
\begin{proof}
  The proof is done by induction in $N=n+2k$, going up in $N$ and at a
  fixed value of $N$, we go up in $k$. For $k\leq1$, we use (\ref{coeffgn0gn1}) to get
    \begin{equation}
      \vert r_{n,0,0}\vert\leq \dfrac{K^{\frac{n}{2}-1}}{30^{\frac{n}{2}-1}}
      \dfrac{1}{n-1}\binom{3(\frac{n}{2}-1)}{\frac{n}{2}-1}
      \leq \Big(\frac{4K}{15}\Big)^{\frac{n}{2}-1}\frac{1}{n-1}
      \leq \frac{1}{4} K^{\frac{n}{2}-1}\Big(\frac{n-4}{2}\Big)!\;,\quad n\geq 4
    \end{equation}
    and
    \begin{equation}
      \vert r_{n,1,0}\vert\leq\dfrac{K^{\frac{n}{2}-1}}
            {30^{\frac{n}{2}-1}}\dfrac{n-4}{4(n-1)}
            \binom{3(\frac{n}{2}-1)}{\frac{n}{2}-1}
            \leq \Big(\frac{4K}{15}\Big)^{\frac{n}{2}-1}\frac{1}{4}
            \leq \frac{1}{4} K^{\frac{n}{2}-1}\Big(\frac{n-2}{2}\Big)!\;,\quad n\geq 4\;.
    \end{equation}
    We have as well
    \begin{align}
     \vert r_{4,1,1}\vert &=4r_{4,0}\leq\frac{2K}{15}\leq \frac{K}{4}\;, \\  
     \vert r_{n,1,1}\vert &\leq\dfrac{K^{\frac{n}{2}-1}}{30^{\frac{n}{2}-1}}
     \dfrac{3n-4}{2(n-1)}\binom{3(\frac{n}{2}-1)}{\frac{n}{2}-1}\leq
     \Big(\frac{4K}{15}\Big)^{\frac{n}{2}-1}\frac{3}{2}\leq
     \frac{1}{4} K^{\frac{n}{2}-1}\Big(\frac{n-2}{2}\Big)!\;,\quad n\geq 6\;.
    \end{align}
We insert the induction hypothesis in the r.h.s of (\ref{FE_trivial_g_n,k})
\begin{itemize}
    \item We treat the cases $k=2$ and $n\geq 4$. We have
    \begin{equation}
        \begin{split}
          \vert r_{n,2,\nu}\vert &\leq \frac{n-4}{n}\vert r_{n,1,\nu}\vert
          +2\sum_{\rho=0}^{1}
          \sum_{\nu'=\max\lbrace\nu-(2-\rho),0\rbrace}^{\min\lbrace \rho,\nu\rbrace}
          \vert r_{n,\rho,\nu'}f_{2,1-\rho,\nu-\nu'}\vert \\
        &+\sum_{\substack{n_1+n_2=n+2\\
              n_i\geq 4}}\sum_{\rho=0}^{2}
          \sum_{\nu'=\max\lbrace \nu-(2-\rho),0\rbrace}^{\min\lbrace\rho,\nu\rbrace}
          \vert r_{n_1,\rho,\nu'}r_{n_2,2-\rho,\nu-\nu'}\vert \\ &+(n+1)\vert r_{n+2,0,\nu}\vert\;.
        \end{split}
    \end{equation}
We use (\ref{coeffgn0gn1}), (\ref{coefff201b1}) and (\ref{coefff22b1}) to get
\begin{itemize}
    \item $\nu=0$:
    \begin{equation}
    \begin{split}
      \vert r_{n,2,0}\vert &\leq  \frac{n-4}{n}
      \vert r_{n,1,0}\vert+2\vert r_{n,0,0}f_{2,1,0}\vert+\sum_{\substack{n_1+n_2=n+2\\
n_i\geq 4}}\sum_{\rho=0}^{2}\vert r_{n_1,\rho,0}r_{n_2,2-\rho,0}\vert \\
&+(n+1)\vert r_{n+2,0,0}\vert \\
      &\leq \frac{1}{4}K^{\frac{n}{2}-1}
      \Big[\Big(\frac{4}{15}\Big)^{\frac{n}{2}-1}
        +8\Big(\frac{4}{15}\Big)^{\frac{n}{2}-1}\frac{3}{n-1}r_{4,0} \\
&+\frac{1}{4}\sum_{\substack{n_1+n_2=n+2\\
            n_i\geq 4}}\sum_{\rho=0}^{2}\left(\rho+\dfrac{n_1-4}{2}\right)!
        \left(2-\rho+\dfrac{n_2-4}{2}\right)!
        +4\Big(\frac{4}{15}\Big)^{\frac{n}{2}}K\Big] \\
      &\leq \frac{1}{4}K^{\frac{n}{2}-1}\Big(\frac{n}{2}\Big)!\;
      \binom{2}{0}\Big[\frac{2}{15}+\frac{2}{15}\frac{2K}{15}
        +\frac{1}{4}+\Big(\frac{4}{15}\Big)^2\frac{2}{15}\Big] \\
&\leq \frac{1}{4}K^{\frac{n}{2}-1}\Big(\frac{n}{2}\Big)!\;\binom{2}{0}\;.
    \end{split}
    \end{equation}
    \item $\nu=1$
\begin{equation}
    \begin{split}
      \vert r_{n,2,1}\vert &\leq \frac{n-4}{4}\vert r_{n,1,1}\vert
      +2\vert r_{n,0,0}\vert+2\vert r_{n,1,0}\vert+\sum_{\substack{n_1+n_2=n+2\\
n_i\geq 4}}\sum_{\rho=0}^{2}\sum_{\nu'=0}^1\vert r_{n_1,\rho,\nu'}r_{n_2,2-\rho,1-\nu'}\vert\\
      &\leq \frac{1}{4}K^{\frac{n}{2}-1}\Big[6\Big(\frac{4}{15}\Big)^{\frac{n}{2}-1}
        +2\Big(\frac{4}{15}\Big)^{\frac{n}{2}-1}\frac{4}{n-1}
        +2\Big(\frac{4}{15}\Big)^{\frac{n}{2}-1} \\
&+\frac{1}{4}\sum_{\substack{n_1+n_2=n+2\\
            n_i\geq 4}}\sum_{\rho=0}^{2}
        \sum_{\nu'=0}^1 \left(\rho+\dfrac{n_1-4}{2}\right)!
        \left(2-\rho+\dfrac{n_2-4}{2}\right)!\;
        \binom{\rho}{\nu'}\binom{2-\rho}{1-\nu'}\Big] \\
      &\leq \frac{1}{4}K^{\frac{n}{2}-1}\Big(\frac{n}{2}\Big)!\;
      \binom{2}{1}\Big[\frac{2}{5}+\frac{2}{3}\frac{4}{15}
        +\frac{1}{2}\frac{4}{15}+\frac{1}{4}\Big]
      \leq \frac{1}{4}K^{\frac{n}{2}-1}\Big(\frac{n}{2}\Big)!\;\binom{2}{1}\;,
    \end{split}
\end{equation}
where we used the Vandermonde formula
\begin{equation}\label{Vandermondeformula}
  \sum_{\nu'=0}^\nu\binom{a}{\nu'}\binom{b}{\nu-\nu'}
  =\binom{a+b}{\nu}\;,\quad \nu,a,b\in\N_0\;, \nu\leq a+b\;.
\end{equation}
\item $\nu=2$: we have first
\begin{equation}
  \vert r_{4,2,2}\vert\leq 2r_{4,0,0}\vert f_{2,1,2}\vert
  +2\vert r_{4,1,1}\vert \vert f_{2,0,1}\vert\leq \frac{1}{3}K\leq\frac{1}{4}K \;2!\;. 
\end{equation}
Then for $n\geq 6$ we have
\begin{equation}
\begin{split}
  \vert r_{n,2,2}\vert &\leq 2\vert r_{n,0,0}\vert+2\vert r_{n,1,1}\vert
  +\sum_{\substack{n_1+n_2=n+2\\
      n_i\geq 4}}\sum_{\rho=0}^{2}
  \sum_{\nu'=0}^2\vert r_{n_1,\rho,\nu'}r_{n_2,2-\rho,2-\nu'}\vert \\
  &\leq \frac{1}{4}K^{\frac{n}{2}-1}\Big(\frac{n}{2}\Big)!\;
  \binom{2}{2}\Big[\frac{4}{3(n-1)}
    \Big(\frac{4}{15}\Big)^{\frac{n}{2}-1}+2\Big(\frac{4}{15}\Big)^{\frac{n}{2}-1}
    +\frac{1}{4}\Big]\\
&\leq \frac{1}{4}K^{\frac{n}{2}-1}\Big(\frac{n}{2}\Big)!\;\binom{2}{2}\;.
\end{split}
\end{equation}
\end{itemize}
\item $n=4$ and $k\geq 1$:
  First we see that $r_{4,1,\nu}$ satisfy the bounds as claimed. The case $k=2$ is already
  treated. For $k=3$, we have using (\ref{coeffgn0gn1}), (\ref{coefff201b1})
  and (\ref{coefff22b1})
\begin{align}
  \vert r_{4,3,0}\vert \leq\frac{4}{3}\Big[r_{4,0,0}\vert f_{2,2,0}\vert+\frac{5}{2}
    \vert r_{6,1,0}\vert\Big]\leq \frac{4}{3}\Big[\frac{K}{30}\frac{K}{20}
    +\frac{5}{2}\frac{1}{2}K^2\frac{1}{300}\Big]\leq\frac{1}{4}K \;3!\;, \\
  \vert r_{4,3,1}\vert\leq \frac{4}{3}\Big[\vert r_{4,1,1} f_{2,1,0}\vert
    +\vert r_{4,2,0} f_{2,0,1}\vert+r_{4,0,0}\vert f_{2,2,1}\vert\Big]
  +\frac{10}{3}\vert r_{6,1,1}\vert \leq  \frac{1}{4}K\; 3!\binom{3}{1}\;, \\
  \vert r_{4,3,2}\vert\leq\frac{4}{3}\Big[r_{4,0,0}\vert f_{2,2,2}\vert
    +\vert r_{4,1,1} f_{2,1,1}\vert+\vert r_{4,2,1}f_{2,0,1}\vert\Big]
  \leq \frac{1}{4}K \; 3!\;\binom{3}{2}\;,\\
  \vert r_{4,3,3}\vert\leq\frac{4}{3}\Big[r_{4,0,0}\vert f_{2,2,3}\vert
    +\vert r_{4,1,1} f_{2,1,2}\vert+\vert r_{4,2,2}f_{2,0,1}\vert\Big]
  \leq \frac{1}{4}K \; 3!\;\binom{3}{3}\;.
\end{align}
Then, for $k\geq 2$ we have, following the proof of Lemma
\ref{BPHZlemmagnkf2kwhole} and (\ref{Vandermondeformula})
\begin{equation}
    \begin{split}
      \vert r_{4,k+2,\nu}\vert &\leq \dfrac{4}{4+2k}\frac{1}{4}
      \Big[3k!+(k+1)!\Big]\binom{k+2}{\nu}+\dfrac{10}{k+2}
      \frac{K^2}{4}(k+1)!\;\binom{k}{\nu} \\
      &\leq \frac{1}{4}K\; (k+2)!\;\binom{k+2}{\nu}
      \Big[\frac{1}{4}+\frac{1}{4}+\frac{5K}{8}\Big]
      \leq\frac{1}{4}K\; (k+2)!\;\binom{k+2}{\nu}\;.
    \end{split}
\end{equation}
\item $n\geq 6$ and $k\geq 1$: We obtain
\begin{equation}
    \begin{split}
      \vert r_{n,k+2,\nu}\vert &\leq\dfrac{n-4}{n+2k}\vert r_{n,k+1,\nu}\vert
      +\dfrac{2n}{n+2k}\sum_{\rho=0}^{k+1}
      \sum_{\nu'=\max\lbrace\nu-(k+2-\rho),0\rbrace}^{\min\lbrace \rho,\nu\rbrace}
      \vert r_{n,\rho,\nu'}f_{2,k+1-\rho,\nu-\nu'}\vert \\
        &+\dfrac{n}{n+2k}\sum_{\substack{n_1+n_2=n+2\\
          n_i\geq 4}}\sum_{\rho=0}^{k+2}
      \sum_{\nu'=\max\lbrace \nu-(k+2-\rho),0\rbrace}^{\min\lbrace\rho,\nu\rbrace}
      \vert r_{n_1,\rho,\nu'}r_{n_2,k+2-\rho,\nu-\nu'}\vert \\
      &+\dfrac{n(n+1)}{n+2k}\vert r_{n+2,k,\nu}\vert \\
      &\leq \frac{1}{4}K^{\frac{n}{2}-1}\Bigg[\dfrac{(n-4)}{(n+2k)}
        \left(k+1+\dfrac{n-4}{2}\right)!\;\binom{k+1}{\nu} \\
        &+\dfrac{2n}{(n+2k)}\sum_{\rho=0}^{k+1}\left(\rho
        +\dfrac{n-4}{2}\right)!\;\vert k-\nu\vert!\;\binom{k+2}{\nu} \\
&+\dfrac{n}{4(n+2k)}\sum_{\substack{n_1+n_2=n+2\\
            n_i\geq 4}}\sum_{\rho=0}^{k+2}\left(\rho+\dfrac{n_1-4}{2}\right)!
        \left(k+2-\rho+\dfrac{n_2-4}{2}\right)!\;\binom{k+2}{\nu} \\
&+\dfrac{n(n+1)K}{(n+2k)}\left(k+\dfrac{n-2}{2}\right)!\;\binom{k}{\nu}\Bigg] \\
      &\leq \frac{K^{\frac{n}{2}-1}}{4}\left(k+2+\dfrac{n-4}{2}\right)!\;\binom{k+2}{\nu} \\
      &\times\Bigg[\dfrac{2(n-4)}{(n+2k)^2}+\dfrac{8n}{(n+2k)^2}+
        \dfrac{n^2}{4(n+2k)^2}+\dfrac{2n(n+1)}{(n+2k)^2}K\Bigg] \\
&\leq\frac{1}{4}K^{\frac{n}{2}-1}\left(k+2+\dfrac{n-4}{2}\right)!\;\binom{k+2}{\nu} \;.
        \end{split}
    \end{equation}
\end{itemize}

For $f_{2,k}$, we proceed by induction in $k$. The bounds are satisfied for
$k\leq 2$. For $k\geq 2$ we have
\begin{equation}
    \begin{split}
      \vert f_{2,k+1,\nu}\vert &\leq\frac{1}{k+1}\Big(\frac{3}{4} K\;
      k!\;\binom{k}{\nu}+ ( k-1) !\;\binom{k+1}{\nu}+
      \sum_{\rho=0}^k \vert \rho-1\vert!\; \vert k-\rho-1\vert!\;\binom{k+2}{\nu}\Big) \\
      &\leq k!\;\binom{k+2}{\nu}\frac{1}{k+1}
      \Big(\frac{3}{4} K+\frac{1}{k}+\frac{4}{k}\Big) \\
    &\leq k!\;\binom{k+2}{\nu}\;.
\end{split}
\end{equation}
\end{proof}

\BPHZbn*
\begin{proof}
  The proof is done by induction in $q\geq 1$. For $q\leq 4$, the bounds
  can be checked by hand. They obviously hold for $q\leq 2$. We have from
  (\ref{bnpolynomialinduction}) and Lemma \ref{BPHZgnkf2kpolynomials2}
    \begin{equation}
    \left\{
     \begin{array}{ll}
    \vert b_{3,0}\vert=\frac{r_{4,0}}{6}\leq\frac{1}{3}\Big(\frac{3}{4}\Big)^1\\
    \vert b_{3,1}\vert\leq\frac{2}{3}r_{4,0}+\frac{1}{18}\leq\frac{1}{3}
    \Big(\frac{3}{4}\Big)^1\; 3\\
    \vert b_{3,2}\vert=\frac{1}{6}\leq \frac{1}{3}\Big(\frac{3}{4}\Big)^1\; 3\\
    \vert b_{3,3}\vert=\frac{1}{9}\leq \frac{1}{3}\Big(\frac{3}{4}\Big)^1
    \end{array}
    \right.\;,\quad\quad
    \left\{
     \begin{array}{ll}
       \vert b_{4,0}\vert\leq\frac{4}{64}+\frac{3r_{4,0}}{128}\leq\frac{1}{4}
       \Big(\frac{3}{4}\Big)^2\\
    \vert b_{4,1}\vert\leq\frac{16+1+8}{64}\leq\frac{1}{4}\Big(\frac{3}{4}\Big)^2\; 4\\
    \vert b_{4,2}\vert\leq\frac{24+4}{64}\leq \frac{1}{4}\Big(\frac{3}{4}\Big)^2\; 6\\
    \vert b_{4,3}\vert\leq\frac{16}{64}\leq \frac{1}{4}\Big(\frac{3}{4}\Big)^2\;4\\
    \vert b_{4,4}\vert\leq\frac{4}{64}\leq\frac{1}{4}\Big(\frac{3}{4}\Big)^2
    \end{array}
    \right.\;.     
    \end{equation}

    We insert the induction hypothesis in the r.h.s. of (\ref{b_n_induction}).
    For $q\geq 4$ we use Lemma \ref{BPHZgnkf2kpolynomials2} to obtain
\begin{equation}\label{BPHZtoprove}
  \frac{(q-1)!}{(q+1)^{q-1}}\leq \Big(\dfrac{3}{4}\Big)^q\frac{1}{5}\;,
  \quad q\geq 4\;.
\end{equation}
We also have
\begin{equation}\label{BPHZsumbn}
\begin{split}
  \sum_{\rho=2}^{q+1}\vert b_{\lbrace\frac{q+1}{\rho},\nu\rbrace}\vert\frac{1}{\rho^q}
  &\leq\binom{q+1}{\nu}\frac{1}{q+1}\Big[\sum_{\substack{\rho=2}}^q\frac{1}{\rho^{q-1}}
    +\frac{1}{(q+1)^{q-1}}\Big] \\
   &\leq \binom{q+1}{\nu}\frac{1}{q+1}\Big[\zeta(q-1)-1+\frac{1}{(q+1)^{q-1}}\Big]   \\
   &\leq \binom{q+1}{\nu}\frac{1}{q+1}\Big[\dfrac{4}{2^{q}}+\frac{1}{(q+1)^{q-1}}\Big]\;.
\end{split}
\end{equation}
Therefore from (\ref{BPHZtoprove}) and (\ref{BPHZsumbn}) we have
\begin{equation}
    \begin{split}
      \vert b_{q+1,\nu}\vert&\leq \frac{1}{q+1}\Big(\dfrac{3}{4}\Big)^{q-1}
      \binom{q+1}{\nu}\frac{3}{20}+\frac{1}{q+1}\binom{q+1}{\nu}
      \Big[\dfrac{4}{2^{q}}+\frac{1}{(q+1)^{q-1}}\Big] \\ 
        &\leq \frac{1}{q+1}\Big(\dfrac{3}{4}\Big)^{q-1}\binom{q+1}{\nu}\;,
    \end{split}
\end{equation}
where we used
\begin{equation}
  \frac{3}{20}+3\Big(\frac{2}{3}\Big)^q+\Big(\frac{4}{3(q+1)}\Big)^{q-1}
  \leq 1\;,\quad q\geq 4\;.
\end{equation}
\end{proof}

\printbibliography

\end{document}